\documentclass[twocolumn]{aastex701}
\usepackage{amsmath}
\allowdisplaybreaks
\usepackage{graphicx}
\usepackage{mathtools}
\usepackage{enumerate}
\usepackage{tabularx}
\usepackage{array}
\usepackage{float}
\usepackage{xspace}
\usepackage{booktabs}   
\usepackage{tabularx}   
\usepackage[version=4]{mhchem} 
\graphicspath{{./}{figures/}}

\begin{document}
\shorttitle{Alkali Metallicity and Mineral Clouds on Jupiter}
\title{Alkali Metallicity, Mineral Clouds, and Deep Atmospheric Variability on Jupiter}

\author[orcid=0000-0002-8706-6963,sname='North America']{Xi Zhang}
\affiliation{Department of Earth and Planetary Sciences, University of California, Santa Cruz, CA 95064, USA}
\email[show]{xiz@ucsc.edu} 

\author[orcid=0000-0002-6666-5457,sname='North America']{Jiheng Hu}
\affiliation{Department of Climate and Space Sciences and Engineering, University of Michigan, Ann Arbor, MI 48109 USA}
\email{jihenghu@umich.edu} 

\author[orcid=0000-0002-8280-3119,sname='North America']{Cheng Li}
\affiliation{Department of Climate and Space Sciences and Engineering, University of Michigan, Ann Arbor, MI 48109 USA}
\email{chengli@umich.edu}

\author[orcid=,sname='North America']{Quentin Williams}
\affiliation{Department of Earth and Planetary Sciences, University of California, Santa Cruz, CA 95064, USA}
\email{qwilliam@ucsc.edu}


\begin{abstract}

The bulk elemental abundances of Jupiter provide critical insights into its formation history and interior structure. Recent observations by the Juno Microwave Radiometer (MWR) reveal a deep Jovian atmosphere significantly depleted in electrons, implying an alkali metal (Na, K) abundance of $10^{-1}$--$10^{-5}\times$ solar. This depletion stands in sharp contrast to the supersolar volatile enrichments measured by the Galileo probe. We propose that this apparent depletion arises from mineral cloud-induced processes deep in the atmosphere. We explore two physical mechanisms using thermochemical and microphysical modeling. In the ``chemical sequestration" scenario, vigorous vertical mixing lofts deep refractory condensates (e.g., spinel) into the 1000--2000 bar region, where they react to form alkali feldspars (albite) and feldspathoids (leucite), efficiently sequestering gaseous Na and K. In the ``dust-catalyzed recombination" scenario, the bulk alkali inventory remains gaseous, but the free electron density is suppressed by dust--plasma interactions. Thermally emitted alkali ions from the surfaces of micron-sized iron and silicate grains significantly increase the cation density, driving rapid recombination of free electrons. Both mechanisms allow for a bulk solar or even supersolar alkali inventory while suppressing the electron density to match Juno observations. Analyzing an extended dataset of MWR observations with 61 perijoves, we detect spatial variability in the deep atmosphere that suggests modulation by mineral clouds. Our findings challenge the traditional rainout framework, unveiling a deep ``mineralogical zone" in Jupiter shaped by dynamics and heterogeneous chemistry, resembling the photospheres of hot exoplanets and brown dwarfs.

\end{abstract}


\section{Introduction} \label{sec:intro}
The bulk elemental abundances of Jupiter impose strong constraints on the processes that shaped both the planet's formation and the evolution of the broader solar system (e.g., \citealt{lunineOriginJupiter2004,mousisJupitersFormationVicinity2019,guillotGiantPlanetsInsideOut2023}). Observations by the Galileo probe mass spectrometer revealed that volatile species such as carbon (C), nitrogen (N), sulfur (S), phosphine (P), and the noble gases (except neon) are supersolar in abundance \citep{wongUpdatedGalileoProbe2004}. The oxygen abundance, however, has remained elusive. Galileo probe entered a 6.57\textdegree N ``hotspot" and measured a depleted water abundance of $0.46\times$ solar, while recent Juno measurements suggest an equatorial water abundance of about $4.9\times$ solar, with a broad range of 1.5--8.3$\times$ solar \citep{liSuperadiabaticTemperatureGradient2024}. 

The refractory elements, such as sodium (Na), potassium (K), iron (Fe), silicon (Si), magnesium (Mg), and aluminum (Al) were not constrained, as they are expected to condense into mineral clouds well below the 100-bar level (e.g., \citealt{lewisObservabilitySpectroscopicallyActive1969,fegleyChemicalModelsDeep1994}). A major recent breakthrough came from the Microwave Radiometer (MWR) onboard NASA's Juno orbiter. MWR monitors Jupiter's thermal radiance at six wavelengths from 0.6 to 22 GHz \citep{janssenMWRMicrowaveRadiometer2017}. The 0.6 GHz channel (50 cm) can probe pressures as deep as roughly 1000--2000 bar. This region lies below the condensation levels of several salt clouds, such as NaCl at around 400 bar and KCl at around 700 bar \citep{bhattacharyaHighlyDepletedAlkali2023}, making it possible for MWR to indirectly constrain the abundances of alkali metals (Na and K) by measuring the electron abundance, which is the dominant source of free-free opacity at 0.6 GHz near 1000 bar \citep{bhattacharyaHighlyDepletedAlkali2023}. Under the corresponding temperature--pressure conditions, most free electrons are expected to originate from the thermal ionization of Na and K gases. By constraining the abundances of electrons and alkali metals, MWR offers a unique window into the refractory inventory of Jupiter's deep atmosphere and interior.

Using MWR brightness temperatures and limb-darkening measurements, \cite{bhattacharyaHighlyDepletedAlkali2023} inferred the deep electron abundance. Assuming simple thermochemical equilibrium between thermal ionization and electron recombination with Na and K, and applying the Saha equation, they derived Na and K abundances of only $\sim 10^{-2}$--$10^{-5}\times$ solar -- far lower than the supersolar volatile enrichments measured by Galileo probe. \cite{aglyamovAlkaliMetalDepletion2025} revisited the electron chemistry in the deep atmosphere beyond the Saha equation, noting that electrons can also recombine with neutral gas-phase species to form anions such as \ce{Cl^-} and \ce{HS^-}, which they identified as the dominant negative charge carriers. With these processes included, their calculations showed that the MWR 0.6 GHz channel data can be matched with an alkali abundance on the order of $0.1\times$ solar. While this value is higher than the \cite{bhattacharyaHighlyDepletedAlkali2023} estimate, it remains significantly depleted relative to volatile species such as carbon and nitrogen.

In this study, we attempt to resolve this puzzle by exploring the chemical and cloud processes operating in the deep atmosphere. We find that two plausible scenarios can reproduce the MWR 0.6 GHz channel observations while still allowing for a solar bulk abundance of alkali metals in Jupiter's interior.

In the ``chemical sequestration" scenario, strong vertical mixing allows alkali metals to form mineral clouds below kilobar, such as feldspar (\ce{NaAlSi3O8}) and leucite (\ce{KAlSi2O6}), leading to substantial depletion of gaseous Na and K in the 1000--2000 bar region. In the ``dust-catalyzed recombination" scenario, the alkali metals are not sequestered into the mineral clouds below 2000 bar, but the free electron density is efficiently suppressed via dust--plasma interactions. In this regime, the iron and silicate grains act as a catalyst for charge removal: by thermally emitting alkali ions, the grains enhance the bulk cation density, driving the rapid depletion of electrons through gas-phase recombination.

Below, we first introduce the MWR data and our Jupiter model setup in Section \ref{sec:data}, followed by the chemical sequestration scenario in Section \ref{sec:cloud_model} and the dust-catalyzed recombination scenario in Section \ref{sec:dust_plasma}. In Section \ref{sec:variability}, we discuss possible observational signatures from Juno to differentiate our proposed scenarios from previous studies suggesting depleted alkali metals. We conclude by discussing the broader implications of this study in Section \ref{sec:conclusion}.

\begin{figure*}
  \centering \includegraphics[width=0.99\textwidth]{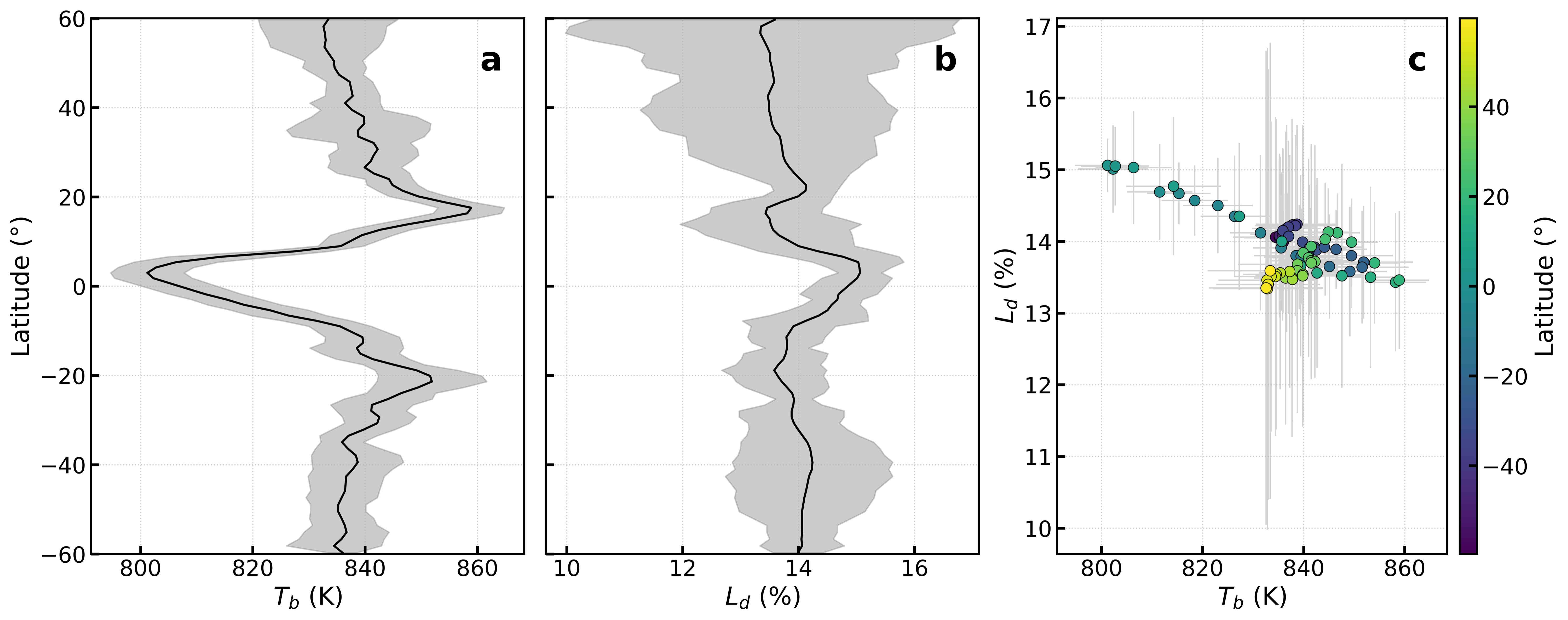}
  \caption{The MWR 0.6 GHz channel observations of Jupiter: \textbf{(a)} nadir-view brightness temperature ($T_b$) and \textbf{(b)} limb darkening ($L_d$) as a function of planetocentric latitude. The data were averaged over the first 12 perijoves; uncertainties are shaded in gray. \textbf{(c)} The correlation between nadir-view $T_b$ and $L_d$, color coded by latitude.} 
  \label{fig:tbld}
\end{figure*}

\section{Juno Observations and Atmospheric Structure} \label{sec:data}

\subsection{Juno MWR data} \label{subsec:mwr_data}
In this study, we focus on Juno MWR data from the first 61 perijove passes, with particular emphasis on the first 12 perijoves where the latitudinal coverage is most complete and the calibration is most robust for limb-darkening analysis.

The raw antenna temperatures ($T_A$) first underwent a series of quality-assurance checks, including flagging and removing non-atmospheric emissions. Following procedures discussed in \citep{oyafusoAngularDependenceSpatial2020,zhangResidualStudyTesting2020}, we removed contamination from lightning, synchrotron radiation, and auroral emissions. Additionally, we flagged and masked data containing large discrete features, such as the Great Red Spot, to ensure the derived profiles represent the background zonal mean atmosphere. 

To derive the pencil-beam atmospheric brightness temperature ($T_b$) from the antenna measurements, a regularized least square approach is used \citep{oyafusoAngularDependenceSpatial2020}. This process models the emission-angle dependence of the brightness temperature, which has been shown to be adequately described by the following form:
\begin{equation}
    T_b(\mu) = \xi(\mu) [a + b(1-\mu) + c(1-\mu)^2],
    \label{eq:limb_darkening_model}
\end{equation}
where $\mu$ is the cosine of the emission angle $\theta$, and the coefficients $a$ (nadir brightness), $b$, and $c$ are determined at 0.5-degree resolution latitude grids. The correction factor $\xi(\mu)$ is equal to 1 for emission angles less than 45$^\circ$, and it is empirically determined based on atmospheric modeling for larger emission angles \citep{oyafusoAngularDependenceSpatial2020}. The limb darkening ($L_d$) is subsequently defined as the percentage difference between the brightness temperature at a $45^\circ$ emission angle and the nadir view: $L_d = (T_b(0^\circ) - T_b(45^\circ))/T_b(0^\circ) \times 100\%$. 

Finally, we corrected the nadir $T_b$ for variations in local gravity due to Jupiter's oblateness. Since a larger gravitational acceleration (e.g., at the poles) yields a smaller photon path length for a given pressure interval, the observed brightness temperature varies with latitude even for a homogeneous atmosphere. We therefore scaled all measurements to an equivalent equatorial gravitational acceleration to isolate the intrinsic atmospheric variability. See more data-processing details in \cite{liSuperadiabaticTemperatureGradient2024}.

The averaged latitudinal profiles of $T_b$ and $L_d$ from perijove 1 to perijove 12 and between $\pm 60$\textdegree{} for the 0.6 GHz channel are presented in Figure \ref{fig:tbld}, together with the correlation plot of nadir-view $T_b$ versus $L_d$. Strong latitudinal variations are evident from the equator to the midlatitudes: $T_b$ is lowest at the equator, around 800 K, and rises to roughly 840–850 K toward the midlatitudes. The $L_d$ is approximately 14\% across all latitudes but exhibits a peak at the equator, showing an inverse trend relative to the nadir-view $T_b$, though with larger uncertainties. 

\subsection{Deep Atmospheric Structure} \label{subsec:deep_tp}
\begin{figure*}
   \centering  \includegraphics[width=0.9\textwidth]{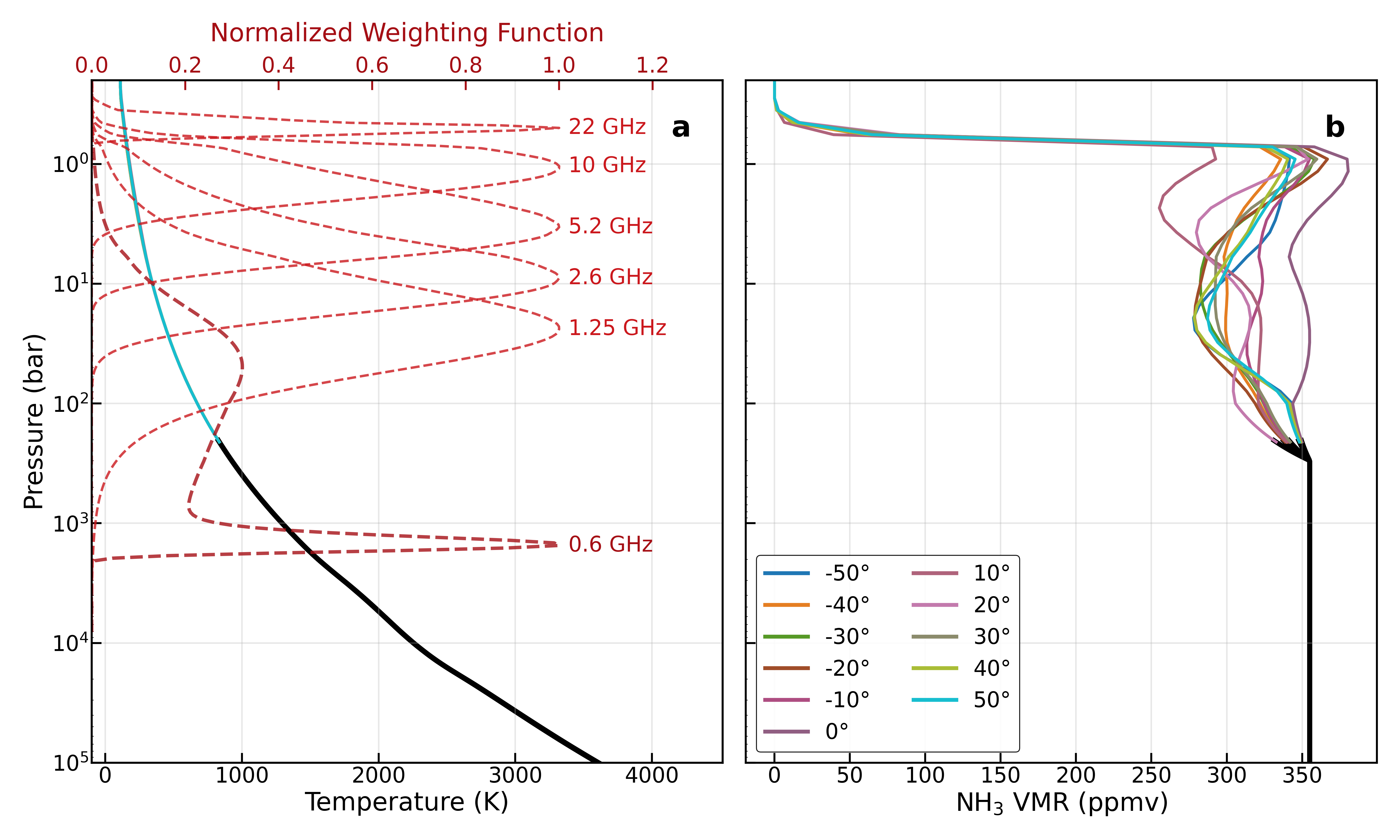} 
  \caption{Atmospheric structure of Jupiter from 1 to $10^5$ bar and the Juno MWR sensitivity. \textbf{(a)} Temperature profiles at different latitudes (solid lines, overlapping with each other) and the normalized weighting functions of the six MWR channels (red dashed lines). The 0.6 GHz channel probes the deepest levels, peaking between 1000 and 2000 bar. The uncertainty in the deep temperature profile is estimated to be about 10 K at 2000 bar due to propagated retrieval errors from the upper atmosphere. \textbf{(b)} Vertical profiles of ammonia volume mixing ratio at different latitudes retrieved from Juno. The deep atmosphere ($>100$ bar), highlighted with thick black lines, is not constrained by MWR channels from 1.25 to 22 GHz and is assumed to follow an adiabat with vertically constant composition of ammonia and water.} 
  \label{fig:TP}
\end{figure*}
Although the 0.6 GHz channel is primarily sensitive to the deep atmosphere, its weighting function has two broad peaks (Figure \ref{fig:TP}): a secondary peak located around 80 bar and a primary peak near 1000--2000 bar. Consequently, the upper atmosphere still contributes to the signal and can shape the latitudinal signatures observed in the 0.6 GHz channel. In fact, most of the observed latitudinal variations may originate from the upper atmosphere. Explaining the latitudinal profiles of $T_b$ and $L_d$ therefore requires knowledge of the latitude--vertical distributions of the upper- atmospheric temperature and opacity from \ce{NH3} and \ce{H2O}.

We constrain the atmosphere above 200 bar using data from MWR channels 2--6 (1.2--22 GHz), as indicated by their weighting functions (Figure \ref{fig:TP}a). Using the two-stage differential fitting method \citep{liLonglastingDeepEffect2023}, \citet{liDistributionAmmoniaJupiter2017,liWaterAbundanceJupiters2020} retrieved the vertical profiles of ammonia volume mixing ratio (VMR) down to approximately 100 bar (Figure \ref{fig:TP}b). Additionally, we adopt the equatorial water abundance derived from the observed superadiabatic profile at a few bars \citep{liSuperadiabaticTemperatureGradient2024} and apply this profile across all latitudes. While recent theoretical work suggests deep water abundance may vary latitudinally due to rotation \citep{geNonuniformWaterDistribution2025}, the opacity of water vapor at 0.6 GHz is negligible compared to that of ammonia and free electrons \citep{bhattacharyaHighlyDepletedAlkali2023}, justifying this assumption for our analysis.

Below 200 bar, where MWR channels 2--6 lose sensitivity, we assume a horizontally homogeneous deep atmosphere as a baseline. This assumption will be revisited in Section \ref{sec:variability}. We adopt vertically constant volume mixing ratios of 355 ppm for ammonia and 2573 ppm for water \citep{liSuperadiabaticTemperatureGradient2024}. We follow \cite{bhattacharyaHighlyDepletedAlkali2023} and \cite{aglyamovAlkaliMetalDepletion2025} to assume an adiabatic temperature profile in the deep atmosphere. We used a non-ideal equation of state (EOS) for a hydrogen--helium--metal (H--He--Z) mixture. We adopt the H/He EOS from \cite{chabrierNewEquationState2019}, including non-ideal mixing corrections to the entropy and density for arbitrary helium fractions following \cite{howardAccountingNonidealMixing2023}. The EOS for heavy elements (represented by water) is approximated using the AQUA table \citep{haldemannAQUACollectionH2O2020}. We adopt $Y=0.23$ and $Z=0.03$ for Jupiter and numerically derive the adiabatic gradient from the mixture entropy table, $S(P, T, Y, Z)$, using the approach and scripts detailed in \cite{tejadaarevaloEquationsStateThermodynamics2024}. The full atmospheric profiles of temperature and \ce{NH3} are presented in Figure \ref{fig:TP}. 

Given the temperature structure and assumed elemental metallicity, we run chemical models that include cloud formation in the deep atmosphere to determine the alkali metal gas abundances, cloud distributions, and the resulting free electron profile. This electron profile, together with the \ce{NH3} and \ce{H2O} distributions, is then passed through the Juno MWR radiance model to compute the nadir-view $T_b$ and $L_d$ for each latitude for comparison with the Juno MWR 0.6 GHz channel data (Figure \ref{fig:tbld}). 

\begin{figure*}
  \centering \includegraphics[width=0.9\textwidth]{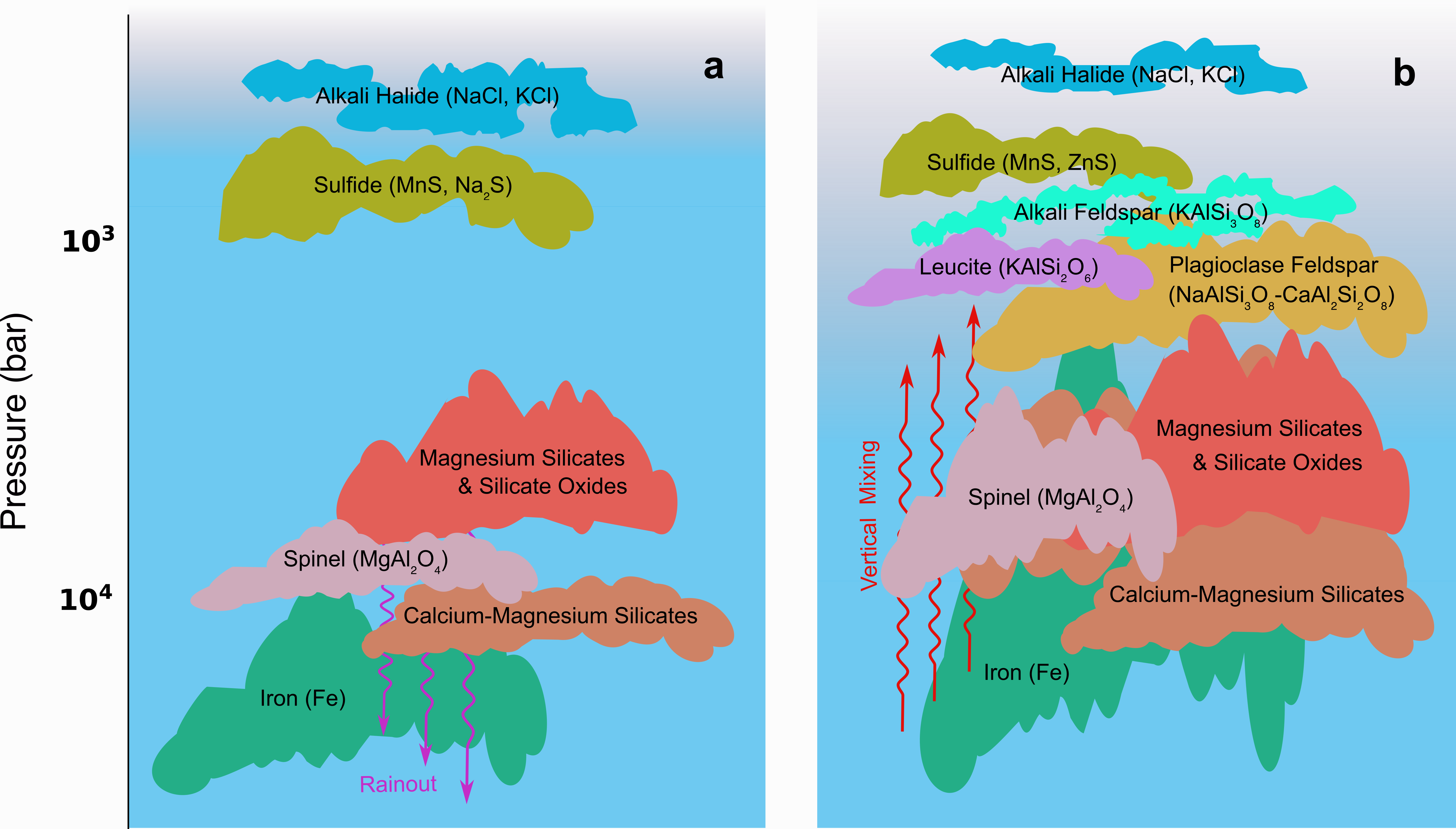} 
  \caption{Schematic view of cloud distributions in the deep atmosphere of Jupiter under different physical assumptions. \textbf{(a)} The rainout scenario: clouds precipitate out of the system immediately upon formation. This permanently restricts refractory elements (Al, Fe, Ca, Mg) at depths $P > 2000$ bar, leaving the upper atmosphere depleted. \textbf{(b)} The equilibrium scenario: vigorous vertical mixing transports deep condensates upward, allowing them to interact chemically with the gas at lower pressures. This facilitates the formation of complex minerals like feldspars and leucite in the sensitivity window of the MWR 0.6 GHz channel (1000--2000 bar).} \label{fig:cloud_schematic}
\end{figure*}

We simulate the top-of-atmosphere microwave radiances using the High-performance Atmospheric Radiation Package (HARP) \citep{liHighperformanceAtmosphericRadiation2018}. Our radiative transfer model was detailed in \cite{bhattacharyaHighlyDepletedAlkali2023}, where the complex refractive index for microwaves propagating through an unmagnetized cold plasma is calculated using the Appleton--Hartree equation \citep{helliwellWhistlersRelatedIonospheric2014}. We introduce two specific modifications to this baseline model for the present study. 

First, \cite{bhattacharyaHighlyDepletedAlkali2023} and \cite{aglyamovAlkaliMetalDepletion2025} focused mainly on interpreting the $T_b$--$L_d$ correlation (Figure \ref{fig:tbld}c) rather than the explicit latitudinal variations. They adopted an approach that scaled the \ce{NH3} vertical profile to account for latitudinal variations, introducing additional uncertainty. In this work, we mitigate this uncertainty by directly using the retrieved temperature and \ce{NH3} distributions from Juno above 200 bar. The retrieved temperature uncertainty is typically $\sim$2–5 K \citep{liSuperadiabaticTemperatureGradient2024}. Because the deeper atmosphere is modeled as an adiabat anchored at this level, this uncertainty propagates to greater depths ($\sim$1000–2000 bar), reaching approximately 10 K. While this temperature uncertainty does propagate into the simulated MWR observations, it is likely smaller than the uncertainty arising from the \ce{NH3} opacity discussed below.

Second, the \ce{NH3} opacity is not measured beyond 500 K (corresponding to about 40 bar on Jupiter) but extrapolated to higher temperatures and pressures using the Ben-Reuven line shape \citep{hanleyNewModelHydrogen2009,bellottiLaboratoryMeasurements5202016,bellottiCorrigendumLaboratoryMeasurements2017}. Fortunately, the sensitivity study in \cite{bhattacharyaHighlyDepletedAlkali2023} found that scaling the \ce{NH3} opacity mainly affects the $T_b$ at the 0.6 GHz channel but not the $L_d$ (see their Figure 4). This implies that $L_d$ provides a stricter constraint on the electron abundance. In this study, we largely follow the opacity formulation of \cite{bhattacharyaHighlyDepletedAlkali2023} but reduce the \ce{NH3} opacity at 0.6 GHz by 10\% to achieve a better fit to $T_b$, which remains within acceptable uncertainty limits.

\section{The Chemical Sequestration Scenario} \label{sec:cloud_model}

In this scenario, we explore the possibility that mineral clouds containing alkali metals form below the 1000 bar level, which was not considered in the traditional picture of Jupiter's deep atmosphere. The atmospheric chemistry for both volatile and refractory species on Jupiter goes back to \cite{lewisObservabilitySpectroscopicallyActive1969} assuming thermochemical equilibrium. Vapor is supplied from the deep interior, and the chemical state at each pressure level is calculated by minimizing the Gibbs free energy of the system at the local temperature. This framework was later improved with more complete thermodynamic data and with vertical mixing included for several volatile species such as CO and HCN to explain their observed disequilibrium abundances (e.g.,  
\citealt{barshayChemicalStructureDeep1978, fegleyChemicalModelsDeep1994, loddersAlkaliElementChemistry1999, loddersChemistryLowMass2006, visscherAtmosphericChemistryGiant2010}).

\begin{figure*}
  \centering \includegraphics[width=0.99\textwidth]{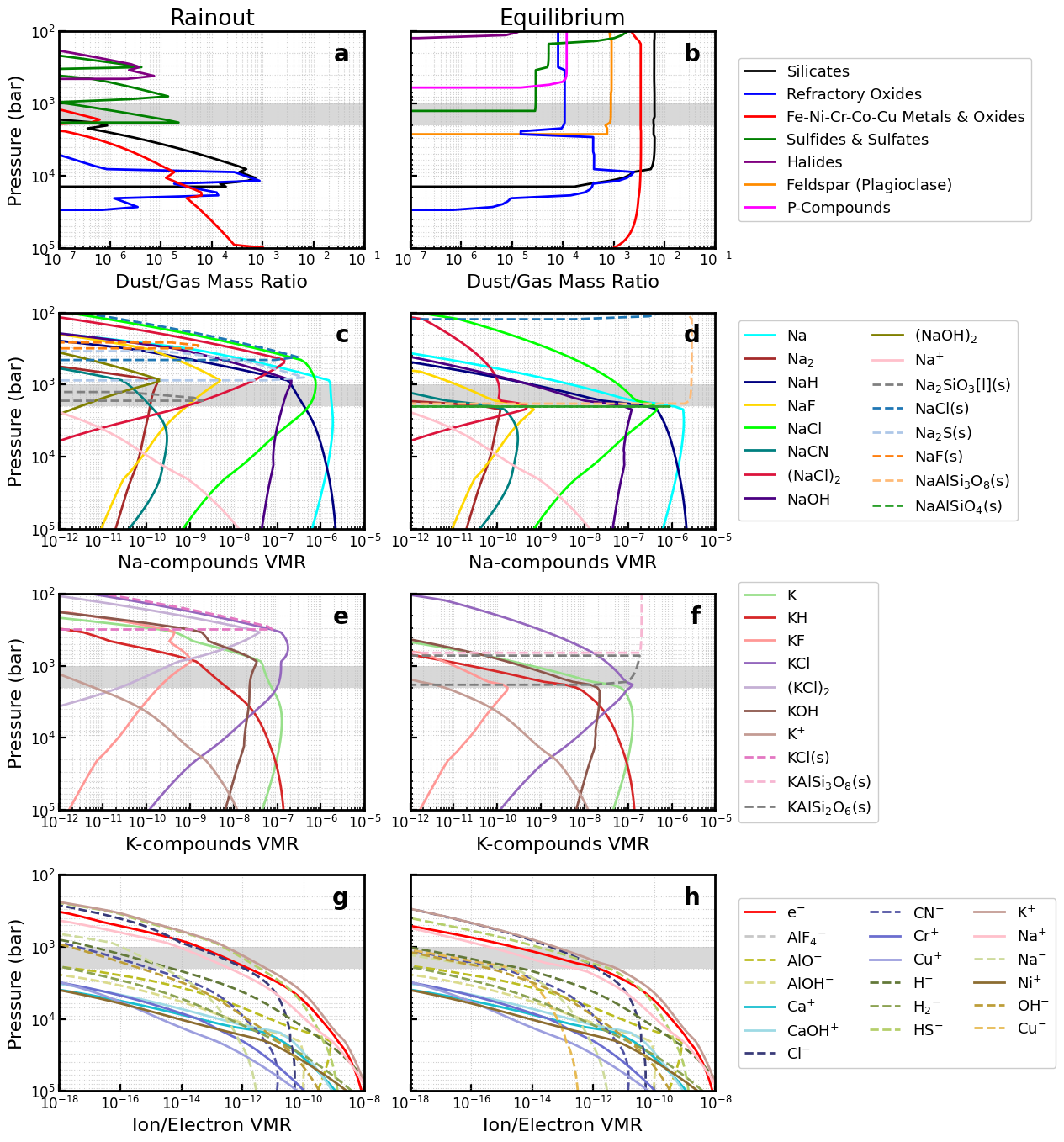} 
  \caption{Vertical profiles of cloud mass, alkali metal speciation, and charge carriers in Jupiter's deep atmosphere under two different transport regimes. \textbf{Left column (rainout):} condensates precipitate immediately upon formation. In this scenario, refractory elements (\ce{Si}, \ce{Mg}, \ce{Al}, \ce{Ca}, \ce{Fe}) are removed deep in the atmosphere ($>2000$ bar), preventing them from reacting with alkali metals at higher altitudes. Consequently, \ce{Na} and \ce{K} remain in their atomic gas phase throughout the 1000--2000 bar region. This results in a high free electron abundance ($e^-$, red dashed line in bottom panel) due to thermal ionization. \textbf{Right column (equilibrium):} vigorous vertical mixing lofts deep refractory condensates, allowing them to react with the gas phase. Here, \ce{K} is efficiently sequestered into leucite (\ce{KAlSi2O6}) and \ce{Na} into albite (plagioclase feldspar, \ce{NaAlSi3O8}) below the kilobar level. The sequestration of alkali metals in the Equilibrium scenario depletes the gas-phase ionization sources, lowering the deep electron density by over an order of magnitude compared to the rainout case. The gray shaded region indicates the approximate pressure range probed by the Juno MWR 0.6 GHz channel.}
  \label{fig:nak}
\end{figure*}

\subsection{The Rainout Chemistry} \label{subsec:rainout}

A key assumption of refractory condensate formation in these traditional giant-planet atmosphere models is ``rainout": whenever a condensate forms, the cloud material is assumed to precipitate and be immediately removed from the system. The remaining vapor continues to mix upward, but it is stripped of the elements locked in the condensate. Consequently, the vertical chemical profile follows a strict condensation sequence where high-temperature condensates form deep clouds and lower-temperature condensates form in the upper atmosphere, with effectively no vertical interaction between the deep cloud layers and the overlying gas. This is similar to the pseudo-adiabatic process known in atmospheric literature. 

A schematic picture of the rainout scenario in Jupiter's deep atmosphere is illustrated in Figure \ref{fig:cloud_schematic}a. Moving from the deep interior ($P > 10^4$ bar) upward, the gas passes through distinct condensation levels. The first major cloud deck to form consists of refractory metals and oxides, primarily iron (Fe) and possibly corundum (\ce{Al2O3}), spinel (\ce{MgAl2O4}), and calcium aluminates, depending on the elemental ratios. Above the refractory oxide layers, magnesium silicates condense at around $10^4$ bar, including enstatite (\ce{MgSiO3}) and forsterite (\ce{Mg2SiO4}). The crucial consequence of the rainout assumption is the permanent sequestration of refractory lithophiles below the 2000 bar level, specifically \ce{Si}, \ce{Mg}, \ce{Al}, \ce{Ca}, and \ce{Fe}. By the time the gas reaches pressures of 2000 bar (the region probed by the MWR 0.6 GHz channel), it is depleted of the reactants necessary to form complex minerals. Sulfides and sulfates form in the kilobar region, and halides with salt clouds such as \ce{NaCl} and \ce{KCl} form above 700 bar. 

We calculate the gas and solid distributions in the rainout scenario using an open-source thermochemical code \texttt{GGchem} \citep{woitkeEquilibriumChemistry1002018}, the same model used in \cite{aglyamovAlkaliMetalDepletion2025}. The model considers a comprehensive database of hundreds of gas-phase species as well as solid and liquid condensates. For Jupiter, we adopt the elemental metallicity following observational constraints where available for C (3.92× solar), N (2.1×), S (3.01×), and P (3.75×) based on the solar abundances in \cite{asplundChemicalMakeupSun2021}. A 2.7× solar enrichment is assumed for other refractory elements (including Mg, Si, Fe, Al, Ca, Ti, Cl, Li, F, V, Cr, Mn, Ni, Zn, Co, Cu, and Ge). Since oxygen on Jupiter remains poorly constrained, here we adopt the value (2.27×) from \cite{liWaterAbundanceJupiters2020}, and modest variations in the oxygen abundance do not significantly affect the results \citep{aglyamovAlkaliMetalDepletion2025}. Na and K abundances are set as solar; the deep molar fraction of Na is about 1 ppm, and K is about 0.1 ppm.

In the rainout scenario, the simulated vertical distributions of gas-phase species and condensates from \texttt{GGchem} are broadly consistent with those in \cite{aglyamovAlkaliMetalDepletion2025}. We grouped more than a hundred condensates into several chemical groups (Table \ref{table:chem}). Their distributions are shown in Figure \ref{fig:nak}a, consistent with the schematic picture in Figure \ref{fig:cloud_schematic}a. We also plot the species containing several important elements, including Na and K compounds, in Figure \ref{fig:nak}, and those containing Si, Mg, Al, Ca, and Fe in Figure \ref{fig:other_species}. 

Because Al and Ca are removed deep in the rainout chemistry, the alkali metals (Na and K) have no reaction partners in the 1000--2000 bar region. They remain in the gas phase as neutral atoms until they reach much lower temperatures above 700 bar where they eventually condense as simple salts (\ce{NaCl}, \ce{KCl}) or sulfides (\ce{Na2S}) (Figure \ref{fig:nak}c and e). The persistence of gaseous alkali metals in the 1000--2000 bar window results in a high free electron abundance via thermal ionization (Figure \ref{fig:nak}g), with a volume mixing ratio greater than $10^{-14}$ at around 1000 bar. Consequently, with solar abundance of Na and K, the electron opacity produces much smaller $T_b$ and $L_d$ than the MWR observations (Figure \ref{fig:cloud_model}). If we reduce the Na and K to 0.1× solar in the model, the rainout chemistry can explain the MWR data well, confirming the results from \cite{aglyamovAlkaliMetalDepletion2025}.

\subsection{The Equilibrium Chemistry} \label{subsec:eqchem}

The rainout scenario assumes that condensates are only affected by gravity but not by atmospheric dynamics. Now we explore an alternative scenario in which vigorous vertical mixing transports condensates in the atmosphere. Vertical mixing is known to play a crucial role in shaping cloud distributions in planetary atmospheres (e.g., \citealt{rossowCloudMicrophysicsAnalysis1978,lunineEvolutionInfraredSpectra1986,ackermanPrecipitatingCondensationClouds2001, hellingExoplanetClouds2019, zhangAtmosphericRegimesTrends2020, gaoAerosolsExoplanetAtmospheres2021}). If atmospheric mixing overcomes the gravitational sedimentation of cloud particles, the deep, high-temperature condensates can be lofted upward, where they react with upper lower-temperature gases to form new minerals not present in the rainout scenario.

We consider a physical limit in which all formed condensates remain within the system and are directly mixed upward to interact with upper-layer gases. In this scenario, precipitation does not occur, and all atmospheric layers maintain uniform elemental abundances, driving thermochemistry solely based on local pressure and temperature conditions. This is referred to as the equilibrium scenario, similar to a reversible-moist-adiabatic process in the atmospheric literature.

The calculation of the equilibrium chemistry via \texttt{GGchem} reveals a fundamentally different cloud structure (Figure \ref{fig:cloud_schematic}b). We find much more abundant cloud species in the upper layers because we do not remove the deep, high-temperature condensates from the system. More importantly, the presence of lofted aluminum and calcium in the 1000--2000 bar region permits the formation of complex aluminosilicates, specifically the feldspar group minerals. The key species formed are alkali feldspars, such as albite (\ce{NaAlSi3O8}), and feldspathoids, such as leucite (\ce{KAlSi2O6}).

\begin{figure*}
  \centering \includegraphics[width=0.99\textwidth]{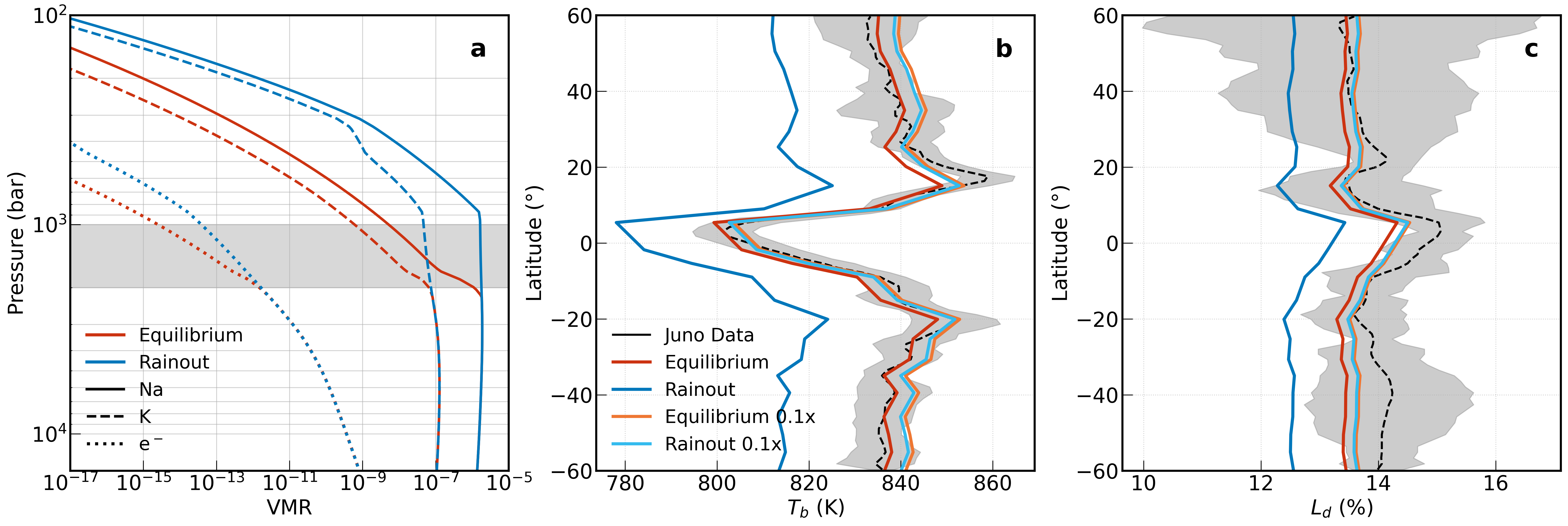} 
  \caption{Comparison of modeled deep atmospheric composition and MWR observables against Juno data. 
  \textbf{(a)} Vertical profiles of simulated gas-phase sodium (solid lines), potassium (dashed lines), and free electrons (dotted lines) in different cases. The blue lines represent the Rainout scenario, where alkalis remain gaseous, leading to high electron densities. The red lines represent the equilibrium scenario, where alkalis are sequestered into feldspars and leucite, depleting the electron density in the MWR sensitive region (gray shaded band, 1000--2000 bar). 
  \textbf{(b)} Nadir-view brightness temperature ($T_b$) and \textbf{(c)} limb darkening ($L_d$) at 0.6 GHz as a function of latitude. The black dashed lines and gray shading indicate Juno data and uncertainties. }
  \label{fig:cloud_model}
\end{figure*}

The formation of these minerals acts as a chemical trap for alkali metals. As shown in the vertical profiles in Figure \ref{fig:nak} (right column), gaseous potassium is efficiently sequestered into solid leucite starting near 2000 bar. Sodium is similarly removed into an albite solid solution. The potassium depletion is important because potassium has the lowest ionization potential (4.34 eV) of the abundant metals. By locking potassium into leucite and sodium into feldspar, the reservoir of easily ionizable atoms is depleted, leading to a free electron density that is about an order of magnitude lower than in the rainout case (Figure \ref{fig:nak}h and Figure \ref{fig:cloud_model}a). This depleted electron profile provides a successful explanation for the reduced opacity required to match the Juno MWR brightness temperatures at the 0.6 GHz channel (Figure \ref{fig:cloud_model}).

However, the viability of the chemical sequestration scenario hinges on two key physical factors: the dynamic ability of the atmosphere to loft deep condensates through vertical mixing and the kinetic feasibility of forming complex minerals from these seeds via heterogeneous chemistry. We explore these possibilities below.

\subsection{Vertical Mixing vs. Sedimentation} \label{subsec:kzz}

To assess the dynamical plausibility of vertical cloud mixing, we compare the timescale of vertical mixing against gravitational settling. We adopt the internal heat flux for Jupiter, $F_{\mathrm{int}} = 7.485~\mathrm{W~m^{-2}}$ \citep{liLessAbsorbedSolar2018}. In the deep, optically thick atmosphere, the temperature profile follows a dry adiabat. With an adiabatic index $\nabla_{\mathrm{ad}} \approx 0.28$ (assuming ideal gas), the temperature profile scales as
\begin{equation}
    T(P) \approx 1280~\mathrm{K} \left( \frac{P}{1000~\mathrm{bar}} \right)^{0.28},
\end{equation}
which is roughly consistent with the deep atmospheric structure in Figure \ref{fig:TP}. Vertical transport in Jupiter's deep atmosphere is governed by turbulent convection, but its efficiency is strongly modulated by the planet's rapid rotation. Near the equator (slow rotation regime), the Coriolis force is negligible. We estimate the convective velocity scale, $w_{\mathrm{eq}}$, using mixing length theory (e.g., \citealt{clayton-1968,showmanScalingLawsConvection2011,wangNewInsightsJupiters2015,zhangAtmosphericRegimesTrends2020}):
\begin{equation}
    w_{\mathrm{eq}} \approx \left( \frac{F_{\mathrm{int}} R_{\mathrm{gas}}}{\mu \rho c_p} \right)^{1/3} \approx 0.4~\mathrm{m~s^{-1}} \left( \frac{P}{1000~\mathrm{bar}} \right)^{-0.24},
\end{equation}
where $c_p$ is the specific heat capacity. The corresponding eddy diffusion coefficient is $K_{zz} \approx \frac{1}{3} w_{\mathrm{eq}} H_p$, where $H_p = R_{\mathrm{gas}}T / \mu g$ is the pressure scale height. Since $H_p \propto T \propto P^{0.28}$, $K_{zz} \sim 3 \times 10^8 ~\mathrm{cm^2~s^{-1}}$ is nearly constant with depth. 

At mid-to-high latitudes, vertical motion is affected by rotation (rapid rotation regime). The characteristic mixing length becomes limited by the Coriolis timescale rather than the scale height (e.g., \citealt{golitsyn-1980,golitsyn-1981,showmanScalingLawsConvection2011,wangNewInsightsJupiters2015,zhangAtmosphericRegimesTrends2020}). The convective velocity scales as
\begin{equation}
    w_{\mathrm{rot}} \approx \left( \frac{\alpha g F_{\mathrm{int}}}{\rho c_p \Omega} \right)^{1/2} \approx 0.05~\mathrm{m~s^{-1}} \left( \frac{P}{1000~\mathrm{bar}} \right)^{-0.5}.
\end{equation}
Consequently, the eddy diffusion coefficient drops by about one order of magnitude to $K_{zz} \approx 10^7~\mathrm{cm^2~s^{-1}}$. Our estimates are consistent with typical values of Jupiter from \citep{wangNewInsightsJupiters2015}. The estimated $K_{zz}\approx 10^7-10^8~\mathrm{cm^2~s^{-1}}$ is capable of maintaining the observed disequilibrium species such as CO (e.g., \citealt{prinnCarbonMonoxideJupiter1977}).

We compare these lofting velocities to the terminal sedimentation velocity ($v_f$) of condensate particles determined using Stokes flow. Assuming a gas dynamic viscosity of $\eta \approx 3 \times 10^{-5}~\mathrm{Pa~s}$ at 1000 bar that scales as $\eta \propto T^{0.7}$, and a particle density of $\rho_p \approx 3000~\mathrm{kg~m^{-3}}$, the settling speed is given by
\begin{align}
    v_f &= \frac{2 (\rho_p - \rho_{\mathrm{gas}}) g r^2}{9 \eta} \nonumber \\
        &\approx 0.06~\mathrm{m~s^{-1}} \left( \frac{r}{10~\mu\mathrm{m}} \right)^2 
        \left( \frac{P}{1000~\mathrm{bar}} \right)^{-0.2}.
\end{align}
We note that for larger particles (r $\sim 10~\mu\mathrm{m}$) at these high pressures, the Reynolds number exceeds unity, entering the transition regime. Applying empirical drag corrections reduces the terminal velocity by roughly a factor of two compared to Stokes flow, making the particles even more susceptible to convective lofting. Thus, the above equation represents a conservative upper bound on sedimentation.

By equating the lofting velocities with the sedimentation velocity, we find that mixing can readily loft particles with radii up to $r \sim 30~\mu\mathrm{m}$ at the equator and $10~\mu\mathrm{m}$ at midlatitudes. Thus, micron- to sub-millimeter-sized particles can be effectively transported from the deep formation region of spinel and silicates ($>3000 \, \mathrm{bar}$) into the feldspar-stability window ($1000$--$2000 \, \mathrm{bar}$). This indicates that the transport of refractory seeds required for our scenario is dynamically plausible for typical cloud particle sizes.

\subsection{Possible Chemical Pathways} \label{subsec:mineral_chem}

While our thermochemical calculations predict that alkali feldspars and feldspathoids are the stable equilibrium phases at 1000--2000 bar, their reaction pathways remain uncertain. In a dynamic atmosphere, the formation of alkali feldspars is a competition between chemical kinetics and gravitational settling. Even if vigorous mixing brings deep refractory seeds (like spinel) into the upper regions, these grains must efficiently transform into feldspars before they sediment out of the stability window.

In a solid-state regime, the transition from refractory oxide seeds (spinel) to complex aluminosilicates presents a kinetic challenge. This conversion requires the breaking of strong \ce{Al-O} and \ce{Mg-O} bonds to shift aluminum from octahedral (sixfold) coordination in spinel to tetrahedral (fourfold) coordination in feldspar, likely via a nucleation and growth process. Such reactions are likely inefficient on atmospheric timescales.

Efficient heterogeneous reaction pathways are possible if the formation of feldspars proceeds via a solution-mediated mechanism. In this scenario, the transformation does not occur by mechanically rearranging atoms within the solid crystal. Instead, the unstable solid dissolves into a liquid solvent, and the stable mineral crystallizes from that liquid. This is enabled by the formation of transient alkali-silicate-rich melts. We note that the general idea of liquids mediating vapor--solid reactions is not new: refractory silicate melts have been proposed to be important in the mineralogic evolution of chondrules under the markedly lower pressures of solar nebula condensation (\citealt{yonedaCondensationCaOMgOAl2O3SiO2Liquids1995,ebelGibbsEnergyMinimization2000}).

At pressures of 1000--2000 bar, the system is rich in two potent fluxing agents: alkali metals (\ce{Na}, \ce{K}) and water vapor. It is well established in experimental petrology that the presence of alkalis and volatiles significantly depresses the solidus temperature of silicate systems (e.g., \citealt{hackPhaseRelationsInvolving2007}). Indeed, the rainout scenario model results (Figure \ref{fig:nak}) indicate that sodium can segregate into a liquid silicate phase (\ce{Na2SiO3(l)}) rather than a solid.

Moreover, the \texttt{GGchem} model likely underestimates the stability and abundance of these liquid phases. Typically, natural silicate melts are complex, multicomponent solutions, and the database used in our calculations relies primarily on end-member melting points and does not fully incorporate the thermodynamics of multicomponent silicate melts \citep{woitkeEquilibriumChemistry1002018}. In realistic geological fluids, the mixing of multiple components (\ce{Na}, \ce{K}, \ce{Ca}, \ce{Mg}, \ce{Al}, \ce{Si}) significantly lowers the Gibbs free energy relative to pure end-members, expanding the stability field of the liquid (e.g., \citealt{ghiorsoChemicalMassTransfer1995,baleFactSageThermochemicalSoftware2009}). As a simple illustration of this effect, the addition of alumina to the \ce{Na2SiO3} system can lower the melting temperature from near 1360 K to as low as 1183 K \citep{utlakThermodynamicAssessmentPseudoternary2018}. Therefore, the pure \ce{Na2SiO3} liquid predicted by our rainout model serves as a proxy for a more chemically complex, stable ``dirty'' melt that would readily form on grain surfaces in this regime, and this is expected to have a substantially larger stability field along the Jovian adiabat than that of the pure sodium silicate liquid shown in Figure \ref{fig:other_species}.

In the presence of melts, gaseous silica (\ce{SiO}) and alkalis do not react directly with the crystal lattice. Instead, they dissolve into the surface melt, creating a highly reactive, hydrous silicate solvent. The liquid phase captures vapor more efficiently than solid surfaces and facilitates rapid diffusion \citep{karglFormationchannelsFastion2006,zhangDiffusionDataSilicate2010}. The presence of a condensed liquid also juxtaposes a higher density of reactants with the crystal surface relative to the vapor phase, enhancing the rate of chemical reactivity. Thus, the chemically unstable spinel seed can first dissolve into the melt. As the melt becomes enriched in aluminum from spinel dissolution, the thermodynamically stable feldspars can nucleate and grow.

For the formation of anorthite (\ce{CaAl2Si2O8}), the stoichiometry describes the reaction of the solid seed with the dissolved melt components:
\begin{align}
\ce{MgAl2O4(s) &+ [CaO + 3SiO2]_{(melt)} \nonumber \\ 
 &\longrightarrow CaAl2Si2O8(s) + MgSiO3(s)}.
\end{align}
Here, the reactants in parentheses represent species derived from the gas phase (e.g., via \ce{SiO(g) + H2O(g) -> SiO2(l) + H2(g)}) that have dissolved into the melt.

Similarly, the sequestration of sodium and potassium is driven by the crystallization of progressively more albite-enriched plagioclase and leucite from the melt. The sodic nature of the transient liquid facilitates the precipitation of plagioclase (the solid solution of anorthite and albite). Potassium, sterically hindered from entering the compact plagioclase lattice, concentrates in the residual melt until it nucleates into leucite (\ce{KAlSi2O6}). This crystallization process effectively pumps alkalis from the liquid phase into the solid interior of the growing cloud particle:
\begin{align}
\ce{MgAl2O4(s) &+ [K2O + 5SiO2]_{(melt)} \nonumber \\ 
 &\longrightarrow {2}KAlSi2O6(s) + MgSiO_3(s)}.
\end{align}
Consequently, rapid vertical mixing in Jupiter not only transports the necessary elemental ingredients but also likely maintains the thermodynamic conditions for transient liquid phases that relax the kinetics of these mineral-forming reactions. We acknowledge that these specific pathways are speculative. Future laboratory studies on alkali-silicate reaction kinetics at high vapor pressures are required to validate these hypothetical pathways and their efficiency.

\subsection{Sensitivity to Elemental Abundances} \label{subsec:elemental_sensitivity}

As shown in Figure \ref{fig:cloud_model}, the difference between rainout and equilibrium chemistry depends on the alkali metallicity. For solar metallicity, equilibrium chemistry results in Na, K, and electron abundances that are an order of magnitude lower than the rainout model, with only equilibrium chemistry explaining the MWR data. However, at $0.1\times$ solar alkali abundance, the discrepancy between the two scenarios vanishes, and both can explain the MWR observations (Figure \ref{fig:cloud_model}). This occurs because, in the equilibrium scenario, the solar abundance of Si and Al is sufficient to sequester Na and K into condensates, effectively capping the gas-phase abundance at roughly $0.1\times$ solar regardless of the initial reservoir. In contrast, the rainout scenario depletes Si and Al in the upper layers, leaving gaseous Na and K unaffected and preserving their original metallicity.

The validity of our proposed feldspar and leucite cloud mechanism should also place strong constraints on the deep elemental inventory of Jupiter. We performed sensitivity tests reducing the abundances of key refractory elements to assess their impact on alkali depletion. We found that reducing the abundance of either silicon or aluminum to $0.1\times$ solar inhibits the formation of feldspars and leucite; in these metal-poor scenarios, the required level of sequestration of Na and K is insufficient, leaving them in the gas phase and resulting in high electron densities that fail to explain the Juno observations. Conversely, reducing the magnesium abundance to $0.1\times$ or even $0.01\times$ solar does not hinder alkali depletion. In Mg-poor conditions, the deep aluminum carrier simply shifts from spinel to other refractory oxides (e.g., corundum (\ce{Al2O3}) or Ca-aluminates), which are still lofted and serve as effective seeds for reacting with gaseous silica and alkalis. This implies that our proposed mechanism might be robust to the specific mineralogy of the seed, provided that the deep atmosphere is sufficiently enriched in aluminum and silicon. Therefore, the depletion signal observed by Juno MWR indirectly points to a deep atmosphere that retains at least solar-like abundances of Si and Al.

\section{The Dust-Catalyzed Recombination Scenario} \label{sec:dust_plasma}

In the previous section, we demonstrated that vigorous vertical mixing, coupled with heterogeneous chemistry, can lead to the sequestration of alkali metals into feldspar and leucite clouds, thereby reducing the electron density. However, if the formation of feldspar and leucite clouds is kinetically inefficient, alkali metals may remain abundant in the gas phase below the 1000 bar level. In such a case, the electron density resulting from their thermal ionization would be too high to be consistent with the MWR signal. Here, we explore an alternative mechanism that does not focus on reducing the source of electrons but rather on increasing the sink. We propose that free electrons generated by thermal ionization of solar metallicity alkali metals could still be efficiently suppressed by recombination with the thermally emitted alkali ions from other mineral condensates.

\begin{figure}
  \centering \includegraphics[width=0.47\textwidth]{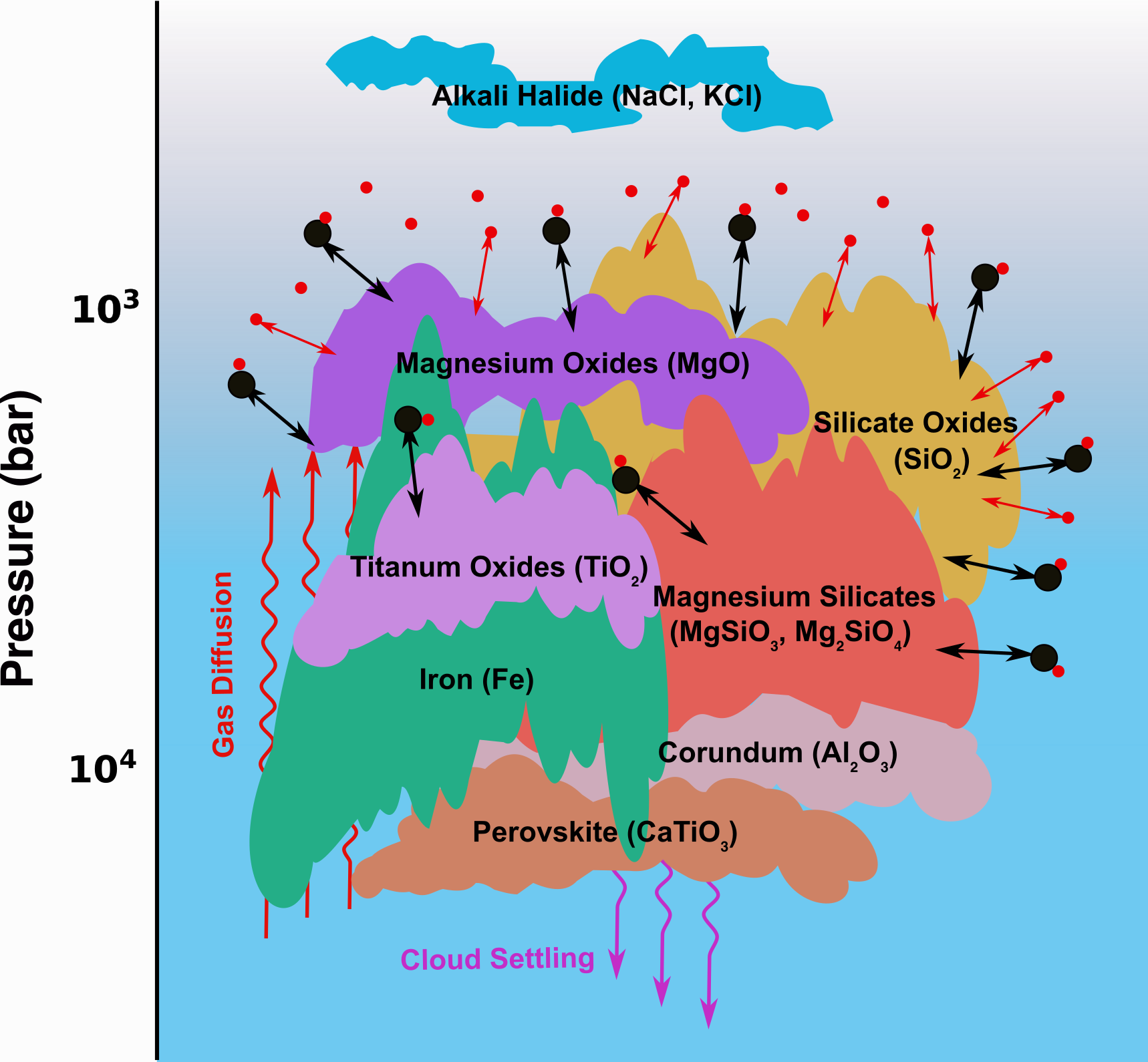} 
  \caption{Schematic representation of the dust-catalyzed recombination scenario in the deep atmosphere ($1000$--$3000$ bar). It relies on the physical presence of a dense, extended cloud deck composed of iron (\ce{Fe}), silicates (\ce{MgSiO3}, \ce{Mg2SiO4}), and refractory oxides (\ce{Al2O3}, \ce{TiO2}). Vertical gas diffusion counteracts gravitational settling, sustaining a cloud distribution with high grain surface area. The black dots represent positive ions, and the red dots represent free electrons. The arrows indicate key dust--plasma interaction processes, including ion capture and emission, as well as electron capture and thermionic emission. Through these pathways, free electron densities are efficiently suppressed by recombining with the thermally emitted alkali ions from dust particle surfaces, depleting the free electron density without requiring the chemical removal of sodium or potassium.}
  \label{fig:el_schematic}
\end{figure}

The interaction between free electrons and dust grains is a fundamental process in dusty plasmas and has been studied extensively in the context of interstellar medium \citep[e.g.,][]{draineCollisionalChargingInterstellar1987,weingartnerPhotoelectricEmissionInterstellar2001,draineAstrophysicsDustCold2004}, protoplanetary disks \citep[e.g.,][]{ilgnerIonisationFractionProtoplanetary2006, okuzumiNUMERICALMODELINGCOAGULATION2009,baiMAGNETOROTATIONALINSTABILITYDRIVENACCRETIONPROTOPLANETARY2011,fujiiFASTACCURATECALCULATION2011, deschHIGHTEMPERATUREIONIZATIONPROTOPLANETARY2015,ivlevIONIZATIONDUSTCHARGING2016} and brown dwarf atmospheres \citep[e.g.,][]{hellingIonisationDischargeCloudforming2016}, where dust grains act as massive charge carriers to affect the ionization balance. In the context of Jupiter's deep atmosphere, the ionization fraction is typically calculated assuming purely homogeneous gas-phase equilibrium (as in \texttt{GGChem}). However, the presence of dust introduces heterogeneous pathways for charge sources and sinks. At a high temperature, dust grains do not merely act as passive sinks for electrons; they actively emit adsorbed alkali ions into the gas phase. This surface-driven ion emission significantly enhances the local cation density, thereby accelerating the rate of gas-phase recombination. If deep mineral clouds persist in the 1000--2000 bar region, this dust-catalyzed recombination mechanism can suppress the free electron density, effectively reducing the 0.6 GHz opacity without requiring a subsolar elemental abundance of sodium or potassium. A cartoon of this scenario is illustrated in Figure \ref{fig:el_schematic}. To test the viability, we couple a microphysical cloud model with a kinetic model for charging to simulate the multicomponent dusty plasma.

\subsection{Microphysical Cloud Modeling} \label{subsec:exolyn}

To characterize the surface area available for dust--plasma interaction, we need to move beyond the thermochemical equilibrium predictions of \texttt{GGchem}. Cloud microphysical and transport processes are required to accurately predict the particle size distribution and number density. Currently, such a model does not exist for the deep atmosphere of Jupiter. However, microphysical-transport models for high-temperature refractory clouds have been extensively developed in the field of exoplanets and brown dwarfs over the past several decades (e.g., \citealt{lunineEvolutionInfraredSpectra1986, ackermanPrecipitatingCondensationClouds2001, hellingDustBrownDwarfs2008,leeModellingLocalGlobal2015,powellFormationSilicateTitanium2018, ohnoMicrophysicalModelingMineral2018, hellingExoplanetClouds2019, ormelARCiSFrameworkExoplanet2019,gaoAerosolCompositionHot2020, zhangAtmosphericRegimesTrends2020,huangExoLynGoldenMean2024}). In this study, we employ a 1D Eulerian cloud microphysical code, \texttt{ExoLyn} \citep{huangExoLynGoldenMean2024}, to simulate the deep clouds on Jupiter.

\texttt{ExoLyn} solves the advection--diffusion equation for condensate mass and number density, incorporating nucleation, condensation growth, evaporation, gravitational settling, and vertical mixing within a self-consistent framework. In the model, cloud particles consist of nuclei that serve as condensation sites surrounded by multiple condensate species. Nucleation initiates the formation of cloud particles, which subsequently grow as vapor deposits onto their surfaces under conditions of vapor supersaturation. These particles interact dynamically with the atmosphere through processes such as settling, diffusion, and mixing, redistributing mass, number density, and surface area available for phenomena like dust-catalyzed recombination. While bin-scheme models are capable of arbitrary size distributions (e.g., \citealt{powellFormationSilicateTitanium2018, gaoAerosolCompositionHot2020}), \texttt{ExoLyn} provides a computationally efficient approach to estimate bulk cloud properties, specifically the mean particle size and total mass loading that are the key variables governing the heterogeneous surface area essential for interaction with ions and electrons.

\texttt{ExoLyn} incorporates chemical reactions to drive condensation, enabling the formation of multicomponent clouds from multispecies vapors. For the deep clouds on Jupiter, the cloud formation pathways considered in the model include magnesium silicates, iron species, and refractory oxides:
\begin{align}
    \ce{SiO_{(g)} + H2O_{(g)}} &\longrightarrow \ce{SiO2_{(s)} + H2_{(g)}} \\
    \ce{Mg_{(g)} + H2O_{(g)}} &\longrightarrow \ce{MgO_{(s)} + H2_{(g)}} \\
    \ce{TiO_{(g)} + H2O_{(g)}} &\longrightarrow \ce{TiO2_{(s)} + H2_{(g)}} \\
    \ce{Fe_{(g)}} &\longrightarrow \ce{Fe_{(s)}} \\
    \ce{Fe_{(g)} + H2O_{(g)}} &\longrightarrow \ce{FeO_{(s)} + H2_{(g)}} \\
    \ce{Fe_{(g)} + H2S_{(g)}} &\longrightarrow \ce{FeS_{(s)} + H2_{(g)}} \\
    \ce{2Fe_{(g)} + 3H2O_{(g)}} &\longrightarrow \ce{Fe2O3_{(s)} + 3H2_{(g)}} \\
    \ce{2Al_{(g)} + 3H2O_{(g)}} &\longrightarrow \ce{Al2O3_{(s)} + 3H2_{(g)}} \\
    \ce{Mg_{(g)} + SiO_{(g)} + 2H2O_{(g)}} &\longrightarrow \ce{MgSiO3_{(s)} + 2H2_{(g)}} \\
    \ce{2Mg_{(g)} + SiO_{(g)} + 3H2O_{(g)}} &\longrightarrow \ce{Mg2SiO4_{(s)} + 3H2_{(g)}}.
\end{align}

The cloud condensation and chemistry described in the \texttt{ExoLyn} model are considerably simpler than the thermochemical equilibrium cloud model \texttt{GGchem} presented in Section \ref{sec:cloud_model}. However, the goal of this scenario is different from that of the previous model. In the chemical sequestration scenario, the focus is on alkali metal gas depletion within complex mineral clouds, making the specific chemistry and mineralogical identity of the condensate essential. The mineral clouds must provide compatible crystal lattices and chemical pathways that enable the sequestration of alkali atoms into the clouds. In contrast, the dust-catalyzed recombination scenario primarily depends on the total available surface area of the cloud particles, rather than their specific mineralogical composition, although the composition does dictate the work function and surface binding energies that regulate charge emission. In a solar-composition atmosphere, magnesium and iron are substantially more abundant than aluminum or calcium, meaning that the bulk of the condensate mass, and consequently the geometric cross section, are dominated by iron and magnesium silicates. As a first-order estimate of this effect, the microphysical model can be restricted to these dominant mass reservoirs to capture the macroscopic surface area available for plasma interaction.

Using \texttt{ExoLyn}, we explicitly simulate the gas and cloud species on Jupiter from 100 to $10^5$ bar. We initialize the model using the temperature profile from Section \ref{sec:data} and a constant vertical eddy diffusivity, $K_{zz} = 10^8~\mathrm{cm^2~s^{-1}}$, as described in Section \ref{subsec:kzz}. The gas-phase equilibrium abundances at the bottom of the domain ($10^5~\mathrm{bar}$) are set using the results from the rainout scenario in \texttt{GGchem} (Section \ref{sec:cloud_model}).

Condensation nuclei provide the essential seeds for cloud particle growth. Following the standard exoplanet configuration in \texttt{ExoLyn}, we adopt a fixed nuclei bulk density of 2.8 $\mathrm{g~cm^{-3}}$ and a radius of 1 nm. However, the production rate of condensation nuclei ($S_n$) acts as a critical free parameter that requires careful consideration. The nucleation profile depends on factors such as the composition and supersaturation of precursor species (e.g., \citealt{gailSeedParticleFormation2013, leeDustBrownDwarfs2015}), as well as the size and geometry of the nuclei (e.g., \citealt{powellDepletionGaseousCO2022}). Consequently, $S_n$ remains highly uncertain due to its dependence on these microscopic processes.
\begin{figure*}
  \centering \includegraphics[width=0.9\textwidth]{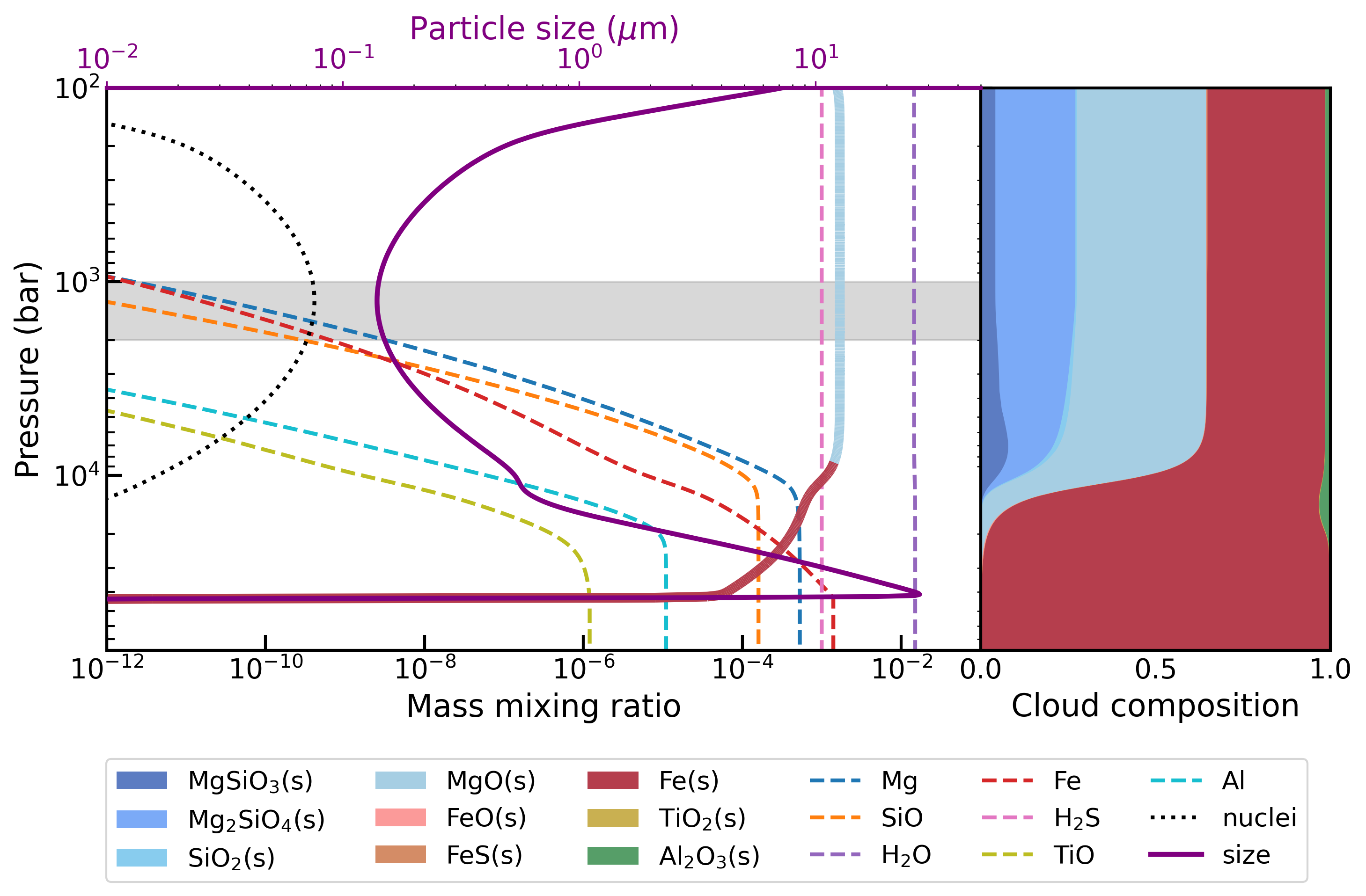} 
  \caption{Vertical structure of the deep Jovian cloud layers simulated using the \texttt{ExoLyn} microphysical model assuming a column-integrated nuclei production rate of $10^{-4}$ $\mathrm{g}~\mathrm{cm}^{-2}~\mathrm{s}^{-1}$. Left Panel: Vertical profiles of mass mixing ratios (bottom axis) for condensable gases (thin dashed lines), condensation nuclei (dotted black line), and total solid condensates (thick multicolor line). The color of the thick solid line indicates the dominant aerosol species by mass at each pressure level. The solid purple line represents the mean particle radius ($r_p$), corresponding to the top axis scales.  The gray shaded region marks the 1000--2000 bar pressure level, corresponding to the peak sensitivity of the Juno MWR 0.6 GHz channel. 
Right Panel: The cumulative fractional composition of the cloud mass as a function of pressure. The simulation predicts a deep, extended cloud deck dominated by iron (\ce{Fe}), magnesium oxide (\ce{MgO}), and silicates (\ce{MgSiO3}, \ce{Mg2SiO4}) with particle radii of $0.1$--$10~\mu$m, providing large surface area for electron depletion due to dust--plasma interactions.}
  \label{fig:exolyn}
\end{figure*}

To mitigate these uncertainties, \texttt{ExoLyn} parameterizes $S_n$ as a log-normal distribution in pressure defined by three parameters: the pressure of the peak rate ($p_n$), the distributional width ($\sigma_n$), and the total column-integrated nucleation rate ($\dot{\Sigma}_n$). Standard values used for hot Jupiters in \texttt{ExoLyn} are inapplicable here because mineral clouds on Jupiter form at significantly higher pressures ($>1000$ bar) compared to hot exoplanets ($<1$ bar). Previous studies of heterogeneous nucleation for refractory species like \ce{SiO} and \ce{TiO2} suggest that nucleation rates peak in the $1000$--$1500$ K temperature range \citep{gailSeedParticleFormation2013, leeDustBrownDwarfs2015}. In Jupiter's deep atmosphere, this corresponds to pressures of $1000$--$2000$ bar. Accordingly, we set the peak nucleation pressure to $p_n = 2000$ bar with a width of $\sigma_n = 0.5$.

The column-integrated nucleation rate, $\dot{\Sigma}_n$, is relatively poorly constrained. Typical \texttt{ExoLyn} setups adopt $\dot{\Sigma}_n = 10^{-15}~\mathrm{g~cm^{-2}~s^{-1}}$ for cloud-forming regions in hot Jupiters at pressures of order $\sim 10^{-4}$ bar, whereas the pressure regime of Jupiter considered in this study, $\sim 10^{3}$ bar, is roughly 7 orders of magnitude higher. According to classical homogeneous nucleation theory \citep[e.g.,][]{zeldovichTheoryNewPhase1943}, the nucleation rate scales with the square of the monomer number density ($S_n \propto n_{\mathrm{vapor}}^2$), implying a correspondingly higher local vapor density in Jupiter’s deep atmosphere. If the nucleating vapors maintain comparable levels of supersaturation, the resulting nucleation rates in Jupiter’s deep atmosphere could exceed those in hot exoplanet atmospheres by more than 10 orders of magnitude.

In this study, we define a nominal cloud case with $\dot{\Sigma}_n = 10^{-4}~\mathrm{g~cm^{-2}~s^{-1}}$. This corresponds to a specific nucleation rate relative to the hydrogen number density of $S_n/n_H \sim 10^{-14}~\mathrm{s^{-1}}$ at the 2000-bar peak. We note that a column-integrated nucleation rate of $10^{-4} \text{ g cm}^{-2} \text{ s}^{-1}$ exceeds the theoretical maximum upward vapor mass flux of pure \ce{TiO2}, which is estimated to be on the order of $10^{-5} \text{ g cm}^{-2} \text{ s}^{-1}$, assuming a vertical wind speed of about $0.1 \text{ m s}^{-1}$. Therefore, the initial condensate seeds cannot be composed solely of titanium dioxide, but must incorporate more abundant refractory species such as iron and silicates, which are expected to co-condense in this high-density regime. 

The resulting vertical structure of this nominal case is shown in Figure \ref{fig:exolyn}. At pressures larger than $10^4$ bar, the atmosphere is dominated by iron clouds (\ce{Fe}). At lower pressures, the clouds consist of a mixture comprising approximately 50\% iron, 30\% magnesium oxide (\ce{MgO}), and 20\% silicates (\ce{MgSiO3}, \ce{Mg2SiO4}). 

This composition differs significantly from the rainout scenario predicted by thermochemical models such as \texttt{GGchem} (Figure \ref{fig:nak}a). However, it is qualitatively consistent with the equilibrium scenario (Figure \ref{fig:nak}b), where metal clouds, silicates, and refractory oxides are vertically extended and dominate the cloud composition. This structure reflects the influence of strong vertical mixing, although the microphysical model does not include the alkali metals to form feldspars or leucite. The simulated total cloud mass mixing ratio for the nominal case is approximately $10^{-3}$ above the $10^4$ bar level, lower than in the equilibrium scenario, yet significantly higher than in the rainout scenario. The simulated characteristic particle radii ($r_p$) range from $0.1~\mu$m near the nucleation peak at 2000 bar to $\sim 10$--$20~\mu$m at the cloud base. Because the mass mixing ratio profile is roughly constant with pressure, the presence of these smaller, submicron particles implies a much larger total surface area available to efficiently interact with ions and electrons in the 1000--2000 bar region.

\begin{figure*}
  \centering \includegraphics[width=0.9\textwidth]{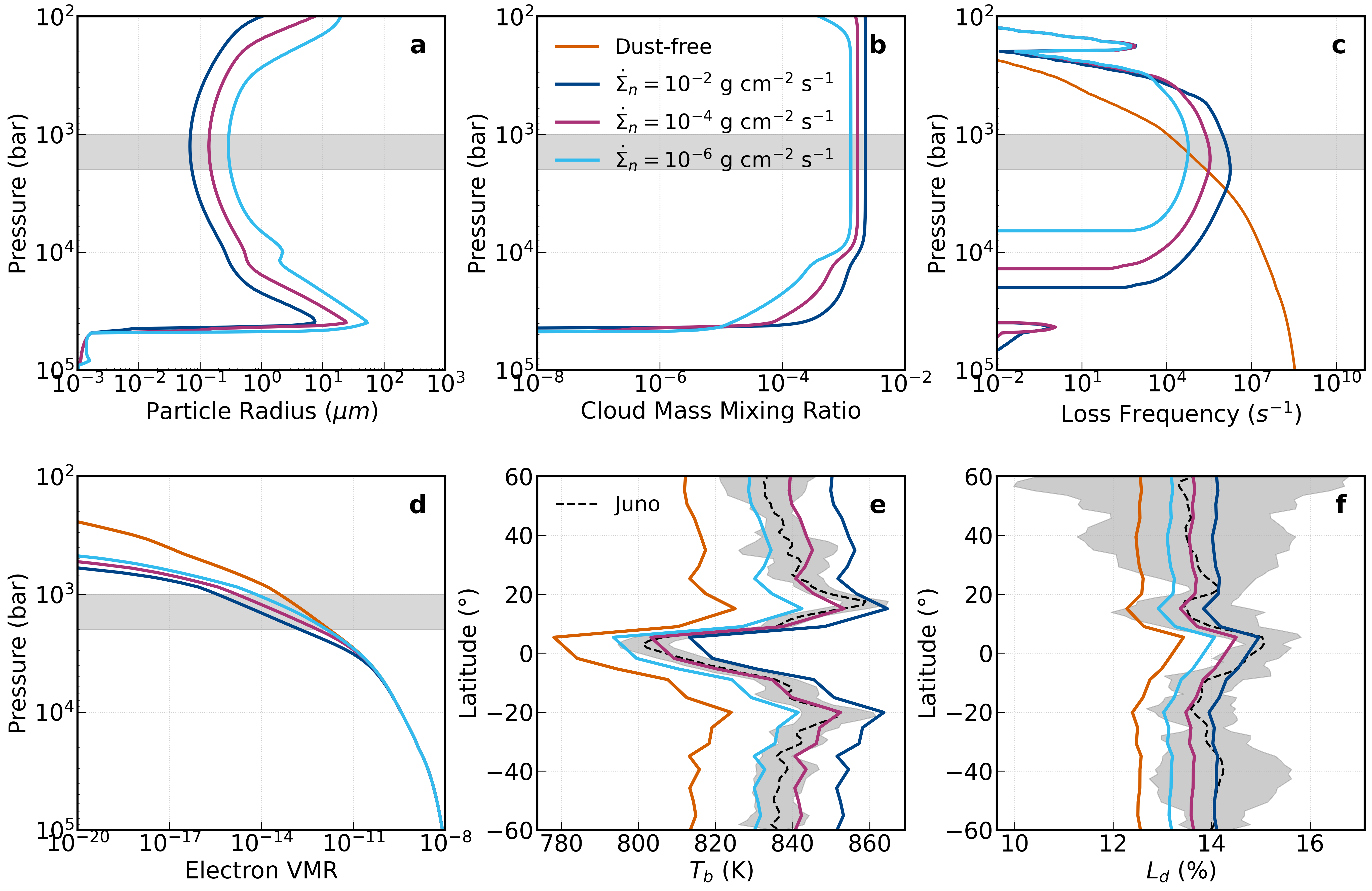} 
  \caption{Impact of cloud microphysics on electron depletion and MWR observables in the dust-catalyzed recombination scenario. 
  \textbf{(a)} Vertical profiles of mean particle radius and \textbf{(b)} cloud mass mixing ratio simulated by \texttt{ExoLyn} for three different nucleation rates, corresponding to column-integrated nucleation rates of $10^{-2}$, $10^{-4}$, and $10^{-6}$ $\mathrm{g}~\mathrm{cm}^{-2}~\mathrm{s}^{-1}$. 
  \textbf{(c)} Electron loss frequencies, where the orange line (``dustfree") represents the baseline loss rate due to gas-phase recombination, and the blue/purple lines depict loss frequencies via attachment to aerosols. Within the MWR sensitivity window (gray band, 1000–2000 bar), aerosol attachment exceeds gas-phase loss by orders of magnitude. 
  \textbf{(d)} Resulting free electron VMR demonstrating deep electron suppression relative to the gas-only case (orange). 
  \textbf{(e)} Nadir-view brightness temperature ($T_b$) and \textbf{(f)} limb darkening ($L_d$) at 0.6 GHz as a function of latitude, with black dashed lines and gray shading indicating Juno data and uncertainties.}
  \label{fig:exolyn_mwr}
\end{figure*}

To assess the sensitivity of the microphysical results to the nucleation rate, we also explore a high nucleation case ($\dot{\Sigma}_n = 10^{-2}~\mathrm{g~cm^{-2}~s^{-1}}$) and a lower nucleation case ($\dot{\Sigma}_n = 10^{-6}~\mathrm{g~cm^{-2}~s^{-1}}$). A comparison of the particle sizes and cloud masses for these cases is shown in Figure \ref{fig:exolyn_mwr}. As $\dot{\Sigma}_n$ increases by 3 orders of magnitude, the mean particle size decreases by approximately a factor of 3, while the cloud mass increases by less than a factor of 2 due to the limitation of the condensable vapor inventory. While other free parameters influence the nucleation profile, the total column-integrated nucleation rate directly governs the magnitude of the particle size. We do not explore the full parameter space here; rather, our objective is to demonstrate the viability of electron depletion in the presence of dust grains, as described below.

\subsection{Dust--Plasma Interaction} \label{subsec:dust_plasma}

To determine the equilibrium electron density in the deep Jovian atmosphere, defined by pressures $P \gtrsim 1000$ bar and temperatures $T \gtrsim 1000$ K, we model the environment as a multicomponent dusty plasma. We adopt the calculation framework of \citet{deschHIGHTEMPERATUREIONIZATIONPROTOPLANETARY2015}, with modifications for the high-density regime. The charge distribution in this system is governed by the interaction between four distinct reservoirs: free electrons with number density $n_e$; positive ions, denoted $n_+$, which are primarily ionized alkali metals (${\rm K}^+$ and ${\rm Na}^+$); gas-phase anions, $n_-$, consisting of species such as ${\rm Cl}^-$ and ${\rm HS}^-$; and charged dust grains with a number density $n_d$. We construct our physical model by first defining the thermodynamic equilibrium of the background gas in the absence of aerosols and subsequently introducing dust grains as active charge sinks (via capture) and sources (via emission).

\subsubsection{Gas-phase Ionization and Recombination} \label{subsec:baseline}

We establish the baseline thermochemical state of the atmosphere from \texttt{GGchem}. This calculation provides the equilibrium number densities of electrons, cations, and anions--denoted $n_{e,0}$, $n_{+,0}$, and $n_{-,0}$, respectively--in a dust-free environment. In this equilibrium state, the thermal production of electrons is exactly balanced by the gas-phase loss rate. We model this loss as an effective recombination process proportional to the product of electron and positive ion densities. By anchoring our kinetic model to this thermodynamic baseline, we derive the fundamental rate constants that govern the system.

The primary source of charge carriers is the thermal ionization of alkali metals. In a dusty environment, grains adsorb neutral atoms, reducing the gas-phase inventory available for ionization. Since the volumetric production rate depends linearly on the number density of neutral atoms, the effective source term $Q$ is depleted relative to the dust-free baseline. As potassium is the main electron contributor, we quantify this by defining a gas-phase depletion factor, $\xi_{\rm gas} \equiv n_{\rm K, gas} / n_{\rm K, tot}$, such that the source term becomes
\begin{equation}
Q = Q_0 \times \xi_{\rm gas}. \label{eq:Q}
\end{equation}
Here, $Q_0$ is the baseline thermal ionization rate in thermodynamic equilibrium, and $\xi_{\rm gas}$ varies from 1 (dust-free limit) to 0 (complete adsorption). The calculation of $\xi_{\rm gas}$ is detailed in Section \ref{subsec:coverage}.

The removal of free electrons proceeds through a two-step mechanism: rapid attachment to electrophilic neutrals (e.g., forming $\mathrm{Cl}^-$, $\mathrm{HS}^-$) followed by mutual neutralization with positive ions. At the high pressures of the deep atmosphere ($P \gtrsim 1000$ bar), charge-exchange reactions ($e^- + \mathrm{X} \leftrightarrow \mathrm{X}^-$) occur on timescales orders of magnitude shorter than the charge annihilation or dust-charging timescales. Consequently, free electrons and gas-phase anions effectively act as a single coupled reservoir of negative charge in local thermodynamic equilibrium. We simplify the treatment of negative carriers by assuming a fixed anion-to-electron ratio, $\beta \equiv n_- / n_e \approx n_{-,0}/n_{e,0}$, anchored to the thermodynamic baseline calculated by \texttt{GGchem}. The total gas-phase loss rate is then modeled as an effective recombination process proportional to the product of the electron and cation densities:
\begin{equation}
    R_{\rm loss} = k_{\rm rec} n_e n_+.
\end{equation}
We treat the effective recombination coefficient $k_{\rm rec}$ by bridging two distinct physical regimes. At lower pressures, recombination is limited by reaction kinetics (three-body recombination). We adopt the rate from \citet{ashtonKineticsCollisionalIonization1973}:
\begin{equation}
    k_{\rm kin} \approx 4.4 \times 10^{-24} \, T^{-1} \, n_{\rm gas} \quad [\mathrm{cm}^3 \, \mathrm{s}^{-1}].
\end{equation}
However, in the deep atmosphere ($P \gtrsim 1000$ bar), the gas density becomes sufficiently high that the mean free path of charge carriers shrinks, and the reaction becomes transport limited. In this regime, charged particles must physically diffuse through the dense neutral gas to find one another. This process is governed by Langevin recombination \citep{langevinRecombinaisonMobilitesIons1903}, where the rate is determined by the mutual drift of oppositely charged particles under their Coulomb attraction:
\begin{equation}
    k_{\rm diff} = 4 \pi \frac{e^2}{k_B T} (D_e + D_i).
\end{equation}
Here, $D_e$ and $D_i$ are the diffusion coefficients for electrons and ions, respectively. We calculate these from kinetic theory as $D_j = \frac{1}{3} \lambda_j v_{\rm th, j}$, where $v_{\rm th, j}$ is the thermal velocity, and $\lambda_j = (n_{\rm gas} \sigma_j)^{-1}$ is the mean free path. 

The diffusion efficiency depends critically on the collisional cross sections $\sigma_j$. For electrons, scattering is governed by short-range quantum mechanical momentum transfer; we assume a constant cross section $\sigma_e \approx 2.0 \times 10^{-15}$ cm$^2$ (e.g., \citealt{yoonCrossSectionsElectron2008}). For ions, the interaction is dominated by the long-range polarization force (Langevin capture) with neutral H$_2$, which leads to a significantly larger, temperature-dependent cross section (e.g., \citealt{vogtScatteringIonsPolarization1954,gioumousisReactionsGaseousMolecule1958}):
\begin{equation}
    \sigma_i(T) \approx \sigma_{i,0} \left( \frac{T_0}{T} \right)^{1/2},
    \label{eq:sigma_i}
\end{equation}
where $\sigma_{i,0} \approx 1.5 \times 10^{-14}$ cm$^2$ at $T_0 = 1500$ K. As we discuss in Section \ref{subsec:capture}, this difference in cross sections results in distinct transport regimes for electrons versus ions.

To capture the transition between the kinetic-limited and diffusion-limited regimes, we define the effective coefficient as the harmonic mean of the two rates:
\begin{equation}
    k_{\rm rec} = \left( \frac{1}{k_{\rm kin}} + \frac{1}{k_{\rm diff}} \right)^{-1}.
\end{equation}
This formulation ensures that our model respects the reaction kinetics at low pressure while correctly suppressing the recombination rate in the high-density fluid regime where diffusive transport is the bottleneck. 

In the dust-free limit, production must exactly balance loss to reach the thermodynamic equilibrium, allowing us to derive the baseline ionization rate as $Q_0 = R_{\rm loss, 0} \equiv k_{\rm rec} n_{e,0} n_{+,0}$, while correctly scaling the net charge destruction rate with the local cation abundance $n_+$ when dust is present. If grains carry a significant negative charge, the cation density $n_+$ must increase relative to the electron density to maintain neutrality, thereby enhancing the gas-phase recombination rate $R_{\rm loss}$ to remove electrons.

\subsubsection{Grain Charge Capture} \label{subsec:capture}

Dust grains alter the ionization balance by capturing charges from the plasma and, at sufficiently high temperatures, emitting them back into the gas. Each grain acquires a surface potential $\psi$ (normalized as $Ze^2 / r_p k_B T$) determined by the balance of four distinct currents: the capture (adsorption) of diffusing electrons ($J_{e,\rm cap}$), the capture of diffusing positive ions ($J_{+,\rm cap}$), the thermionic emission of electrons from the hot grain surface ($J_{\rm th}$), and the thermal desorption (emission) of adsorbed alkali ions ($J_{\rm ion}$). 

The capture of electrons and ions in low-pressure astrophysical environments is traditionally described by Orbital-Motion-Limited (OML) theory (e.g., \citealt{mott-smithTheoryCollectorsGaseous1926, whipplePotentialsSurfacesSpace1981}). OML theory assumes that the mean free path of the plasma species ($\lambda_{\rm mfp}$) is much larger than the grain radius $r_p$, corresponding to a large Knudsen number ($Kn = \lambda_{\rm mfp} / r_p \gg 1$). In this collisionless regime, particles follow ballistic trajectories determined solely by their thermal energy and the grain's electric potential.

However, the high-pressure environment of the deep atmosphere ($P \sim 1000$~bar) renders charge capture diffusion limited. We model the transition from collisionless to continuum transport using the flux-matching method (e.g., \citealt{fuchsMechanicsAerosolsFuchs1964,seinfeldAtmosphericChemistryPhysics2016}). We also separate treatments for electrons and ions because the ion collision cross section is an order of magnitude larger than that of electrons due to the induced dipole interactions described in Eq.~\ref{eq:sigma_i}, resulting in a significantly shorter mean free path for ions. We incorporate this species-dependent diffusion suppression via the Knudsen number, $Kn_j = \lambda_j / r_p$, and the diffusion factor $\mathcal{D}_j$:
\begin{equation}
\mathcal{D}_j(Kn_j) = \frac{Kn_j}{Kn_j + 1}.
\label{eq:diff_factor}
\end{equation}

Thus, for a grain with normalized potential $\psi$, the diffusing currents for electrons ($J_{e, \rm cap}$) and ions ($J_{+, \rm cap}$) are given by
\begin{align}
J_{e, \rm cap} &= n_e s_e v_{e} \pi r_p^2 \exp(-\psi) \times \mathcal{D}_e(Kn_e), \label{eq:cap_e} \\
J_{+, \rm cap} &= n_+ s_+ v_{+} \pi r_p^2 (1 + \psi) \times \mathcal{D}_i(Kn_i), \label{eq:cap_i}
\end{align}
where $v_j = \sqrt{8 k_B T / \pi m_j}$ is the thermal velocity, and $s_j \approx 1$ is the sticking probability. We assume the typical case of a negatively charged grain ($\psi > 0$). Electrons experience a repulsive Coulomb potential and must overcome an energy barrier to be captured; only the high-energy tail of the thermal distribution contributes, giving rise to the Boltzmann factor $\exp(-\psi)$. Positive ions are electrostatically attracted to the grain, which enhances the effective capture cross section through Coulomb focusing, leading to the factor $(1+\psi)$.

\subsubsection{Grain Charge Emission} \label{subsec:emission}

At temperatures $T \gtrsim 800$ K, the thermal energy is sufficient to eject charge carriers from the grain surface. The high gas density affects emission as well as capture. An emitted electron or ion colliding with a gas molecule within a few mean free paths of the surface has a high probability of being backscattered onto the grain. Therefore, the net emission current leaving the vicinity of the grain is also suppressed by the diffusion factor $\mathcal{D}_j(Kn)$ derived in Eq.~\ref{eq:diff_factor} (e.g., \citealt{fuchsMechanicsAerosolsFuchs1964,seinfeldAtmosphericChemistryPhysics2016}).

Heated grains emit electrons via the Richardson--Dushman process \citep{deschHIGHTEMPERATUREIONIZATIONPROTOPLANETARY2015}. The current depends on the grain's work function $W$ and the surface temperature, which we assume is in equilibrium with the gas temperature $T$. The diffusion-corrected thermionic current is modeled as
\begin{equation}
J_{\rm th} = 4\pi r_p^2 \lambda_R \frac{4\pi m_e (k_B T)^2}{h^3} \exp\left( - \frac{W_{\rm eff}}{k_B T} \right) \times \mathcal{D}_e(Kn).
\label{eq:thermionic}
\end{equation}
Here, $\lambda_R$ is the material-specific Richardson constant correction, typically of the order of 0.5 for astrophysical silicates and metals \citep{deschHIGHTEMPERATUREIONIZATIONPROTOPLANETARY2015}. The effective work function, $W_{\rm eff}$, incorporates the electrostatic potential of the grain. For a negatively charged grain, the surface potential barrier aids electron escape, such that $W_{\rm eff} \approx W - |Z|e^2/r_p$. We adopt a nominal work function of $W = 4.5$~eV, which is representative of heterogeneous iron and silicate grains (e.g., \citealt{fomenkoHandbookThermionicProperties1966,kimuraElectricChargingInterstellar1998}).

In addition to electrons, alkali atoms adsorbed on the grain surface can thermally desorb as ions \citep{deschHIGHTEMPERATUREIONIZATIONPROTOPLANETARY2015}. We treat potassium as the dominant species for ion emission due to its lower ionization potential ($IP_{\rm K} = 4.34$~eV) compared to sodium ($IP_{\rm Na} = 5.14$~eV). The ion emission current is proportional to the surface coverage fraction of potassium ($\theta_{\rm K}$):
\begin{equation}
J_{\rm ion} = (4\pi r_p^2 N_{\rm sites} \theta_{\rm K}) \nu_{\rm vib} \exp\left(-\frac{E_{\rm ads}}{k_B T}\right) f_+ \times \mathcal{D}_i(Kn).
\label{eq:ion_emission}
\end{equation}

Here, $N_{\rm sites} \approx 10^{15}$~cm$^{-2}$ is the surface site density, $\nu_{\rm vib} \approx 3.7 \times 10^{13}$~Hz is the lattice vibration frequency \citep{hagstromDesorptionKineticsAtmospheric2000}. We adopt an effective surface binding energy of $E_{\rm ads} \approx 2.5$~eV as the relevant barrier for alkali ion emission \citep{kudriavtsevCalculationSurfaceBinding2005}. This value assumes that charge exchange is dominated by atoms chemisorbed to accessible surface sites, distinguishing them from both the loosely bound multilayer adatoms ($E \approx E_{\rm vap}$) and atoms locked deep within the bulk refractory lattice ($E > 3$~eV, \citealt{deschHIGHTEMPERATUREIONIZATIONPROTOPLANETARY2015}). The quantity $f_+$ is the ion fraction of the desorbing flux, determined by the Saha--Langmuir equation \citep{deschHIGHTEMPERATUREIONIZATIONPROTOPLANETARY2015}:
\begin{equation}
f_+ = \left[ 1 + \frac{g_0}{g_+} \exp\left( \frac{IP_{\rm K} - W_{\rm eff}}{k_B T} \right) \right]^{-1},
\end{equation}
where $g_0/g_+ = 2$ is the ratio of statistical weights for neutral and ionized potassium. The determination of the surface coverage $\theta_{\rm K}$, which accounts for the full range of binding energies, is detailed in the following subsection.

\subsubsection{Alkali Surface Coverage} \label{subsec:coverage}

The equilibrium surface coverage fractions ($\theta_{\rm K}$ and $\theta_{\rm Na}$) are determined by balancing the impinging gas flux against thermal desorption. To account for mass conservation, the impingement flux $J_{{\rm in}, j}$ is coupled to the remaining gas-phase abundance:
\begin{equation}
J_{{\rm in}, j} = \frac{v_{{\rm th}, j}}{4} \left( n_{j, \rm tot} - 4\pi r_p^2 n_d N_{\rm sites} \theta_j \right) \mathcal{D}_i(Kn),
\label{eq:jin}
\end{equation}
where $n_{j, \rm tot}$ is the total metallicity-defined abundance of species $j$, and the term in parentheses represents the depleted gas concentration. The diffusion factor $\mathcal{D}_i(Kn)$ accounts for transport limitations in the high-pressure regime.

\begin{figure*}
  \centering \includegraphics[width=0.99\textwidth]{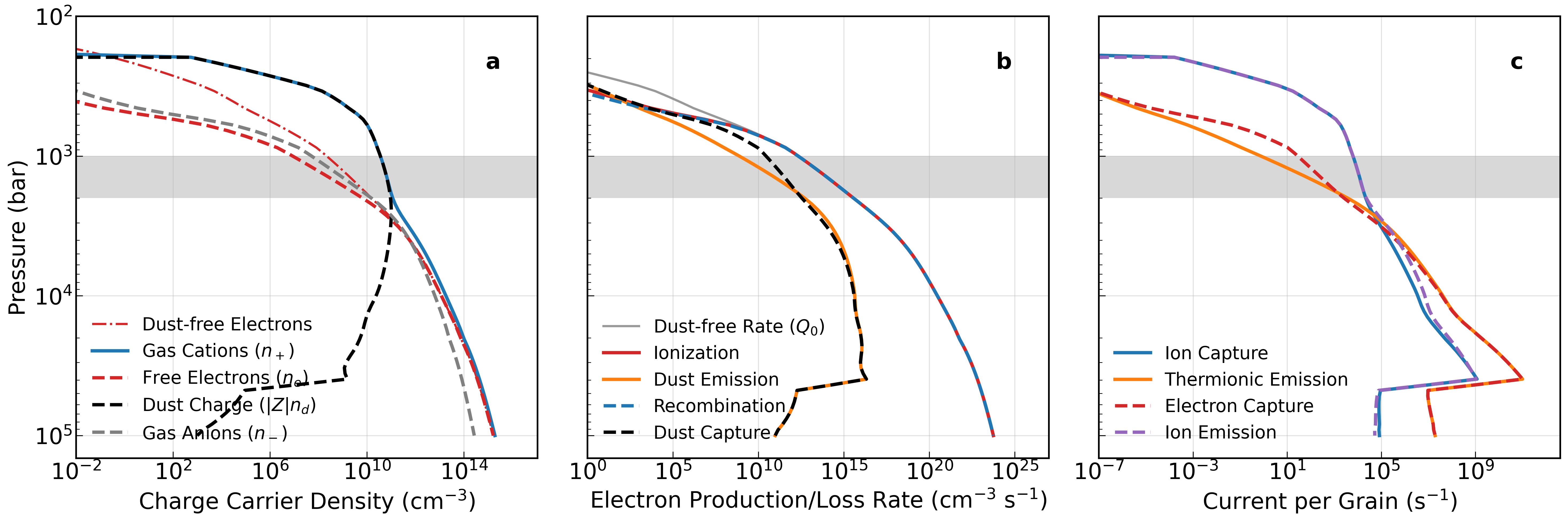} 
  \caption{Vertical profiles of the equilibrium dusty plasma state in the deep Jovian atmosphere, based on the dust cloud properties simulated using \texttt{Exolyn} with $\dot{\Sigma}_n = 10^{-4}~\mathrm{g~cm^{-2}~s^{-1}}$ (the nominal case presented in Figure \ref{fig:exolyn}).
\textbf{(a)} Number densities of charge carriers, showing the depletion of free electrons (red dashed) relative to the dust-free baseline (red dotted--dashed line).
\textbf{(b)} Volumetric electron production (solid line) and loss (dashed ;ome) rates, comparing gas-phase processes with dust-grain interaction terms.
\textbf{(c)} Magnitude of microscopic charging currents to and from a single grain.
The gray shaded region highlights the MWR sensitivity window between 1000 and 2000 bar.}
  \label{fig:dust-grain}
\end{figure*}

We solve for the coverage $\theta_j$ using the competitive Langmuir isotherm formulation (e.g., \citealt{maselPrinciplesAdsorptionReaction1996}):
\begin{equation}
\theta_j = \frac{J_{{\rm in}, j}}{J_{\rm des} + J_{\rm in, K} + J_{\rm in, Na}}.
\end{equation}
The desorption flux is given by $J_{\rm des} = N_{\rm sites} \nu_{\rm vib} \exp(-E_{\rm eff}/k_B T)\mathcal{D}_i(Kn)$. The effective activation energy, $E_{\rm eff}$, varies with the total surface coverage $\Theta_{\rm tot} \equiv \theta_{\rm K} + \theta_{\rm Na}$ to account for saturation effects (e.g., \citealt{yakshinskiyPhotonstimulatedDesorptionNa2004}). We approximate this transition linearly between the chemisorption energy ($E_{\rm ads} \approx 2.5$~eV; \citealt{kudriavtsevCalculationSurfaceBinding2005}) and the bulk vaporization energy ($E_{\rm vap} \approx 0.92$~eV; \citealt{chase1998nist}):
\begin{equation}
E_{\rm eff}(\Theta_{\rm tot}) = E_{\rm ads} (1 - \Theta_{\rm tot}) + E_{\rm vap} \Theta_{\rm tot}.
\end{equation}
We determine $\theta_{\rm K}$ and $\theta_{\rm Na}$ by solving this coupled nonlinear system numerically. The resulting gas-phase depletion factor, $\xi_{\rm gas} = n_{\rm K, gas}/n_{\rm K, tot}$, is then used to scale the bulk ionization source term via Equation \ref{eq:Q}.

\subsubsection{Solution Procedure}

We determine the steady-state electron density $n_e$ and grain potential $\psi$ by numerically solving a system of equations that satisfies three coupled constraints: charge neutrality, global production balance, and current balance.

First, the bulk atmosphere must be electrically neutral. The density of positive ions ($n_+$) must balance all negative charge carriers, including electrons, anions, and charged dust grains. Substituting the relation $n_- = \beta n_e$ derived from the fast-chemistry assumption, the neutrality condition is
\begin{equation}
n_+ = n_e(1 + \beta) + |Z(\psi)| n_d.
\label{eq:neutrality}
\end{equation}

Second, the global source of electrons must equal the sinks. The source term consists of the thermal production $Q$ (scaled by the depletion factor $\xi_{\rm gas}$ via Equation \ref{eq:Q}) and thermionic emission from dust. The sink terms are gas-phase recombination and capture onto dust. The balance equation is given by
\begin{equation}
Q + n_d J_{\rm th}(\psi) = k_{\rm rec} n_e n_+ + n_d J_{e, \rm cap}(\psi).
\label{eq:prod_bal}
\end{equation}
While ion emission ($J_{\rm ion}$) does not explicitly appear in Eq.~\ref{eq:prod_bal} because it produces ${\rm K}^+$ rather than electrons, it strongly influences the balance indirectly by driving $\psi$ to large negative values, which in turn suppresses the electron capture term $J_{e, \rm cap}$. Substituting Eq.~\ref{eq:neutrality} into Eq.~\ref{eq:prod_bal} allows us to express the production balance as a quadratic equation for $n_e$:
\begin{equation}
\mathcal{A} n_e^2 + \mathcal{B}(\psi) n_e - \mathcal{C}(\psi) = 0,
\label{eq:ne}
\end{equation}
with coefficients defined as
\begin{align}
\mathcal{A} &= k_{\rm rec} (1 + \beta), \\
\mathcal{B}(\psi) &= k_{\rm rec} |Z(\psi)| n_d + n_d J_{e, \rm cap}(\psi)/n_e, \\
\mathcal{C}(\psi) &= Q + n_d J_{\rm th}(\psi).
\end{align}

Finally, the grain potential $\psi$ is determined by the condition that the net current to the grain is zero:
\begin{equation}
J_{\rm net}(\psi) = [J_{e, \rm cap}(\psi) + J_{\rm ion}(\psi)] - [J_{+, \rm cap}(\psi) + J_{\rm th}(\psi)] = 0.
\end{equation}
We iteratively solve this system using a root-finding algorithm to determine the unique $\psi$ that satisfies $J_{\rm net}(\psi) = 0$ and the corresponding electron density $n_e(\psi)$ from Equation \ref{eq:ne}.

\subsection{Charge Simulation Results} \label{sec:charge-res}

Figure \ref{fig:dust-grain} presents the equilibrium state of the dusty plasma, illustrating the charge carrier densities, production/loss rates, and microscopic grain currents. These results correspond to the nominal cloud properties simulated using \texttt{ExoLyn} with a nucleation rate of $\dot{\Sigma}_n = 10^{-4}~\mathrm{g~cm^{-2}~s^{-1}}$ (as shown in Figure \ref{fig:exolyn}).

We identify a distinct regime shift in the electrical structure at approximately 3000~bar (Figure \ref{fig:dust-grain}a). In the deep atmosphere ($P > 3000$~bar), the charge balance follows the dust-free thermodynamic baseline; the system is governed by free electrons ($n_e$) and gas-phase cations ($n_+$), while the contribution of charged dust grains ($|Z|n_d$) is negligible. However, at pressures lower than 3000~bar, the system transitions to a dust-dominated regime. Here, negatively charged grains become the primary negative charge reservoir, balancing the gas-phase cations. Consequently, the free electron density is significantly depleted relative to the dust-free baseline, dropping by about an order of magnitude within the Juno MWR sensitivity window (1000--2000~bar).

The mechanism driving this depletion is revealed by the electron sources and sinks (Figure \ref{fig:dust-grain}b) and microscopic currents (Figure \ref{fig:dust-grain}c). Even in the presence of dust, the global source and sink terms are still dominated by gas-phase ionization ($Q$) and recombination ($k_{\rm rec}n_e n_+$); direct electron capture onto grains is a secondary effect. Instead, the depletion is driven indirectly by dust ion emission. As shown in Figure \ref{fig:dust-grain}c, the currents in the deep atmosphere ($P > 3000$~bar) are dominated by the emission and capture of ions. The vigorous desorption of alkali ions (${\rm K}^+$) from the grain surface significantly enhances the gas-phase cation density $n_+$. To maintain ionization equilibrium ($Q \approx k_{\rm rec} n_e n_+$) with a fixed thermal source $Q$, this enhanced cation population forces a compensatory reduction in the free electron density. Thus, the dust acts as a catalyst for recombination by pumping positive ions into the gas phase.

We can quantify the efficiency of this depletion by analyzing the effective electron loss frequency (Figure \ref{fig:exolyn_mwr}c). In a dust-free gas (orange line), the loss frequency is defined by recombination with cations, $\nu_{\rm gas} = k_{\rm rec} n_{+,0}$, which decreases naturally toward lower pressures. In the dusty scenario, we define an effective dust-induced loss frequency, $\nu_{\rm dust} \approx k_{\rm rec} (n_+ - n_{+,0})$, which accounts for the extra cations provided by the dust. In our nominal case (magenta line), $\nu_{\rm dust}$ exceeds $\nu_{\rm gas}$ by more than an order of magnitude in the 1000--2000~bar region, confirming the dust control of the electron lifetime at these levels.

This dust-induced loss cuts off sharply at approximately $10^4$~bar (Figure \ref{fig:exolyn_mwr}c). This termination occurs significantly deeper than the cloud base. Although particle size increases with depth due to condensation, the cutoff is governed by surface chemistry rather than dust cloud mass. As temperatures rise in the deep atmosphere, the surface coverage of potassium atoms sharply drops. Adsorbed atoms gain enough thermal energy to overcome the desorption barrier, effectively cleaning the grain surface and shutting off the ion emission.

The efficiency of electron sink depends on the total surface area of the dust cloud, which is governed by the nucleation rate. A higher nucleation rate yields smaller characteristic particle radii ($r_p$) for a given mass, resulting in a larger total surface area available for more ion emission. In our simulations, the dust loss frequency intersects and exceeds the gas-phase loss frequency in the 1000--2000 bar region. For the high nucleation case, the dust loss rate can exceed the gas loss rate by orders of magnitude, explaining why electrons are efficiently depleted in this pressure window (Figure \ref{fig:exolyn_mwr}d).

The resulting equilibrium electron profiles for the three \texttt{ExoLyn} microphysical cases are compared with the gas-phase equilibrium (rainout scenario in \texttt{GGchem}) in Figure \ref{fig:exolyn_mwr}d. In the nominal microphysical case ($\dot{\Sigma}_n = 10^{-4}~\mathrm{g~cm^{-2}~s^{-1}}$), the electron density drops by an order of magnitude in the 1000--2000 bar region compared to the gas-only equilibrium. This depletion effectively increases the transparency of the deep atmosphere, allowing the model to match the MWR observations at 0.6 GHz.

The magnitude of depletion increases with the nucleation rate. In the high nucleation case ($\dot{\Sigma}_n = 10^{-2}~\mathrm{g~cm^{-2}~s^{-1}}$), the electron density is suppressed by a factor of $\sim$100. Because micron-sized grains have negligible opacity at 0.6 GHz compared to free electrons, this process effectively renders the deep atmosphere transparent to microwave radiation without requiring subsolar alkali abundances. We find that the high nucleation case produces a limb-darkening profile consistent with the data within uncertainties, although the nadir brightness temperature is slightly higher than observed. We note, however, that the ammonia opacity remains uncertain at these pressures; a modest adjustment to the ammonia absorption could reconcile the modeled $T_b$ values with observations. On the other hand, the low nucleation case ($\dot{\Sigma}_n = 10^{-6}~\mathrm{g~cm^{-2}~s^{-1}}$) produces insufficient electron depletion to explain the observed limb darkening. These results demonstrate that, if the deep atmosphere is sufficiently turbulent to sustain a population of suspended submicron aerosols, dust-catalyzed recombination offers a viable physical mechanism to reconcile the solar composition of Jupiter with the MWR measurements.

\section{Implications for Deep Atmospheric Variability} \label{sec:variability}

Both the alkali sequestration and dust-catalyzed recombination scenarios predicate the depletion of free electrons on the presence of mineral clouds. Since cloud formation is an inherently dynamic process driven by convection and global circulation, we expect the resulting electron abundances at the kilobar level to be spatially heterogeneous. This contrasts with the uniform distribution expected if the bulk alkali abundances were genuinely subsolar throughout Jupiter's interior. Such inhomogeneities and dynamical events should manifest as observable brightness temperature variations in the deep atmosphere, which are probed by the MWR 0.6 GHz channel. 

Here, we present a preliminary analysis of an extended Juno MWR dataset spanning 61 perijoves to search for this deep atmospheric variability. We focused on the nadir-view brightness temperatures ($T_b$), processed following the same gravity correction and limb-darkening removal procedures described in Section \ref{sec:data}.

\subsection{Differential Frequency-Scaling Method} \label{subsec:f12}

Discerning deep variability observationally requires separating the signal originating in the deep atmosphere ($P > 100$ bar) from the variability of the upper atmosphere (the ``weather layer", $P \lesssim 100$ bar), which significantly contaminates the 0.6 GHz channel. A comprehensive retrieval framework using all MWR channels is beyond the scope of this study. As an approximation method toward revealing potential variability in Jupiter's deep atmosphere, we employ a differential analysis method using MWR channel 1 (0.6 GHz) and channel 2 (1.25 GHz). 

The brightness temperature fluctuation $\delta T_b$ observed in a given channel $\nu$ can be expressed to first order as the integral of the weighting function $W(P, \nu)$ and the perturbation of the microwave source function $\delta S(P)$ (arising from variations in temperature or absorber mixing ratio):
\begin{equation}
    \delta T_b(\nu) = \int_{\text{top}}^{\text{bottom}} W(P, \nu) \, \delta S(P) \, d\ln P.
\end{equation}
As shown in Figure \ref{fig:TP}a, the weighting function for channel 1 is bimodal, with sensitivity peaks in both the upper atmosphere ($P \sim 80$ bar) and the deep interior ($P \sim 1500$ bar). Conversely, channel 2 is sensitive primarily to the upper atmosphere (peaking near 30 bar) and is effectively opaque to the kilobar level.

\begin{figure*}
    \centering
    \includegraphics[width=1\linewidth]{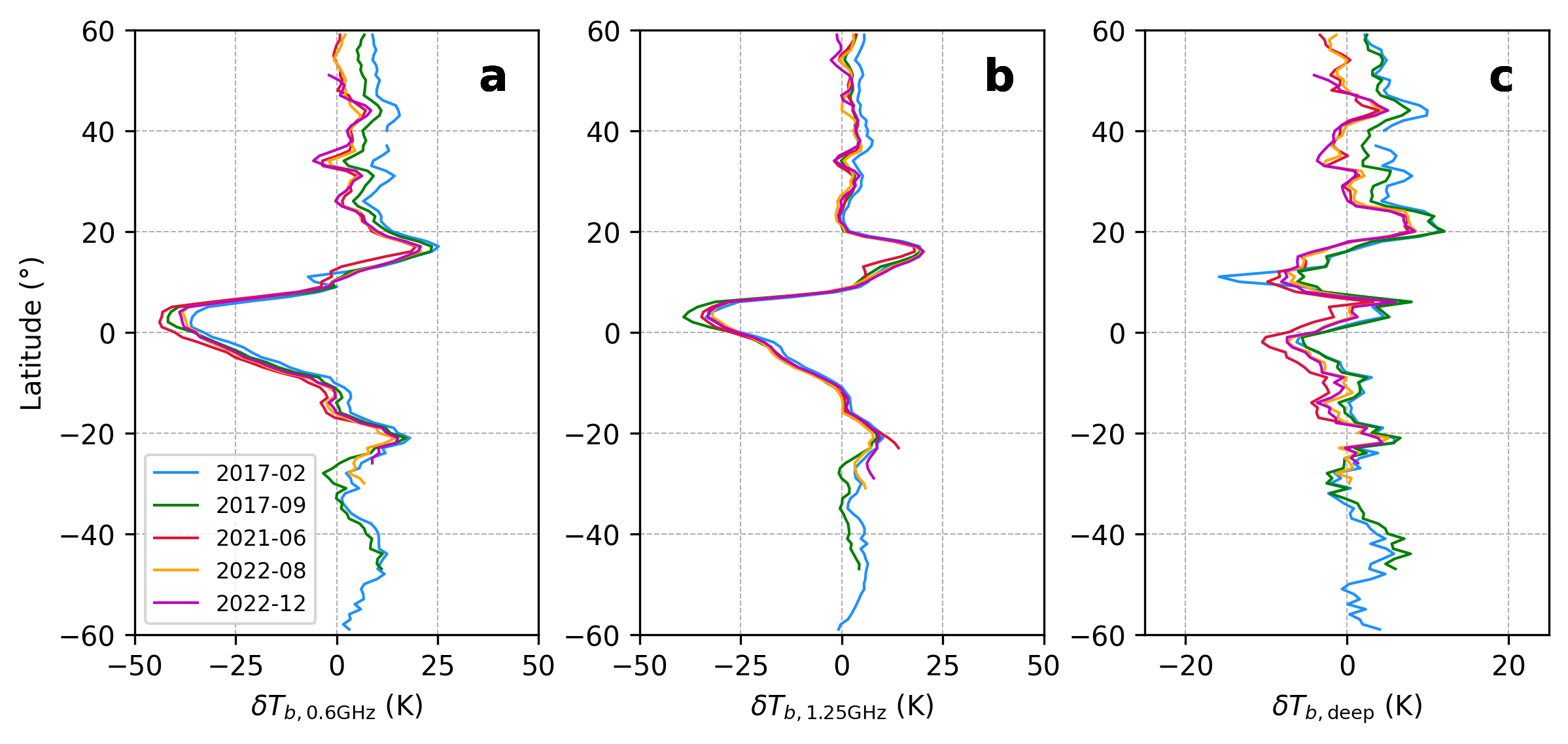}
    \caption{Latitudinal profiles of brightness temperature anomalies for selected perijoves. \textbf{(a)} Channel 1 (0.6 GHz) anomalies. \textbf{(b)} Channel 2 (1.25 GHz) anomalies. \textbf{(c)} The minimum deep-layer anomalies ($\delta T_{b,\mathrm{deep}}$), obtained by subtracting the scaled channel 2 signal from channel 1 (Equation~\ref{eq:deep_residual}), representing the atmospheric structure below 100 bar. Anomalies are derived relative to the global mean brightness temperature.}
    \label{fig:lat_variability}
\end{figure*}

We assume that atmospheric variations in the upper atmosphere are vertically coherent over the sensitivity range of channels 1 and 2. Under this assumption, the upper-atmospheric contribution to the channel 1 signal can be approximated as a linear scaling of the channel 2 signal. We define the deep brightness temperature anomaly, $\delta T_{b,\mathrm{deep}}$, as:
\begin{equation}
    \delta T_{b,\mathrm{deep}} \approx \delta T_{b,0.6~\mathrm{GHz}} - f_{12} \, \delta T_{b,1.25~\mathrm{GHz}},
    \label{eq:deep_residual}
\end{equation}
where the scaling factor $f_{12}$ represents the sensitivity of channel 1 to upper-layer perturbations probed by channel 2. We define $f_{12}$ as the ratio of the brightness temperature responses to a perturbation in the atmospheric state vector $\mathbf{X}_{\mathrm{upper}}$ (e.g., ammonia or temperature) confined to $P < 100$ bar:
\begin{equation}
    f_{12} = \frac{\partial T_{b,0.6~\mathrm{GHz}} / \partial \mathbf{X}_{\mathrm{upper}}}{\partial T_{b,1.25~\mathrm{GHz}} / \partial \mathbf{X}_{\mathrm{upper}}}.
\end{equation}

Forward modeling indicates that $f_{12}$ is not a unique constant; it depends on the nature and depth of the perturbation. Perturbing temperature versus ammonia, or varying the pressure level of the perturbation, yields a range of theoretical values for $f_{12}$ spanning approximately $0.4$ (when the perturbation is restricted to $P < 20$ bars) to $1.4$ (when the perturbation penetrates deeper than 20 bars). Given this uncertainty, a single fixed scaling factor cannot robustly quantify the exact amplitude of the deep atmospheric variability.

Instead, we adopt an approach to establish a \textit{lower limit} on the variability. Our goal is to determine if any residual variability remains in channel 1 after subtracting the upper-atmosphere signal probed by channel 2. We proceed as follows:
\begin{enumerate}
    \item For each perijove, we calculate the $T_b$ anomaly for both channels by subtracting the global mean (weighted average across latitudes) from the zonally averaged latitudinal profile. By analyzing the anomalies relative to the mean, we mitigate the influence of calibration uncertainty. 
    \item We survey the parameter space of $f_{12}$ from 0.4 to 1.4. For each test value, we compute the deep residual $\delta T_{b,\mathrm{deep}}$ using Equation \ref{eq:deep_residual}.
    \item We calculate the standard deviation of $\delta T_{b,\mathrm{deep}}$ across latitudes as a metric of variability.
    \item Across all possible values of $f_{12}$, we use the minimum standard deviation as the lower bound for the deep atmospheric variability.
\end{enumerate}
We find that a scaling factor of $f_{12}=1.2$ minimizes the standard deviation of the residual signal. If this minimized residual still exhibits significant spatial structure, it provides robust evidence that the deep atmosphere of Jupiter is variable.

\subsection{Deep Atmospheric Variability}

\begin{figure*}
    \centering
    \includegraphics[width=0.7\linewidth]{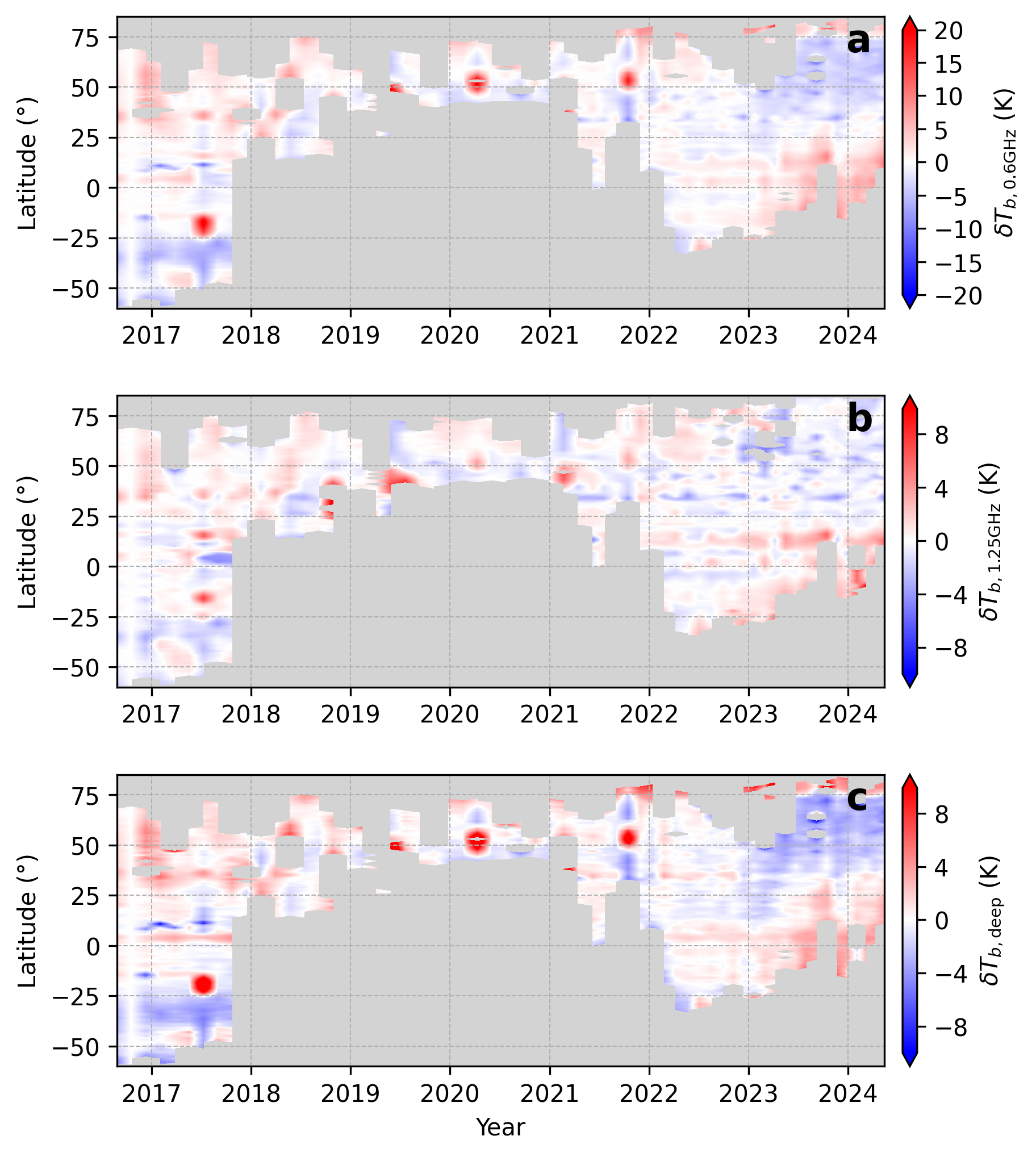}
\caption{Latitude–time map of MWR brightness temperature anomalies, $\delta T_b$, across 61 perijoves. \textbf{(a)} channel 1 (0.6 GHz) anomalies. \textbf{(b)} channel 2 (1.25 GHz) anomalies. \textbf{(c)} The minimum deep-layer anomalies. The color scale denotes deviations in kelvin.}
    \label{fig:temporal_variability}
\end{figure*}

Both channels 1 and 2 exhibit substantial spatial variability. The latitudinal brightness temperature anomalies, $T_b$, relative to the global mean are shown in Figure \ref{fig:lat_variability} for a representative subset of perijoves. The observed variability exceeds the instrumental noise level of approximately $0.5$ K \citep{janssenMicrowaveRemoteSensing2005}. The pronounced negative anomalies at low latitudes are primarily driven by enhanced ammonia abundances in the equatorial zone relative to the mid- and high-latitude regions \citep{liDistributionAmmoniaJupiter2017}. Figure \ref{fig:lat_variability}c shows the minimum deep-layer anomaly, $\delta T_{b,\mathrm{deep}}$, computed using Equation \ref{eq:deep_residual} with $f_{12}=1.2$. Even in this conservative limit, the equatorial and low-latitude regions (within $\pm 20^\circ$) exhibit significant variations ranging from $-10$ to $+10$ K with banded pattens, while the midlatitude variations are milder, on the order of a few kelvin. Since the primary weighting function peak of channel 1 is located at $\sim 1500$ bar, this residual variability (excluding channel 2 contributions) predominantly originates from the kilobar region.

The Juno data may be subject to systematic temporal drift, which prevents a conclusive assessment of true temporal variability. In addition, because each perijove samples a different longitude, it is challenging to unambiguously separate temporal variability from longitudinal variability. Nevertheless, we can evaluate the stability of the spatial features. Figure \ref{fig:lat_variability} already shows results from several representative perijoves. We further examine all zero-centered $\delta T_b$ measurements from the full set of 61 perijoves in Figure \ref{fig:temporal_variability}. 

The minimum $\delta T_{b,\mathrm{deep}}$ anomaly map (Figure \ref{fig:temporal_variability}c) indicates that the overall amplitude of the latitudinal variability remains largely invariant with time and longitude. Several discrete features appear episodically and are sampled only during individual perijove flybys. For example, a localized anomaly between $40^\circ$–$50^\circ$ N in 2020 exhibits a strong signal in channel 1 but not in channel 2, suggesting that some deep perturbations do not penetrate into the overlying shallow layers. However, we find no compelling evidence for coherent or systematic evolution in these discrete features over the duration of the Juno mission. To first order, the latitudinal variability in the deep atmosphere appears stable over time.

This persistent deep latitudinal variability could arise from variations in either physical temperature, composition or electron density. Gravity measurements from Juno have constrained that Jupiter's zonal jets extend deep into the atmosphere, reaching depths of approximately $10^5$ bar ($3000$ km below the cloud top) and decaying inward with a cylindrical geometry \citep[e.g.,][]{guillotSuppressionDifferentialRotation2018, kaspiJupitersAtmosphericJet2018, kaspiObservationalEvidenceCylindrically2023}. To maintain these flows, thermal wind balance requires density perturbations along isobaric surfaces, which manifest as latitudinal temperature gradients. 

\citet{liuPredictionsThermalGravitational2013a} calculated the resulting temperature variations for deep jets of varying penetration depths, assuming the jets rotate on cylinders within Jupiter. The jet depth consistent with Juno observations (e.g., the case with $r_c = 0.94~R_J$) predicts a deep thermal structure featuring a warmer equator and colder midlatitudes, with amplitudes of only a few kelvin. However, the variation amplitude is much smaller than that observed at low latitudes in our derived minimum $\delta T_{b,\mathrm{deep}}$ profile (Figure \ref{fig:lat_variability}c). Furthermore, the $\delta T_{b,\mathrm{deep}}$ pattern remains relatively flat beyond $20^\circ$ latitude, which is inconsistent with the steady poleward cooling trend expected from deep jet dynamics.

Therefore, we conclude that temperature variations alone cannot explain the observed signal. We attribute the deep atmospheric variability to substantial spatial variations in electron density, likely driven by the nonuniform distribution of deep mineral clouds, or a variation of water concentration that would influence the formation of mineral clouds. This conclusion aligns with our theoretical framework: regions of varying vertical mixing strength, analogous to the cloudless belts and ammonia cloudy zones in the upper troposphere, would sustain different cloud properties. These spatial differences can arise either via the formation of feldspar and leucite clouds or through the lofting of submicron particles, ultimately producing spatially variable electron depletion. Future high-resolution mapping of the deep atmosphere could further distinguish between a uniform heavy-element depletion and the patchy, cloud-driven depletion mechanisms proposed here.

\begin{figure*}
  \centering \includegraphics[width=0.7\textwidth]{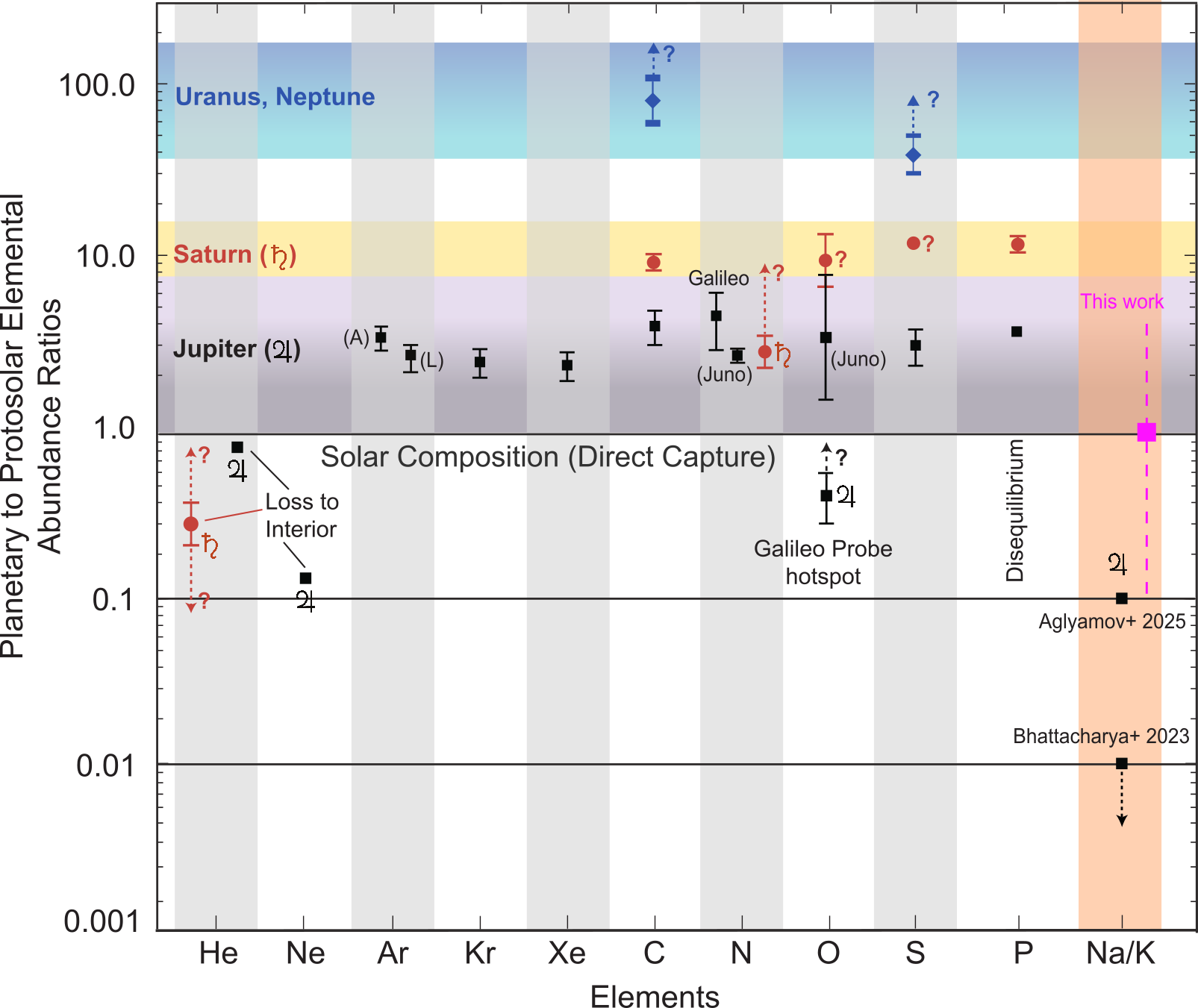}
  \caption{Elemental abundance ratios in Jupiter, Saturn, Uranus, and Neptune compared to protosolar values. This figure is adapted from Figure 3 in \cite{atreyaDeepAtmosphereComposition2020} and includes alkali metal constraints from \cite{bhattacharyaHighlyDepletedAlkali2023}, \cite{aglyamovAlkaliMetalDepletion2025}, and this work. Refer to \cite{atreyaDeepAtmosphereComposition2020} for details on other elements.} 
  \label{fig:element}
\end{figure*}

\section{Conclusion and Discussion} \label{sec:conclusion}

The apparent depletion of alkali metals in Jupiter's deep atmosphere, as inferred from Juno MWR data, presents a significant challenge to standard formation models that predict supersolar enrichments for all heavy elements. In this work, we have demonstrated that this observed depletion does not necessarily imply a low bulk abundance of sodium and potassium. Instead, it can be explained by the sequestration of these species and/or their associated free electrons into deep mineral clouds. We proposed two viable and non-mutually exclusive mechanisms:

1. Chemical sequestration via vertical mixing and mineral formation: strong vertical mixing lofts deep refractory condensates (e.g., spinel, perovskite) into the 1000--2000 bar region. Heterogeneous reactions with the gas phase facilitate the formation of alkali feldspars (albite) and feldspathoids (leucite). This process chemically locks \ce{Na} and \ce{K} into solids, thereby lowering the free electron density.

2. Electron depletion via dust-catalyzed recombination: even if alkali metals remain abundant in the gas phase, the free electron density can be efficiently suppressed by recombining with thermally emitted alkali ions from the surfaces of deep iron and silicate clouds. This mechanism requires the deep atmospheric clouds to exist as sub-micron particles characterized by high number densities of sub-micron particles, a condition that implies high nucleation rates and vigorous turbulent mixing.

As a proof of concept, we ran thermochemical and microphysical models assuming a solar alkali metallicity. We showed that both mechanisms can successfully reproduce the brightness temperature and limb-darkening observations from the Juno MWR 0.6 GHz channel. These results reconcile the MWR observations with the Galileo probe measurements, supporting a formation scenario where Jupiter is globally enriched in heavy elements via icy planetesimal accretion. 

Using newly obtained constraints on refractory elements in giant planets, we have updated the classic heavy elemental abundance diagram shown in Figure \ref{fig:element}, originally adapted from Figure 3 in \citealt{atreyaDeepAtmosphereComposition2020}), to incorporate alkali metals based on constraints from \cite{bhattacharyaHighlyDepletedAlkali2023}, \cite{aglyamovAlkaliMetalDepletion2025}, and this work. We conclude that Juno MWR observations are consistent with solar metallicity alkali. However, we did not rigorously test the upper and lower limits due to large parameter uncertainties in mineral chemistry and cloud microphysics. Instead, we illustrate a physically motivated plausible range using dashed lines in Figure \ref{fig:element}. The lower bound of $0.1\times$ solar represents the threshold indicated by rainout chemistry (Figure \ref{fig:nak}; \citealt{aglyamovAlkaliMetalDepletion2025}). The upper bound extends to $\sim 3\times$ solar to highlight that supersolar alkali enrichments, which are expected to track sulfur and phosphorus abundances under standard planetesimal or pebble accretion scenarios, remain theoretically plausible if the deep atmosphere sustains a highly efficient dust-catalyzed electron sink.

Both our proposed mechanisms rely intimately on the presence of clouds in the deep atmosphere, which are expected to be dynamic and inhomogeneous. We analyzed the complete dataset of 61 Juno perijoves to investigate potential temporal and spatial variations in the MWR 0.6 GHz signal. By differentiating the 0.6 GHz variations from the 1.25 GHz variations, we filtered out upper-atmosphere signals to focus on variability at the kilobar level. 

We detected a clear signal of spatial (latitudinal) variability in the deep atmosphere. The amplitude of this variation appears more consistent with electron abundance fluctuations, likely driven by the heterogeneous distribution of mineral clouds rather than the small temperature gradients associated with deep zonal jets. The latitudinal variability appears consistent over the Juno mission duration. Future work utilizing a more robust retrieval framework is required to fully quantify these data uncertainties and further resolve the origin of Jupiter's deep atmospheric structure.

Our study implies that the deep atmosphere of Jupiter is chemically active and vertically connected. This challenges the traditional `rainout scenario, which assumes a stratified condensation sequence where layers are effectively isolated by gravitational settling. By estimating the vertical mixing timescale, we found that convective transport is sufficient to loft particles as large as 10 microns against gravity. Consequently, vertical mixing, which was neglected in previous thermochemical frameworks of Jupiter's deep clouds, plays a crucial role in shaping the chemistry and cloud distribution, bringing the Jovian paradigm closer to the standard picture of hot exoplanet and brown dwarf atmospheres.

There are caveats to our investigation. In the chemical sequestration scenario, the equilibrium chemistry represents an end-member case assuming mixing is efficient enough to maintain chemical contact between species. On Jupiter, deep iron condensates likely do not mix efficiently all the way to the upper atmosphere ($P <$ 100 bar), which prevents the complete depletion of volatile sulfur (as \ce{FeS}) and remains consistent with the \ce{H2S} measurements by the Galileo probe (see Figure \ref{fig:element}). Moreover, cloud and condensate formation is a kinetic process, and non-equilibrium ``dirty” grains or core--mantle structures may form rather than pure mineral phases (e.g., \citealt{hellingDustBrownDwarfs2008}). The kinetic pathway to form complex aluminosilicates at 1000--2000 bar must be sufficiently fast to overcome gravitational settling. We have proposed that alkali metals and water vapor act as potent fluxing agents to create transient, chemically complex liquid phases that accelerate these heterogeneous reactions via a solution-mediated mechanism, but laboratory measurements of mineral kinetics in hydrogen-dominated, high-pressure environments are urgently needed.

For the dust-catalyzed recombination scenario, the primary uncertainty lies in the nucleation rate. The efficiency of electron removal scales with the total surface area of the cloud, which depends on the particle size distribution. Without a precise understanding of the nucleation pathways for deep refractory species (e.g., \ce{Fe}, \ce{SiO}), the microphysical properties of the cloud deck remain unconstrained. Furthermore, the coupling between plasma chemistry and cloud microphysics may be regulated by grain charging. As demonstrated in studies of protoplanetary disks and brown dwarf atmospheres, electrostatic repulsion can create a Coulomb barrier that inhibits coagulation (e.g., \citealt{okuzumiELECTRICCHARGINGDUST2009,bach-mollerAggregationChargingMineral2024}), while extreme charging may lead to the electrostatic disruption of grains (e.g., \citealt{hellingIONIZATIONATMOSPHERESBROWN2013, starkInhomogeneousCloudCoverage2015}). Both processes serve to suppress grain growth, maintaining a population of small particles with large surface areas. Future investigations into the formation of nucleation seeds and charge--microphysics feedbacks are required to further refine this scenario. 

Finally, our findings establish a strong connection between Jupiter and the study of hot exoplanets and brown dwarfs. Observations from JWST have revealed silicate features in VHS 1256 b \citep{milesJWSTEarlyreleaseScience2023} and quartz features in WASP-17b \citep{grantJWSTTSTDREAMSQuartz2023}, confirming the presence of dynamic mineral clouds in substellar photospheres. In late-T to early-Y brown dwarfs ($T_{\rm eff} \approx 600$--$1000$ K), a decreasing trend in potassium abundance has historically been interpreted as evidence for the rainout scenario, where deep silicate settling leaves alkalis to condense as salt clouds \citep[e.g.,][]{lineUniformAtmosphericRetrieval2017, zaleskyUniformRetrievalAnalysis2022}. However, our equilibrium models demonstrate that the formation of feldspars and feldspathoids leaves residual alkali vapor, which eventually forms salt clouds at lower pressures, regardless. Thus, the presence of salts is not a unique signature of rainout. Moreover, the surface gravity of brown dwarfs is significantly higher than that of Jupiter, which would enhance cloud settling and modify the vertical distribution of condensates. A comparative analysis of brown dwarfs and Jupiter's deep atmosphere, supported by future JWST and radio observations alongside cloud simulations, offers a promising avenue to unveil the universal physics of mineral clouds in substellar atmospheres.

Although Jupiter is classified as a gas giant, the region between $10^2$ and $10^5$ bar is dominated by mineral chemistry. To some extent, this region--spanning a vertical extent of $\sim$3000 km with basal temperatures of $3000$--$4000$ K--shares chemical similarities with Earth's mantle, albeit driven by vigorous gas-phase convection rather than sluggish solid-state flow. The microwave observation is thus not merely a radiative probe of temperature and elemental metallicity, but a mineralogical diagnostic of the complex interplay among atmospheric dynamics, cloud microphysics, heterogeneous chemistry, and even plasma processes in this deep ``mineralogical zone'' of Jupiter.

\begin{acknowledgments}
We thank James Owen for discussions on dust--plasma interaction, Kazumasa Ohno for discussions on planetary formation and microphysics, Roberto Tejada Arevalo and Yubo Su for providing the EOS data and scripts, Helong Huang for assistance with the \texttt{Exolyn} setup, and Fabiano Oyafuso and Zhimeng Zhang for providing the MWR data. X.Z. is supported by the National Science Foundation Astronomy and Astrophysics Research grant (AAG) 2307463, the NASA Exoplanet Research grant (XRP) 80NSSC22K0236, and the NASA Interdisciplinary Consortia for Astrobiology Research grant (ICAR) 80NSSC21K0597. C.L. and J.H. are supported by NASA’s Juno project NNM06AA75C and a subaward to the University of Michigan with project No. Q99063JAR. Q.W. is supported by the National Science Foundation Division of Earth Sciences grant (EAR) 2017294.
\end{acknowledgments}

\begin{table*}[ht!]
    \caption{Chemical Species in Condensation Groups \label{tab:condensates}}
    \centering
    \small
    \renewcommand{\arraystretch}{1.3}
    \setlength{\tabcolsep}{6pt}
    \newcommand{\wrap}[1]{\parbox{11cm}{\raggedright #1}}
    \begin{tabular}{ll}
        \toprule
        \textbf{Condensation Group} & \textbf{Member Species} \\
        \midrule
        \textbf{Silicates} & 
        \wrap{\ce{SiO}, \ce{SiO2}, \ce{MgSiO3}, \ce{Mg2SiO4}, \ce{Mn2SiO4}, \ce{NaAlSiO4}, \ce{Fe2SiO4}, \ce{Na2SiO3}, \ce{CaMgSi2O6}, \ce{KAlSiO4}, \ce{CaSiO3}, \ce{Ca3Fe2Si3O12}, \ce{Mn3Al2Si3O12}, \ce{Ca2SiO4}, \ce{KAlSi2O6}, \ce{NaAlSi2O6}, \ce{MnSiO3}, \ce{NaFeSi2O6}, \ce{NaCrSi2O6}, \ce{K2SiO3}, \ce{Ca2Al2SiO7}, \ce{Ca3Al2Si3O12}, \ce{CaAl2SiO6}, \ce{Ca3Si2O7}, \ce{CaTiSiO5}, \ce{Ca2MgSi2O7}} \\
        \midrule
        \textbf{Feldspar (Plagioclase)} & 
        \wrap{\ce{KAlSi3O8}, \ce{CaAl2Si2O8}, \ce{NaAlSi3O8}} \\
        \midrule
        \textbf{Phyllosilicates} & 
        \wrap{\ce{Mg3Si2O9H4}, \ce{NaMg3AlSi3O12H2}, \ce{Mg3Si4O12H2}, \ce{FeAl2SiO7H2}, \ce{KFe3AlSi3O12H2}, \ce{CaAl2Si2O10H4}, \ce{Fe3Si2O9H4}, \ce{KMg3AlSi3O12H2}, \ce{CaAl2Si4O16H8}, \ce{Al2Si2O9H4}} \\
        \midrule
        \textbf{Refractory Oxides} & 
        \wrap{\ce{Al2O3}, \ce{TiO2}, \ce{Ti2O3}, \ce{Ti4O7}, \ce{MgTi2O5}, \ce{TiO}, \ce{Ti3O5}, \ce{CaO}, \ce{ZrO2}, \ce{WO3}, \ce{VO}, \ce{V2O3}, \ce{V2O4}, \ce{V2O5}} \\
        \midrule
        \textbf{Fe-Ni-Cr-Co-Cu Metals and Oxides} & 
        \wrap{\ce{Fe}, \ce{Ni}, \ce{Cr}, \ce{CrN}, \ce{FeO}, \ce{Fe2TiO4}, \ce{Fe2O3}, \ce{Fe3O4}, \ce{Co}, \ce{Cu}, \ce{Zn}} \\
        \midrule
        \textbf{Sulfides and Sulfates} & 
        \wrap{\ce{S}, \ce{S2}, \ce{S8}, \ce{FeS}, \ce{FeS2}, \ce{MnS}, \ce{Ni3S2}, \ce{MgS}, \ce{CaS}, \ce{Na2S}, \ce{NiS}, \ce{NiS2}, \ce{CaSO4}, \ce{FeSO4}, \ce{MgSO4}, \ce{K2SO4}, \ce{Na2SO4}, \ce{ZnS}, \ce{CoSO4}, \ce{CuSO4}, \ce{ZnSO4}} \\
        \midrule
        \textbf{Carbon and Carbides} & 
        \wrap{\ce{C}, \ce{CaCO3}, \ce{CaMgC2O6}, \ce{CaFeC2O6}, \ce{NaAlCO5H2}, \ce{SiC}, \ce{VC}, \ce{TiC}} \\
        \midrule
        \textbf{Halides} & 
        \wrap{\ce{NaCl}, \ce{KCl}, \ce{LiCl}, \ce{AlCl3}, \ce{CaCl2}, \ce{AlF3}, \ce{CaF2}, \ce{KF}, \ce{NaF}, \ce{FeF2}, \ce{FeCl3}} \\
        \midrule
        \textbf{Phosphorus Compounds} & 
        \wrap{\ce{Ca5P3O12F}, \ce{Ca5P3O13H}, \ce{Mg3P2O8}, \ce{P4O10}, \ce{P4S3}} \\
        \bottomrule
    \end{tabular}
    \label{table:chem}
\end{table*}

\begin{figure*}
  \centering \includegraphics[width=0.99\textwidth]{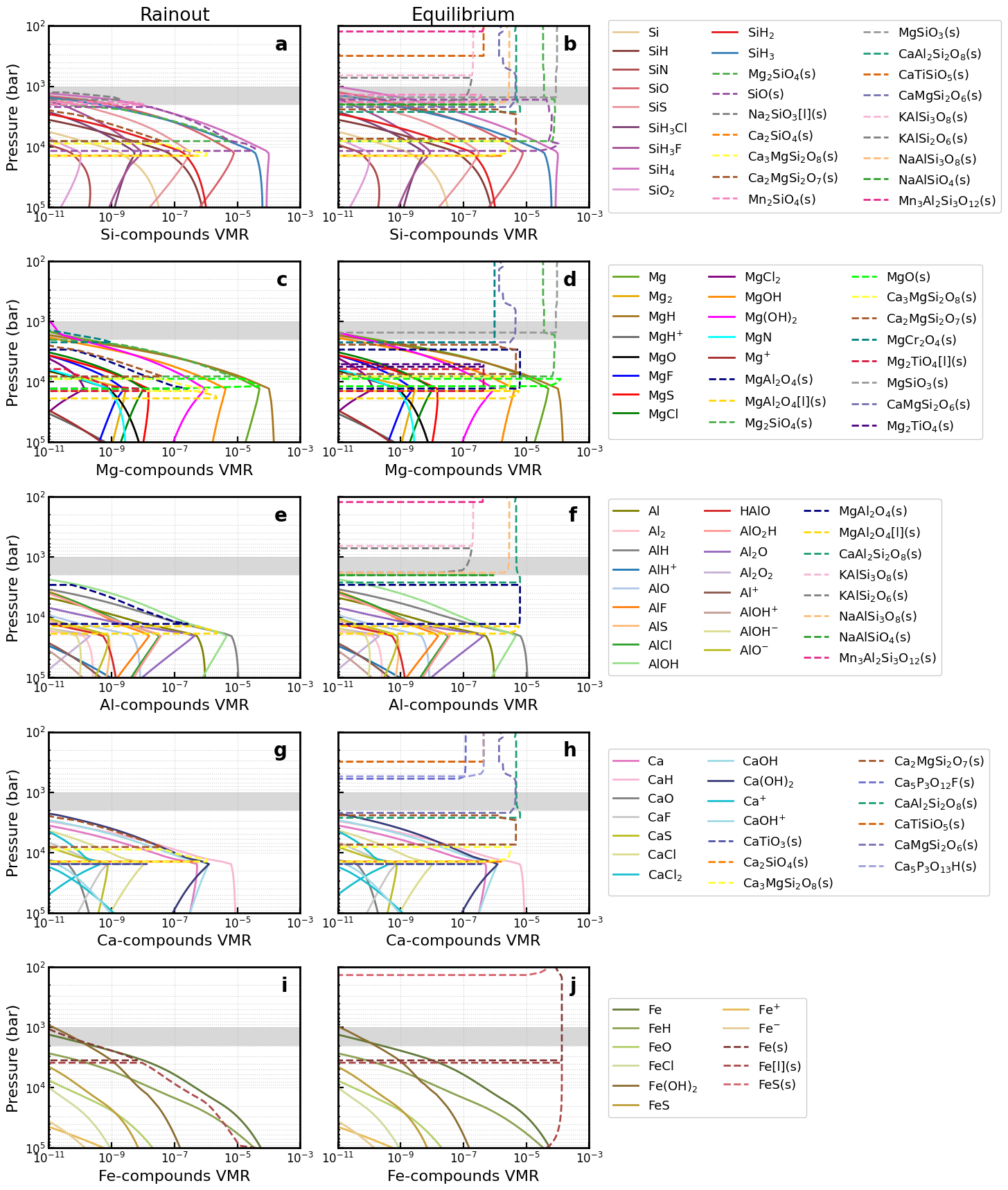} 
  \caption{Vertical distribution of refractory species (\ce{Si}, \ce{Mg}, \ce{Al}, \ce{Ca}, \ce{Fe}) that serve as the ``building blocks" for deep clouds. \textbf{Left column (rainout):} refractory species condense in distinct, separated layers. aluminum (\ce{Al}) condenses primarily as spinel (\ce{MgAl2O4}) and calcium (\ce{Ca}) as titanates/silicates at pressures $>2500$ bar (bottom panels). Because these seeds settle out before reaching the silicate condensation level, the middle atmosphere contains magnesium silicates (enstatite \ce{MgSiO3}) but is completely depleted of \ce{Al} and \ce{Ca} gas. \textbf{Right column (equilibrium):} strong vertical mixing maintains a supply of \ce{Al}, \ce{Ca}, and \ce{Fe} condensates in the upper layers. The availability of aluminum (third row, right) is the critical factor that enables the formation of alkali feldspars and leucite seen in Figure \ref{fig:nak}. Note that pure corundum (\ce{Al2O3}) does not form; instead, aluminum is hosted in spinel deep in the atmosphere and transitions to anorthite (\ce{CaAl2Si2O8}) and alkali feldspars at lower pressures ($<2000$ bar). The horizontal gray band marks the MWR sensitivity window at the 0.6 GHz channel.}
\label{fig:other_species}
\end{figure*}


\begin{thebibliography}{}
\expandafter\ifx\csname natexlab\endcsname\relax\def\natexlab#1{#1}\fi
\providecommand{\url}[1]{\href{#1}{#1}}
\providecommand{\dodoi}[1]{doi:~\href{http://doi.org/#1}{\nolinkurl{#1}}}
\providecommand{\doeprint}[1]{\href{http://ascl.net/#1}{\nolinkurl{http://ascl.net/#1}}}
\providecommand{\doarXiv}[1]{\href{https://arxiv.org/abs/#1}{\nolinkurl{https://arxiv.org/abs/#1}}}

\bibitem[{A.~S. Ackerman \& M.~S. Marley(2001)Ackerman \&
  Marley}]{ackermanPrecipitatingCondensationClouds2001}
Ackerman, A.~S., \& Marley, M.~S. 2001, \bibinfo{title}{Precipitating
  {{Condensation Clouds}} in {{Substellar Atmospheres}},} The Astrophysical
  Journal, 556, 872, \dodoi{10.1086/321540}

\bibitem[{Y.~S. Aglyamov {et~al.}(2025)Aglyamov, Atreya, Bhattacharya, Li,
  Levin, Bolton, \& Wong}]{aglyamovAlkaliMetalDepletion2025}
Aglyamov, Y.~S., Atreya, S.~K., Bhattacharya, A., {et~al.} 2025,
  \bibinfo{title}{Alkali Metal Depletion in the Deep {{Jovian}} Atmosphere:
  {{The}} Role of Anions,} Icarus, 425, 116334,
  \dodoi{10.1016/j.icarus.2024.116334}

\bibitem[{A.~F. Ashton \& A.~N. Hayhurst(1973)Ashton \&
  Hayhurst}]{ashtonKineticsCollisionalIonization1973}
Ashton, A.~F., \& Hayhurst, A.~N. 1973, \bibinfo{title}{Kinetics of Collisional
  Ionization of Alkali Metal Atoms and Recombination of Electrons with Alkali
  Metal Ions in Flames,} Combustion and Flame, 21, 69,
  \dodoi{10.1016/0010-2180(73)90008-4}

\bibitem[{M. Asplund {et~al.}(2021)Asplund, Amarsi, \&
  Grevesse}]{asplundChemicalMakeupSun2021}
Asplund, M., Amarsi, A.~M., \& Grevesse, N. 2021, \bibinfo{title}{The Chemical
  Make-up of the {{Sun}}: {{A}} 2020 Vision,} Astronomy and Astrophysics, 653,
  A141, \dodoi{10.1051/0004-6361/202140445}

\bibitem[{S.~K. Atreya {et~al.}(2020)Atreya, Hofstadter, In, Mousis, Reh, \&
  Wong}]{atreyaDeepAtmosphereComposition2020}
Atreya, S.~K., Hofstadter, M.~H., In, J.~H., {et~al.} 2020,
  \bibinfo{title}{Deep {{Atmosphere Composition}}, {{Structure}}, {{Origin}},
  and {{Exploration}}, with {{Particular Focus}} on {{Critical}} in Situ
  {{Science}} at the {{Icy Giants}},} Space Science Reviews, 216, 18,
  \dodoi{10.1007/s11214-020-0640-8}

\bibitem[{N. {Bach-M{\o}ller} {et~al.}(2024){Bach-M{\o}ller}, Helling,
  J{\o}rgensen, \& Enghoff}]{bach-mollerAggregationChargingMineral2024}
{Bach-M{\o}ller}, N., Helling, C., J{\o}rgensen, U.~G., \& Enghoff, M.~B. 2024,
  \bibinfo{title}{Aggregation and {{Charging}} of {{Mineral Cloud Particles}}
  under {{High-energy Irradiation}},} The Astrophysical Journal, 962, 87,
  \dodoi{10.3847/1538-4357/ad13ef}

\bibitem[{X.-N.
  Bai(2011)Bai}]{baiMAGNETOROTATIONALINSTABILITYDRIVENACCRETIONPROTOPLANETARY2011}
Bai, X.-N. 2011, \bibinfo{title}{{{MAGNETOROTATIONAL-INSTABILITY-DRIVEN
  ACCRETION IN PROTOPLANETARY DISKS}},} The Astrophysical Journal, 739, 50,
  \dodoi{10.1088/0004-637X/739/1/50}

\bibitem[{C.~W. Bale {et~al.}(2009)Bale, B{\'e}lisle, Chartrand, Decterov,
  Eriksson, Hack, Jung, Kang, Melan{\c c}on, Pelton, Robelin, \&
  Petersen}]{baleFactSageThermochemicalSoftware2009}
Bale, C.~W., B{\'e}lisle, E., Chartrand, P., {et~al.} 2009,
  \bibinfo{title}{{{FactSage}} Thermochemical Software and Databases --- Recent
  Developments,} Calphad, 33, 295, \dodoi{10.1016/j.calphad.2008.09.009}

\bibitem[{S.~S. Barshay \& J.~S. Lewis(1978)Barshay \&
  Lewis}]{barshayChemicalStructureDeep1978}
Barshay, S.~S., \& Lewis, J.~S. 1978, \bibinfo{title}{Chemical Structure of the
  Deep Atmosphere of {{Jupiter}},} Icarus, 33, 593,
  \dodoi{10.1016/0019-1035(78)90192-6}

\bibitem[{A. Bellotti {et~al.}(2016)Bellotti, Steffes, \&
  Chinsomboom}]{bellottiLaboratoryMeasurements5202016}
Bellotti, A., Steffes, P.~G., \& Chinsomboom, G. 2016,
  \bibinfo{title}{Laboratory Measurements of the 5-20 Cm Wavelength Opacity of
  Ammonia, Water Vapor, and Methane under Simulated Conditions for the Deep
  Jovian Atmosphere,} Icarus, 280, 255, \dodoi{10.1016/j.icarus.2016.07.013}

\bibitem[{A. Bellotti {et~al.}(2017)Bellotti, Steffes, \&
  Chinsomboon}]{bellottiCorrigendumLaboratoryMeasurements2017}
Bellotti, A., Steffes, P.~G., \& Chinsomboon, G. 2017,
  \bibinfo{title}{Corrigendum to ``{{Laboratory}} Measurements of the 5--20~Cm
  Wavelength Opacity of Ammonia, Water Vapor, and Methane under Simulated
  Conditions for the Deep Jovian Atmosphere'' [{{Icarus}} 280 (2016)
  255--267],} Icarus, 284, 491, \dodoi{10.1016/j.icarus.2016.11.006}

\bibitem[{A. Bhattacharya {et~al.}(2023)Bhattacharya, Li, Atreya, Steffes,
  Levin, Bolton, Guillot, Gupta, Ingersoll, Lunine, Orton, Oyafuso, Waite,
  Bellotti, \& Wong}]{bhattacharyaHighlyDepletedAlkali2023}
Bhattacharya, A., Li, C., Atreya, S.~K., {et~al.} 2023, \bibinfo{title}{Highly
  {{Depleted Alkali Metals}} in {{Jupiter}}'s {{Deep Atmosphere}},} The
  Astrophysical Journal Letters, 952, L27, \dodoi{10.3847/2041-8213/ace115}

\bibitem[{G. Chabrier {et~al.}(2019)Chabrier, Mazevet, \&
  Soubiran}]{chabrierNewEquationState2019}
Chabrier, G., Mazevet, S., \& Soubiran, F. 2019, \bibinfo{title}{A {{New
  Equation}} of {{State}} for {{Dense Hydrogen}}--{{Helium Mixtures}},} The
  Astrophysical Journal, 872, 51, \dodoi{10.3847/1538-4357/aaf99f}

\bibitem[{M.~W. Chase \& J.~A. N.~A. Force(1998)Chase \& Force}]{chase1998nist}
Chase, M.~W., \& Force, J. A. N.~A. 1998, \bibinfo{title}{{{NIST-JANAF}}
  Thermochemical Tables,}

\bibitem[{D.~D. Clayton(1968)Clayton}]{clayton-1968}
Clayton, D.~D. 1968, Principles of {{Stellar Evolution}} and
  {{Nucleosynthesis}} (McGraw-Hill, New York)

\bibitem[{S.~J. Desch \& N.~J. Turner(2015)Desch \&
  Turner}]{deschHIGHTEMPERATUREIONIZATIONPROTOPLANETARY2015}
Desch, S.~J., \& Turner, N.~J. 2015, \bibinfo{title}{{{HIGH-TEMPERATURE
  IONIZATION IN PROTOPLANETARY DISKS}},} The Astrophysical Journal, 811, 156,
  \dodoi{10.1088/0004-637X/811/2/156}

\bibitem[{B.~T. Draine(2004)Draine}]{draineAstrophysicsDustCold2004}
Draine, B.~T. 2004, \bibinfo{title}{Astrophysics of {{Dust}} in {{Cold
  Clouds}},} in The {{Cold Universe}}: {{Saas-Fee Advanced Course}} 32 2002
  {{Swiss Society}} for {{Astrophysics}} and {{Astronomy}}, ed. A.~W. Blain,
  F.~Combes, B.~T. Draine, D.~Pfenniger, \& Y.~Revaz (Berlin, Heidelberg:
  Springer), 213--304, \dodoi{10.1007/3-540-31636-1_3}

\bibitem[{B.~T. Draine \& B. Sutin(1987)Draine \&
  Sutin}]{draineCollisionalChargingInterstellar1987}
Draine, B.~T., \& Sutin, B. 1987, \bibinfo{title}{Collisional {{Charging}} of
  {{Interstellar Grains}},} The Astrophysical Journal, 320, 803,
  \dodoi{10.1086/165596}

\bibitem[{D.~S. Ebel {et~al.}(2000)Ebel, Ghiorso, Sack, \&
  Grossman}]{ebelGibbsEnergyMinimization2000}
Ebel, D.~S., Ghiorso, M.~S., Sack, R.~O., \& Grossman, L. 2000,
  \bibinfo{title}{Gibbs Energy Minimization in Gas + Liquid + Solid Systems,}
  Journal of Computational Chemistry, 21, 247,
  \dodoi{10.1002/(SICI)1096-987X(200003)21:4<247::AID-JCC1>3.0.CO;2-J}

\bibitem[{B.~J. Fegley \& K. Lodders(1994)Fegley \&
  Lodders}]{fegleyChemicalModelsDeep1994}
Fegley, B.~J., \& Lodders, K. 1994, \bibinfo{title}{Chemical Models of the Deep
  Atmospheres of {{Jupiter}} and {{Saturn}},} Icarus, 110, 117,
  \dodoi{10.1006/icar.1994.1111}

\bibitem[{V.~S. Fomenko \& G.~V. Samsonov(1966)Fomenko \&
  Samsonov}]{fomenkoHandbookThermionicProperties1966}
Fomenko, V.~S., \& Samsonov, G.~V., eds. 1966, Handbook of {{Thermionic
  Properties}} (Boston, MA: Springer US), \dodoi{10.1007/978-1-4684-7293-6}

\bibitem[{N.~A. Fuchs(1964)Fuchs}]{fuchsMechanicsAerosolsFuchs1964}
Fuchs, N.~A. 1964, The Mechanics of Aerosols, by {{N}}. {{A}}. {{Fuchs}}.
  {{Translated}} from the {{Russian}} by {{R}}. {{E}}. {{Daisley}} and {{Marina
  Fuchs}}. {{Translation}} Edited by {{C}}. {{N}}. {{Davies}}., rev. and enl.
  ed. edn. (Oxford: Pergamon Press; distributed in the Western Hemisphere by
  Macmillan, New York)

\bibitem[{Y.~I. Fujii {et~al.}(2011)Fujii, Okuzumi, \&
  Inutsuka}]{fujiiFASTACCURATECALCULATION2011}
Fujii, Y.~I., Okuzumi, S., \& Inutsuka, S.-i. 2011, \bibinfo{title}{A {{FAST
  AND ACCURATE CALCULATION SCHEME FOR IONIZATION DEGREES IN PROTOPLANETARY AND
  CIRCUMPLANETARY DISKS WITH CHARGED DUST GRAINS}},} The Astrophysical Journal,
  743, 53, \dodoi{10.1088/0004-637X/743/1/53}

\bibitem[{H.-P. Gail {et~al.}(2013)Gail, Wetzel, Pucci, \&
  Tamanai}]{gailSeedParticleFormation2013}
Gail, H.-P., Wetzel, S., Pucci, A., \& Tamanai, A. 2013, \bibinfo{title}{Seed
  Particle Formation for Silicate Dust Condensation by {{SiO}} Nucleation,}
  Astronomy \& Astrophysics, 555, A119, \dodoi{10.1051/0004-6361/201321807}

\bibitem[{P. Gao {et~al.}(2021)Gao, Wakeford, Moran, \&
  Parmentier}]{gaoAerosolsExoplanetAtmospheres2021}
Gao, P., Wakeford, H.~R., Moran, S.~E., \& Parmentier, V. 2021,
  \bibinfo{title}{Aerosols in {{Exoplanet Atmospheres}},} Journal of
  Geophysical Research: Planets, 126, e2020JE006655,
  \dodoi{10.1029/2020JE006655}

\bibitem[{P. Gao {et~al.}(2020)Gao, Thorngren, Lee, Fortney, Morley, Wakeford,
  Powell, Stevenson, \& Zhang}]{gaoAerosolCompositionHot2020}
Gao, P., Thorngren, D.~P., Lee, G. K.~H., {et~al.} 2020,
  \bibinfo{title}{Aerosol Composition of Hot Giant Exoplanets Dominated by
  Silicates and Hydrocarbon Hazes,} Nature Astronomy, 4, 1,
  \dodoi{10.1038/s41550-020-1114-3}

\bibitem[{H. Ge {et~al.}(2025)Ge, Li, Zhang, Ingersoll, \&
  Chen}]{geNonuniformWaterDistribution2025}
Ge, H., Li, C., Zhang, X., Ingersoll, A.~P., \& Chen, S. 2025,
  \bibinfo{title}{Nonuniform Water Distribution in {{Jupiter}}'s Midlatitudes:
  {{Influence}} of Precipitation and Planetary Rotation,} Proceedings of the
  National Academy of Sciences, 122, e2419087122,
  \dodoi{10.1073/pnas.2419087122}

\bibitem[{M.~S. Ghiorso \& R.~O. Sack(1995)Ghiorso \&
  Sack}]{ghiorsoChemicalMassTransfer1995}
Ghiorso, M.~S., \& Sack, R.~O. 1995, \bibinfo{title}{Chemical Mass Transfer in
  Magmatic Processes {{IV}}. {{A}} Revised and Internally Consistent
  Thermodynamic Model for the Interpolation and Extrapolation of Liquid-Solid
  Equilibria in Magmatic Systems at Elevated Temperatures and Pressures,}
  Contributions to Mineralogy and Petrology, 119, 197,
  \dodoi{10.1007/BF00307281}

\bibitem[{G. Gioumousis \& D.~P. Stevenson(1958)Gioumousis \&
  Stevenson}]{gioumousisReactionsGaseousMolecule1958}
Gioumousis, G., \& Stevenson, D.~P. 1958, \bibinfo{title}{Reactions of
  {{Gaseous Molecule Ions}} with {{Gaseous Molecules}}. {{V}}. {{Theory}},} The
  Journal of Chemical Physics, 29, 294, \dodoi{10.1063/1.1744477}

\bibitem[{G.~S. Golitsyn(1980)Golitsyn}]{golitsyn-1980}
Golitsyn, G.~S. 1980, \bibinfo{title}{Geostrophic {{Convection}},} Doklady
  Akademii Nauk SSSR, 251, 1356

\bibitem[{G.~S. Golitsyn(1981)Golitsyn}]{golitsyn-1981}
Golitsyn, G.~S. 1981, \bibinfo{title}{Convection Structure during Fast
  Rotation,} Doklady Akademii Nauk SSSR, 261, 317

\bibitem[{D. Grant {et~al.}(2023)Grant, Lewis, Wakeford, Batalha, Glidden,
  Goyal, Mullens, MacDonald, May, Seager, Stevenson, Valenti, Visscher,
  Alderson, Allen, Ca{\~n}as, Col{\'o}n, Clampin, Espinoza, Gressier, Huang,
  Lin, Long, Louie, {Pe{\~n}a-Guerrero}, Ranjan, Sotzen, Valentine, Anderson,
  Balmer, Bellini, Hoch, Kammerer, Libralato, Mountain, Perrin, Pueyo, Rickman,
  Rebollido, Sohn, {van der Marel}, \& Watkins}]{grantJWSTTSTDREAMSQuartz2023}
Grant, D., Lewis, N.~K., Wakeford, H.~R., {et~al.} 2023,
  \bibinfo{title}{{{JWST-TST DREAMS}}: {{Quartz Clouds}} in the {{Atmosphere}}
  of {{WASP-17b}},} The Astrophysical Journal Letters, 956, L32,
  \dodoi{10.3847/2041-8213/acfc3b}

\bibitem[{T. Guillot {et~al.}(2023)Guillot, Fletcher, Helled, Ikoma, Line, \&
  Parmentier}]{guillotGiantPlanetsInsideOut2023}
Guillot, T., Fletcher, L.~N., Helled, R., {et~al.} 2023, \bibinfo{title}{Giant
  {{Planets}} from the {{Inside-Out}},} in Protostars and {{Planets}} 7, ed.
  S.-I. Inutsuka (Astronomical Society of the Pacific),
  \dodoi{10.26624/BHFW2605}

\bibitem[{T. Guillot {et~al.}(2018)Guillot, Miguel, Militzer, Hubbard, Kaspi,
  Galanti, Cao, Helled, Wahl, \&
  Iess}]{guillotSuppressionDifferentialRotation2018}
Guillot, T., Miguel, Y., Militzer, B., {et~al.} 2018, \bibinfo{title}{A
  Suppression of Differential Rotation in {{Jupiter}}'s Deep Interior,} Nature,
  555, 227

\bibitem[{A.~C. Hack {et~al.}(2007)Hack, Thompson, \&
  Aerts}]{hackPhaseRelationsInvolving2007}
Hack, A.~C., Thompson, A.~B., \& Aerts, M. 2007, \bibinfo{title}{Phase
  {{Relations Involving Hydrous Silicate Melts}}, {{Aqueous Fluids}}, and
  {{Minerals}},} Reviews in Mineralogy and Geochemistry, 65, 129,
  \dodoi{10.2138/rmg.2007.65.5}

\bibitem[{M. Hagstr{\"o}m {et~al.}(2000)Hagstr{\"o}m, Engvall, \&
  Pettersson}]{hagstromDesorptionKineticsAtmospheric2000}
Hagstr{\"o}m, M., Engvall, K., \& Pettersson, J. B.~C. 2000,
  \bibinfo{title}{Desorption {{Kinetics}} at {{Atmospheric Pressure}}:\,
  {{Alkali Metal Ion Emission}} from {{Hot Platinum Surfaces}},} The Journal of
  Physical Chemistry B, 104, 4457, \dodoi{10.1021/jp000311w}

\bibitem[{J. Haldemann {et~al.}(2020)Haldemann, Alibert, Mordasini, \&
  Benz}]{haldemannAQUACollectionH2O2020}
Haldemann, J., Alibert, Y., Mordasini, C., \& Benz, W. 2020,
  \bibinfo{title}{{{AQUA}}: A Collection of {{H2O}} Equations of State for
  Planetary Models,} Astronomy \& Astrophysics, 643, A105,
  \dodoi{10.1051/0004-6361/202038367}

\bibitem[{T.~R. Hanley {et~al.}(2009)Hanley, Steffes, \&
  Karpowicz}]{hanleyNewModelHydrogen2009}
Hanley, T.~R., Steffes, P.~G., \& Karpowicz, B.~M. 2009, \bibinfo{title}{A New
  Model of the Hydrogen and Helium-Broadened Microwave Opacity of Ammonia Based
  on Extensive Laboratory Measurements,} Icarus, 202, 316,
  \dodoi{10.1016/j.icarus.2009.02.002}

\bibitem[{C. Helling(2019)Helling}]{hellingExoplanetClouds2019}
Helling, C. 2019, \bibinfo{title}{Exoplanet {{Clouds}},} Annual Review of Earth
  and Planetary Sciences, 47, 583, \dodoi{10.1146/annurev-earth-053018-060401}

\bibitem[{C. Helling {et~al.}(2016)Helling, Rimmer, {Rodriguez-Barrera}, Wood,
  Robertson, \& Stark}]{hellingIonisationDischargeCloudforming2016}
Helling, C., Rimmer, P.~B., {Rodriguez-Barrera}, I.~M., {et~al.} 2016,
  \bibinfo{title}{Ionisation and Discharge in Cloud-Forming Atmospheres of
  Brown Dwarfs and Extrasolar Planets,} Plasma Physics and Controlled Fusion,
  58, 074003, \dodoi{10.1088/0741-3335/58/7/074003}

\bibitem[{C. Helling {et~al.}(2008)Helling, Woitke, \&
  Thi}]{hellingDustBrownDwarfs2008}
Helling, C., Woitke, P., \& Thi, W.-F. 2008, \bibinfo{title}{Dust in Brown
  Dwarfs and Extra-Solar Planets-{{I}}. {{Chemical}} Composition and Spectral
  Appearance of Quasi-Static Cloud Layers,} Astronomy \& Astrophysics, 485, 547

\bibitem[{{\relax Ch}. Helling {et~al.}(2013)Helling, Jardine, Stark, \&
  Diver}]{hellingIONIZATIONATMOSPHERESBROWN2013}
Helling, {\relax Ch}., Jardine, M., Stark, C., \& Diver, D. 2013,
  \bibinfo{title}{{{IONIZATION IN ATMOSPHERES OF BROWN DWARFS AND EXTRASOLAR
  PLANETS}}. {{III}}. {{BREAKDOWN CONDITIONS FOR MINERAL CLOUDS}},} The
  Astrophysical Journal, 767, 136, \dodoi{10.1088/0004-637X/767/2/136}

\bibitem[{R.~A.
  Helliwell(2014)Helliwell}]{helliwellWhistlersRelatedIonospheric2014}
Helliwell, R.~A. 2014, Whistlers and {{Related Ionospheric Phenomena}}
  (Chelmsford, MA: Courier Corporation)

\bibitem[{S. Howard \& T. Guillot(2023)Howard \&
  Guillot}]{howardAccountingNonidealMixing2023}
Howard, S., \& Guillot, T. 2023, \bibinfo{title}{Accounting for Non-Ideal
  Mixing Effects in the Hydrogen-Helium Equation of State,} Astronomy \&
  Astrophysics, 672, L1, \dodoi{10.1051/0004-6361/202244851}

\bibitem[{H. Huang {et~al.}(2024)Huang, Ormel, \&
  Min}]{huangExoLynGoldenMean2024}
Huang, H., Ormel, C.~W., \& Min, M. 2024, \bibinfo{title}{{{ExoLyn}}: {{A}}
  Golden Mean Approach to Multispecies Cloud Modeling in Atmospheric
  Retrieval,} Astronomy \& Astrophysics, 691, A291,
  \dodoi{10.1051/0004-6361/202451112}

\bibitem[{M. Ilgner \& R.~P. Nelson(2006)Ilgner \&
  Nelson}]{ilgnerIonisationFractionProtoplanetary2006}
Ilgner, M., \& Nelson, R.~P. 2006, \bibinfo{title}{On the Ionisation Fraction
  in Protoplanetary Disks - {{I}}. {{Comparing}} Different Reaction Networks,}
  Astronomy \& Astrophysics, 445, 205, \dodoi{10.1051/0004-6361:20053678}

\bibitem[{A.~V. Ivlev {et~al.}(2016)Ivlev, Akimkin, \&
  Caselli}]{ivlevIONIZATIONDUSTCHARGING2016}
Ivlev, A.~V., Akimkin, V.~V., \& Caselli, P. 2016, \bibinfo{title}{{{IONIZATION
  AND DUST CHARGING IN PROTOPLANETARY DISKS}},} The Astrophysical Journal, 833,
  92, \dodoi{10.3847/1538-4357/833/1/92}

\bibitem[{M.~A. Janssen {et~al.}(2005)Janssen, Hofstadter, Gulkis, Ingersoll,
  Allison, Bolton, Levin, \& Kamp}]{janssenMicrowaveRemoteSensing2005}
Janssen, M.~A., Hofstadter, M.~D., Gulkis, S., {et~al.} 2005,
  \bibinfo{title}{Microwave Remote Sensing of {{Jupiter}}'s Atmosphere from an
  Orbiting Spacecraft,} Icarus, 173, 447, \dodoi{10.1016/j.icarus.2004.08.012}

\bibitem[{M.~A. Janssen {et~al.}(2017)Janssen, Oswald, Brown, Gulkis, Levin,
  Bolton, Allison, Atreya, Gautier, Ingersoll, Lunine, Orton, Owen, Steffes,
  Adumitroaie, Bellotti, Jewell, Li, Li, Misra, Oyafuso, {Santos-Costa},
  Sarkissian, Williamson, Arballo, Kitiyakara, {Ulloa-Severino}, Chen, Maiwald,
  Sahakian, Pingree, Lee, Mazer, Redick, Hodges, Hughes, Bedrosian, Dawson,
  Hatch, Russell, Chamberlain, Zawadski, Khayatian, Franklin, Conley,
  Kempenaar, Loo, Sunada, Vorperion, \&
  Wang}]{janssenMWRMicrowaveRadiometer2017}
Janssen, M.~A., Oswald, J.~E., Brown, S.~T., {et~al.} 2017,
  \bibinfo{title}{{{MWR}}: {{Microwave Radiometer}} for the {{Juno Mission}} to
  {{Jupiter}},} Space Science Reviews, 213, 139,
  \dodoi{10.1007/s11214-017-0349-5}

\bibitem[{F. Kargl {et~al.}(2006)Kargl, Meyer, Koza, \&
  Schober}]{karglFormationchannelsFastion2006}
Kargl, F., Meyer, A., Koza, M.~M., \& Schober, H. 2006,
  \bibinfo{title}{Formation of channels for Fast-Ion Diffusion in Alkali
  Silicate Melts: {{A}} Quasielastic Neutron Scattering Study,} Physical Review
  B, 74, 014304, \dodoi{10.1103/PhysRevB.74.014304}

\bibitem[{Y. Kaspi {et~al.}(2018)Kaspi, Galanti, Hubbard, Stevenson, Bolton,
  Iess, Guillot, Bloxham, Connerney, \& Cao}]{kaspiJupitersAtmosphericJet2018}
Kaspi, Y., Galanti, E., Hubbard, W.~B., {et~al.} 2018,
  \bibinfo{title}{Jupiter's Atmospheric Jet Streams Extend Thousands of
  Kilometres Deep,} Nature, 555, 223, \dodoi{10.1038/nature25793}

\bibitem[{Y. Kaspi {et~al.}(2023)Kaspi, Galanti, Park, Duer, Gavriel, Durante,
  Iess, Parisi, Buccino, Guillot, Stevenson, \&
  Bolton}]{kaspiObservationalEvidenceCylindrically2023}
Kaspi, Y., Galanti, E., Park, R.~S., {et~al.} 2023,
  \bibinfo{title}{Observational Evidence for Cylindrically Oriented Zonal Flows
  on {{Jupiter}},} Nature Astronomy, 7, 1463,
  \dodoi{10.1038/s41550-023-02077-8}

\bibitem[{H. Kimura \& I. Mann(1998)Kimura \&
  Mann}]{kimuraElectricChargingInterstellar1998}
Kimura, H., \& Mann, I. 1998, \bibinfo{title}{The {{Electric Charging}} of
  {{Interstellar Dust}} in the {{Solar System}} and {{Consequences}} for {{Its
  Dynamics}},} The Astrophysical Journal, 499, 454, \dodoi{10.1086/305613}

\bibitem[{Y. Kudriavtsev {et~al.}(2005)Kudriavtsev, Villegas, Godines, \&
  Asomoza}]{kudriavtsevCalculationSurfaceBinding2005}
Kudriavtsev, Y., Villegas, A., Godines, A., \& Asomoza, R. 2005,
  \bibinfo{title}{Calculation of the Surface Binding Energy for Ion Sputtered
  Particles,} Applied Surface Science, 239, 273,
  \dodoi{10.1016/j.apsusc.2004.06.014}

\bibitem[{P. Langevin(1903)Langevin}]{langevinRecombinaisonMobilitesIons1903}
Langevin, P. 1903, \bibinfo{title}{Recombinaison et Mobilites Des Ions Dans Les
  Gaz,} Ann. Chim. Phys, 28, 122

\bibitem[{E. Lee {et~al.}(2015)Lee, Helling, Giles, \&
  Bromley}]{leeDustBrownDwarfs2015}
Lee, E., Helling, C., Giles, H., \& Bromley, S.~T. 2015, \bibinfo{title}{Dust
  in Brown Dwarfs and Extra-Solar Planets - {{IV}}. {{Assessing TiO2}} and
  {{SiO}} Nucleation for Cloud Formation Modelling,} Astronomy \& Astrophysics,
  575, A11, \dodoi{10.1051/0004-6361/201424621}

\bibitem[{G. Lee {et~al.}(2015)Lee, Helling, {Dobbs-Dixon}, \&
  Juncher}]{leeModellingLocalGlobal2015}
Lee, G., Helling, C., {Dobbs-Dixon}, I., \& Juncher, D. 2015,
  \bibinfo{title}{Modelling the Local and Global Cloud Formation on {{HD}}
  189733b,} Astronomy \& Astrophysics, 580, A12,
  \dodoi{10.1051/0004-6361/201525982}

\bibitem[{J.~S.
  Lewis(1969)Lewis}]{lewisObservabilitySpectroscopicallyActive1969}
Lewis, J.~S. 1969, \bibinfo{title}{Observability of Spectroscopically Active
  Compounds in the Atmosphere of {{Jupiter}},} Icarus, 10, 393,
  \dodoi{10.1016/0019-1035(69)90094-3}

\bibitem[{C. Li {et~al.}(2023)Li, {de Pater}, Moeckel, Sault, Butler, {deBoer},
  \& Zhang}]{liLonglastingDeepEffect2023}
Li, C., {de Pater}, I., Moeckel, C., {et~al.} 2023,
  \bibinfo{title}{Long-Lasting, Deep Effect of {{Saturn}}'s Giant Storms,}
  Science Advances, 9, eadg9419, \dodoi{10.1126/sciadv.adg9419}

\bibitem[{C. Li {et~al.}(2018)Li, Le, Zhang, \&
  Yung}]{liHighperformanceAtmosphericRadiation2018}
Li, C., Le, T., Zhang, X., \& Yung, Y.~L. 2018, \bibinfo{title}{A
  High-Performance Atmospheric Radiation Package: {{With}} Applications to the
  Radiative Energy Budgets of Giant Planets,} Journal of Quantitative
  Spectroscopy and Radiative Transfer, 217, 353,
  \dodoi{10.1016/j.jqsrt.2018.06.002}

\bibitem[{C. Li {et~al.}(2017)Li, Ingersoll, Janssen, Levin, Bolton,
  Adumitroaie, Allison, Arballo, Bellotti, \&
  Brown}]{liDistributionAmmoniaJupiter2017}
Li, C., Ingersoll, A., Janssen, M., {et~al.} 2017, \bibinfo{title}{The
  Distribution of Ammonia on {{Jupiter}} from a Preliminary Inversion of
  {{Juno}} Microwave Radiometer Data,} Geophysical Research Letters, 44, 5317,
  \dodoi{10.1002/2017GL073159}

\bibitem[{C. Li {et~al.}(2020)Li, Ingersoll, Bolton, Levin, Janssen, Atreya,
  Lunine, Steffes, Brown, Guillot, Allison, Arballo, Bellotti, Adumitroaie,
  Gulkis, Hodges, Li, Misra, Orton, Oyafuso, {Santos-Costa}, Waite, \&
  Zhang}]{liWaterAbundanceJupiters2020}
Li, C., Ingersoll, A., Bolton, S., {et~al.} 2020, \bibinfo{title}{The Water
  Abundance in {{Jupiter}}'s Equatorial Zone,} Nature Astronomy, 4, 609,
  \dodoi{10.1038/s41550-020-1009-3}

\bibitem[{C. Li {et~al.}(2024)Li, Allison, Atreya, Brueshaber, Fletcher,
  Guillot, Li, Lunine, Miguel, Orton, Steffes, Waite, Wong, Levin, \&
  Bolton}]{liSuperadiabaticTemperatureGradient2024}
Li, C., Allison, M., Atreya, S., {et~al.} 2024, \bibinfo{title}{Super-Adiabatic
  Temperature Gradient at {{Jupiter}}'s Equatorial Zone and Implications for
  the Water Abundance,} Icarus, 414, 116028,
  \dodoi{10.1016/j.icarus.2024.116028}

\bibitem[{L. Li {et~al.}(2018)Li, Jiang, West, Gierasch, {Perez-Hoyos},
  {Sanchez-Lavega}, Fletcher, Fortney, Knowles, Porco, Baines, Fry, Mallama,
  Achterberg, Simon, Nixon, Orton, Dyudina, Ewald, \&
  Schmude}]{liLessAbsorbedSolar2018}
Li, L., Jiang, X., West, R.~A., {et~al.} 2018, \bibinfo{title}{Less Absorbed
  Solar Energy and More Internal Heat for {{Jupiter}},} Nature Communications,
  9, 3709, \dodoi{10.1038/s41467-018-06107-2}

\bibitem[{M.~R. Line {et~al.}(2017)Line, Marley, Liu, Burningham, Morley,
  Hinkel, Teske, Fortney, Freedman, \&
  Lupu}]{lineUniformAtmosphericRetrieval2017}
Line, M.~R., Marley, M.~S., Liu, M.~C., {et~al.} 2017, \bibinfo{title}{Uniform
  {{Atmospheric Retrieval Analysis}} of {{Ultracool Dwarfs}}. {{II}}.
  {{Properties}} of 11 {{T}} Dwarfs,} The Astrophysical Journal, 848, 83,
  \dodoi{10.3847/1538-4357/aa7ff0}

\bibitem[{J. Liu {et~al.}(2013)Liu, Schneider, \&
  Kaspi}]{liuPredictionsThermalGravitational2013a}
Liu, J., Schneider, T., \& Kaspi, Y. 2013, \bibinfo{title}{Predictions of
  Thermal and Gravitational Signals of {{Jupiter}}'s Deep Zonal Winds,} Icarus,
  224, 114, \dodoi{10.1016/j.icarus.2013.01.025}

\bibitem[{K. Lodders(1999)Lodders}]{loddersAlkaliElementChemistry1999}
Lodders, K. 1999, \bibinfo{title}{Alkali {{Element Chemistry}} in {{Cool Dwarf
  Atmospheres}},} The Astrophysical Journal, 519, 793, \dodoi{10.1086/307387}

\bibitem[{K. Lodders \& B. Fegley(2006)Lodders \&
  Fegley}]{loddersChemistryLowMass2006}
Lodders, K., \& Fegley, B. 2006, \bibinfo{title}{Chemistry of {{Low Mass
  Substellar Objects}},} in Astrophysics {{Update}} 2, ed. J.~W. Mason (Berlin,
  Heidelberg: Springer), 1--28, \dodoi{10.1007/3-540-30313-8_1}

\bibitem[{J.~I. Lunine {et~al.}(2004)Lunine, Coradini, Gautier, Owen, \&
  Wuchterl}]{lunineOriginJupiter2004}
Lunine, J.~I., Coradini, A., Gautier, D., Owen, T.~C., \& Wuchterl, G. 2004,
  \bibinfo{title}{The Origin of {{Jupiter}},} in Jupiter. {{The Planet}},
  {{Satellites}} and {{Magnetosphere}}, Vol.~1 (Cambridge University Press),
  19--34

\bibitem[{J.~I. Lunine {et~al.}(1986)Lunine, Hubbard, \&
  Marley}]{lunineEvolutionInfraredSpectra1986}
Lunine, J.~I., Hubbard, W.~B., \& Marley, M.~S. 1986, \bibinfo{title}{Evolution
  and {{Infrared Spectra}} of {{Brown Dwarfs}},} The Astrophysical Journal,
  310, 238, \dodoi{10.1086/164678}

\bibitem[{R.~I. Masel(1996)Masel}]{maselPrinciplesAdsorptionReaction1996}
Masel, R.~I. 1996, Principles of {{Adsorption}} and {{Reaction}} on {{Solid
  Surfaces}} (New York: Wiley-Interscience)

\bibitem[{B.~E. Miles {et~al.}(2023)Miles, Biller, Patapis, Worthen, Rickman,
  Hoch, Skemer, Perrin, Whiteford, Chen, Sargent, Mukherjee, Morley, Moran,
  Bonnefoy, Petrus, Carter, Choquet, Hinkley, {Ward-Duong}, Leisenring,
  {Millar-Blanchaer}, Pueyo, Ray, Sallum, Stapelfeldt, Stone, Wang, Absil,
  Balmer, Boccaletti, Bonavita, Booth, Bowler, Chauvin, Christiaens, Currie,
  Danielski, Fortney, Girard, Grady, Greenbaum, Henning, Hines, Janson, Kalas,
  Kammerer, Kennedy, Kenworthy, Kervella, Lagage, Lew, Liu, Macintosh, Marino,
  Marley, Marois, Matthews, Matthews, Mawet, McElwain, Metchev, Meyer,
  Molliere, Pantin, Quirrenbach, Rebollido, Ren, Schneider, Vasist, Wyatt,
  Zhou, Briesemeister, Bryan, Calissendorff, Cantalloube, Cugno, De~Furio,
  Dupuy, Factor, Faherty, Fitzgerald, Franson, Gonzales, Hood, Howe, Kraus,
  Kuzuhara, Lagrange, Lawson, Lazzoni, Liu, {Llop-Sayson}, Lloyd, Martinez,
  Mazoyer, Quanz, Redai, Samland, Schlieder, Tamura, Tan, Uyama, Vigan, Vos,
  Wagner, Wolff, Ygouf, Zhang, Zhang, \&
  Zhang}]{milesJWSTEarlyreleaseScience2023}
Miles, B.~E., Biller, B.~A., Patapis, P., {et~al.} 2023, \bibinfo{title}{The
  {{JWST Early-release Science Program}} for {{Direct Observations}} of
  {{Exoplanetary Systems II}}: {{A}} 1 to 20 {$\mu$}m {{Spectrum}} of the
  {{Planetary-mass Companion VHS}} 1256--1257 b,} The Astrophysical Journal
  Letters, 946, L6, \dodoi{10.3847/2041-8213/acb04a}

\bibitem[{H.~M. {Mott-Smith} \& I. Langmuir(1926){Mott-Smith} \&
  Langmuir}]{mott-smithTheoryCollectorsGaseous1926}
{Mott-Smith}, H.~M., \& Langmuir, I. 1926, \bibinfo{title}{The {{Theory}} of
  {{Collectors}} in {{Gaseous Discharges}},} Physical Review, 28, 727,
  \dodoi{10.1103/PhysRev.28.727}

\bibitem[{O. Mousis {et~al.}(2019)Mousis, Ronnet, \&
  Lunine}]{mousisJupitersFormationVicinity2019}
Mousis, O., Ronnet, T., \& Lunine, J.~I. 2019, \bibinfo{title}{Jupiter's
  {{Formation}} in the {{Vicinity}} of the {{Amorphous Ice Snowline}},} The
  Astrophysical Journal, 875, 9, \dodoi{10.3847/1538-4357/ab0a72}

\bibitem[{K. Ohno \& S. Okuzumi(2018)Ohno \&
  Okuzumi}]{ohnoMicrophysicalModelingMineral2018}
Ohno, K., \& Okuzumi, S. 2018, \bibinfo{title}{Microphysical {{Modeling}} of
  {{Mineral Clouds}} in {{GJ1214}} b and {{GJ436}} b: {{Predicting Upper
  Limits}} on the {{Cloud-top Height}},} The Astrophysical Journal, 859, 34,
  \dodoi{10.3847/1538-4357/aabee3}

\bibitem[{S. Okuzumi(2009)Okuzumi}]{okuzumiELECTRICCHARGINGDUST2009}
Okuzumi, S. 2009, \bibinfo{title}{{{ELECTRIC CHARGING OF DUST AGGREGATES AND
  ITS EFFECT ON DUST COAGULATION IN PROTOPLANETARY DISKS}},} The Astrophysical
  Journal, 698, 1122, \dodoi{10.1088/0004-637X/698/2/1122}

\bibitem[{S. Okuzumi {et~al.}(2009)Okuzumi, Tanaka, \&
  Sakagami}]{okuzumiNUMERICALMODELINGCOAGULATION2009}
Okuzumi, S., Tanaka, H., \& Sakagami, M.-a. 2009, \bibinfo{title}{{{NUMERICAL
  MODELING OF THE COAGULATION AND POROSITY EVOLUTION OF DUST AGGREGATES}},} The
  Astrophysical Journal, 707, 1247, \dodoi{10.1088/0004-637X/707/2/1247}

\bibitem[{C.~W. Ormel \& M. Min(2019)Ormel \&
  Min}]{ormelARCiSFrameworkExoplanet2019}
Ormel, C.~W., \& Min, M. 2019, \bibinfo{title}{{{ARCiS}} Framework for
  Exoplanet Atmospheres - {{The}} Cloud Transport Model,} Astronomy \&
  Astrophysics, 622, A121, \dodoi{10.1051/0004-6361/201833678}

\bibitem[{F. Oyafuso {et~al.}(2020)Oyafuso, Levin, Orton, Brown, Adumitroaie,
  Janssen, Wong, Fletcher, Steffes, Li, Gulkis, Atreya, Misra, \&
  Bolton}]{oyafusoAngularDependenceSpatial2020}
Oyafuso, F., Levin, S., Orton, G., {et~al.} 2020, \bibinfo{title}{Angular
  {{Dependence}} and {{Spatial Distribution}} of {{Jupiter}}'s
  {{Centimeter-Wave Thermal Emission From Juno}}'s {{Microwave Radiometer}},}
  Earth and Space Science, 7, e2020EA001254, \dodoi{10.1029/2020EA001254}

\bibitem[{D. Powell {et~al.}(2022)Powell, Gao, {Murray-Clay}, \&
  Zhang}]{powellDepletionGaseousCO2022}
Powell, D., Gao, P., {Murray-Clay}, R., \& Zhang, X. 2022,
  \bibinfo{title}{Depletion of Gaseous {{CO}} in Protoplanetary Disks by
  Surface-Energy-Regulated Ice Formation,} Nature Astronomy, 6, 1147,
  \dodoi{10.1038/s41550-022-01741-9}

\bibitem[{D. Powell {et~al.}(2018)Powell, Zhang, Gao, \&
  Parmentier}]{powellFormationSilicateTitanium2018}
Powell, D., Zhang, X., Gao, P., \& Parmentier, V. 2018,
  \bibinfo{title}{Formation of {{Silicate}} and {{Titanium Clouds}} on {{Hot
  Jupiters}},} The Astrophysical Journal, 860, 18,
  \dodoi{10.3847/1538-4357/aac215}

\bibitem[{R.~G. Prinn \& S.~S. Barshay(1977)Prinn \&
  Barshay}]{prinnCarbonMonoxideJupiter1977}
Prinn, R.~G., \& Barshay, S.~S. 1977, \bibinfo{title}{Carbon {{Monoxide}} on
  {{Jupiter}} and {{Implications}} for {{Atmospheric Convection}},} Science,
  198, 1031, \dodoi{10.1126/science.198.4321.1031}

\bibitem[{W.~B. Rossow(1978)Rossow}]{rossowCloudMicrophysicsAnalysis1978}
Rossow, W.~B. 1978, \bibinfo{title}{Cloud Microphysics: {{Analysis}} of the
  Clouds of {{Earth}}, {{Venus}}, {{Mars}} and {{Jupiter}},} Icarus, 36, 1

\bibitem[{J.~H. Seinfeld \& S.~N. Pandis(2016)Seinfeld \&
  Pandis}]{seinfeldAtmosphericChemistryPhysics2016}
Seinfeld, J.~H., \& Pandis, S.~N. 2016, Atmospheric Chemistry and Physics: From
  Air Pollution to Climate Change (John Wiley \& Sons)

\bibitem[{A.~P. Showman {et~al.}(2011)Showman, Kaspi, \&
  Flierl}]{showmanScalingLawsConvection2011}
Showman, A.~P., Kaspi, Y., \& Flierl, G.~R. 2011, \bibinfo{title}{Scaling Laws
  for Convection and Jet Speeds in the Giant Planets,} Icarus, 211, 1258,
  \dodoi{10.1016/j.icarus.2010.11.004}

\bibitem[{C.~R. Stark {et~al.}(2015)Stark, Helling, \&
  Diver}]{starkInhomogeneousCloudCoverage2015}
Stark, C.~R., Helling, C., \& Diver, D.~A. 2015, \bibinfo{title}{Inhomogeneous
  Cloud Coverage through the {{Coulomb}} Explosion of Dust in Substellar
  Atmospheres,} Astronomy \& Astrophysics, 579, A41,
  \dodoi{10.1051/0004-6361/201526045}

\bibitem[{R. Tejada~Arevalo {et~al.}(2024)Tejada~Arevalo, Su, Sur, \&
  Burrows}]{tejadaarevaloEquationsStateThermodynamics2024}
Tejada~Arevalo, R., Su, Y., Sur, A., \& Burrows, A. 2024,
  \bibinfo{title}{Equations of {{State}}, {{Thermodynamics}}, and {{Miscibility
  Curves}} for {{Jovian Planet}} and {{Giant Exoplanet Evolutionary Models}},}
  The Astrophysical Journal Supplement Series, 274, 34,
  \dodoi{10.3847/1538-4365/ad6cd7}

\bibitem[{S.~A. Utlak \& T.~M. Besmann(2018)Utlak \&
  Besmann}]{utlakThermodynamicAssessmentPseudoternary2018}
Utlak, S.~A., \& Besmann, T.~M. 2018, \bibinfo{title}{Thermodynamic Assessment
  of the Pseudoternary {{Na2O}}--{{Al2O3}}--{{SiO2}} System,} Journal of the
  American Ceramic Society, 101, 928, \dodoi{10.1111/jace.15166}

\bibitem[{C. Visscher {et~al.}(2010)Visscher, Lodders, \&
  Fegley~Jr}]{visscherAtmosphericChemistryGiant2010}
Visscher, C., Lodders, K., \& Fegley~Jr, B. 2010, \bibinfo{title}{Atmospheric
  Chemistry in Giant Planets, Brown Dwarfs, and Low-Mass Dwarf Stars. {{III}}.
  {{Iron}}, Magnesium, and Silicon,} The Astrophysical Journal, 716, 1060

\bibitem[{E. Vogt \& G.~H. Wannier(1954)Vogt \&
  Wannier}]{vogtScatteringIonsPolarization1954}
Vogt, E., \& Wannier, G.~H. 1954, \bibinfo{title}{Scattering of {{Ions}} by
  {{Polarization Forces}},} Physical Review, 95, 1190,
  \dodoi{10.1103/PhysRev.95.1190}

\bibitem[{D. Wang {et~al.}(2015)Wang, Gierasch, Lunine, \&
  Mousis}]{wangNewInsightsJupiters2015}
Wang, D., Gierasch, P.~J., Lunine, J.~I., \& Mousis, O. 2015,
  \bibinfo{title}{New Insights on {{Jupiter}}'s Deep Water Abundance from
  Disequilibrium Species,} Icarus, 250, 154,
  \dodoi{10.1016/j.icarus.2014.11.026}

\bibitem[{J.~C. Weingartner \& B.~T. Draine(2001)Weingartner \&
  Draine}]{weingartnerPhotoelectricEmissionInterstellar2001}
Weingartner, J.~C., \& Draine, B.~T. 2001, \bibinfo{title}{Photoelectric
  {{Emission}} from {{Interstellar Dust}}: {{Grain Charging andGas Heating}},}
  The Astrophysical Journal Supplement Series, 134, 263, \dodoi{10.1086/320852}

\bibitem[{E.~C. Whipple(1981)Whipple}]{whipplePotentialsSurfacesSpace1981}
Whipple, E.~C. 1981, \bibinfo{title}{Potentials of Surfaces in Space,} Reports
  on Progress in Physics, 44, 1197, \dodoi{10.1088/0034-4885/44/11/002}

\bibitem[{P. Woitke {et~al.}(2018)Woitke, Helling, Hunter, Millard, Turner,
  Worters, Blecic, \& Stock}]{woitkeEquilibriumChemistry1002018}
Woitke, P., Helling, C., Hunter, G.~H., {et~al.} 2018,
  \bibinfo{title}{Equilibrium Chemistry down to 100 {{K}} - {{Impact}} of
  Silicates and Phyllosilicates on the Carbon to Oxygen Ratio,,} Astronomy \&
  Astrophysics, 614, A1, \dodoi{10.1051/0004-6361/201732193}

\bibitem[{M.~H. Wong {et~al.}(2004)Wong, Mahaffy, Atreya, Niemann, \&
  Owen}]{wongUpdatedGalileoProbe2004}
Wong, M.~H., Mahaffy, P.~R., Atreya, S.~K., Niemann, H.~B., \& Owen, T.~C.
  2004, \bibinfo{title}{Updated {{Galileo}} Probe Mass Spectrometer
  Measurements of Carbon, Oxygen, Nitrogen, and Sulfur on {{Jupiter}},} Icarus,
  171, 153, \dodoi{10.1016/j.icarus.2004.04.010}

\bibitem[{B.~V. Yakshinskiy \& T.~E. Madey(2004)Yakshinskiy \&
  Madey}]{yakshinskiyPhotonstimulatedDesorptionNa2004}
Yakshinskiy, B.~V., \& Madey, T.~E. 2004, \bibinfo{title}{Photon-Stimulated
  Desorption of {{Na}} from a Lunar Sample: Temperature-Dependent Effects,}
  Icarus, 168, 53, \dodoi{10.1016/j.icarus.2003.12.007}

\bibitem[{S. Yoneda \& L. Grossman(1995)Yoneda \&
  Grossman}]{yonedaCondensationCaOMgOAl2O3SiO2Liquids1995}
Yoneda, S., \& Grossman, L. 1995, \bibinfo{title}{Condensation of
  {{CaOMgOAl2O3SiO2}} Liquids from Cosmic Gases,} Geochimica et Cosmochimica
  Acta, 59, 3413, \dodoi{10.1016/0016-7037(95)00214-K}

\bibitem[{J.-S. Yoon {et~al.}(2008)Yoon, Song, Han, Hwang, Chang, Lee, \&
  Itikawa}]{yoonCrossSectionsElectron2008}
Yoon, J.-S., Song, M.-Y., Han, J.-M., {et~al.} 2008, \bibinfo{title}{Cross
  {{Sections}} for {{Electron Collisions}} with {{Hydrogen Molecules}},}
  Journal of Physical and Chemical Reference Data, 37, 913,
  \dodoi{10.1063/1.2838023}

\bibitem[{J.~A. Zalesky {et~al.}(2022)Zalesky, Saboi, Line, Zhang, Schneider,
  Liu, Best, \& Marley}]{zaleskyUniformRetrievalAnalysis2022}
Zalesky, J.~A., Saboi, K., Line, M.~R., {et~al.} 2022, \bibinfo{title}{A
  {{Uniform Retrieval Analysis}} of {{Ultra-cool Dwarfs}}. {{IV}}. {{A
  Statistical Census}} from 50 {{Late-T Dwarfs}},} The Astrophysical Journal,
  936, 44, \dodoi{10.3847/1538-4357/ac786c}

\bibitem[{Y.~B. Zeldovich(1943)Zeldovich}]{zeldovichTheoryNewPhase1943}
Zeldovich, Y.~B. 1943, \bibinfo{title}{On the Theory of New Phase Formation:
  Cavitation,} Acta Physicochem., USSR, 18, 1

\bibitem[{X. Zhang(2020)Zhang}]{zhangAtmosphericRegimesTrends2020}
Zhang, X. 2020, \bibinfo{title}{Atmospheric Regimes and Trends on Exoplanets
  and Brown Dwarfs,} Research in Astronomy and Astrophysics, 20, 099,
  \dodoi{10.1088/1674-4527/20/7/99}

\bibitem[{Y. Zhang {et~al.}(2010)Zhang, Ni, \&
  Chen}]{zhangDiffusionDataSilicate2010}
Zhang, Y., Ni, H., \& Chen, Y. 2010, \bibinfo{title}{Diffusion {{Data}} in
  {{Silicate Melts}},} Reviews in Mineralogy and Geochemistry, 72, 311,
  \dodoi{10.2138/rmg.2010.72.8}

\bibitem[{Z. Zhang {et~al.}(2020)Zhang, Adumitroaie, Allison, Arballo, Atreya,
  Bjoraker, Bolton, Brown, Fletcher, Guillot, Gulkis, Hodges, Ingersoll,
  Janssen, Levin, Li, Li, Lunine, Misra, Orton, Oyafuso, Steffes, \&
  Wong}]{zhangResidualStudyTesting2020}
Zhang, Z., Adumitroaie, V., Allison, M., {et~al.} 2020,
  \bibinfo{title}{Residual Study: {{Testing}} Jupiter Atmosphere Models against
  Juno {{MWR}} Observations,} Earth and Space Science, 7, e2020EA001229,
  \dodoi{10.1029/2020EA001229}

\end{thebibliography}
\end{document}